WEIZMANN INSTITUTE OF SCIENCE

# PROTEIN FLEXIBILITY UPON LIGAND BINDING: DOCKING PREDICTIONS AND STATISTICAL ANALYSIS

Thesis for the degree
Doctor of Philosophy

By
Rafael Najmanovich

May – 2003

This work has been performed under the supervision of

Professor Meir Edelman
&
Dr. Vladimir Sobolev

Plant Sciences Department
Weizmann Institute of Science

And

Professor Eytan Domany

Physics of Complex Systems Department
Weizmann Institute of Science

This work is dedicated to the memory of

Esther Szeinfeld Schiftan, my mother.

# Acknowledgements

I would like to start by thanking Prof. Meir Edelman and Dr. Vladimir Sobolev for their guidance and the high level of scientific independence they trusted in me. A different posture would not have permitted my full development as a free thinker.

I would like to thank Professors Ron Unger and Amnon Horovitz for their valuable comments throughout this work as assigned readers of this work.

I would like to thank Prof. Eytan Domany for making it possible for me to start my studies at the Weizmann Institute under his guidance and thus return back home to Israel. I would also like to thank Eytan for all he taught me directly and indirectly as a scientific role model.

I would like thank the support of past and present group member who have contributed throughout the many discussions we held, to improve this work and make our work environment very friendly and enjoyable. This people are Dr. Brendan McConkey, Dr. Alexander Riskind, Dr. Moshe van Dyke, Vladimir Potapov, Mariana Babor and Eran Eyal. I am especially thankful to Eran Eyal for the many fruitful discussions we had and our close collaboration at work.

I would like to thank the multitude of friends I made while studying at the Weizmann Institute of Science. These people are now in almost all corners of the earth and our close friendship is of the deepest importance to me. I am proud of requiring both hands to count the number of close friends I made during my PhD.

I would like to thank my brother Luis Ariel Najmanovich. The close friendship between Luis and I is something that few brothers can be proud to have. His support during my studies were diverse, subtle and indispensable.

I would like to thank my father, Mario Najmanovich, for his support. When all has been done and said, I know I can always count on my family.

Last, I would like to thank someone who will never know that I reached this day. The person that would be the proudest of all in seeing her son complete this step in his life. I thank my mother, Esther Szeinfeld, for the example of courage and determination and for her unshakable belief in me.

# Abstract

Side chain flexibility is an important factor in ligand binding, being partly responsible for the general effect known as induced fit. In order to determine the extent to which side chain flexibility is involved in ligand binding, a knowledge-based approach was taken. A database was generated composed of pairs of files containing the experimentally determined atomic coordinates of a protein in the presence or absence of a given ligand. The database is used to analyze which side chains undergo side chain conformational changes. Such an analysis has determined that up to 40% of binding site do not present side chain conformational changes. A total of three residues undergoing side chain conformational changes encompass approximately 85% of the binding sites studied. When analyzing the propensities of different amino acids to undergo side chain conformational changes we find that there are considerable differences between different amino acids.

The side chain flexibility scale does not provide enough information in order to accurately predict which specific side chains are likely to undergo side chain conformational changes upon ligand binding. For that purpose, a support vector machine learning approach was used to create a classifier system utilizing information about the solvent accessible area as well as flexibility scale value of each specific side chain to be predicted together with its neighboring side chains. An accuracy level of 70% is reached using this approach.

The fact that a small number of residues undergo side chain conformational changes in the majority of binding sites makes it feasible to introduce side chain flexibility in docking simulations. An algorithm has been developed for introducing side chain flexibility utilizing a hybrid genetic-algorithm/exhaustive-search procedure and a surface complementarity based scoring function. This approach is implemented in the software tool FlexAID. FlexAID is able to perform local docking simulations, that is, in a small area around the binding site, as well as global docking simulations, where no information on the location of the binding site is utilized. Moreover, FlexAID utilizes a rotamer library to create alternative conformations for a list of residues that are exhaustively searched during the docking simulation. The performance of FlexAID in rigid local as well as global simulations has been determined. The accuracy obtained falls in the 70-80% range for both local and global simulations.

The performance of FlexAID has not been determined when considering side chain flexibility. However, a few cases were analyzed giving promising results for both local as well as global flexible docking simulations. The alteration of the scoring function to include a repulsive interaction with the solvent considerably improved the performance of FlexAID and might be associated to the dynamics of the genetic algorithm search procedure. Other improvements in the scoring function are necessary to be able to test the performance of FlexAID when allowing flexibility in the ligand molecule as well.

## TABLE OF CONTENTS

# 1. INTRODUCTION

Docking algorithms are a fundamental bioinformatics tool in the study of interactions between ligands and proteins. The task of a docking algorithm is to determine the structure of the ligand-protein complex given the structure of the protein.  A docking algorithm can be seen conceptually divided into three interconnected components: representation, scoring and searching. The representation component includes the choices of how are the ligand and protein molecules represented as well as the level of flexibility associated to each component of the system. Scoring functions are applied to assess the relative quality of different possible solutions. Finally, The space defined by the choices of representation needs to be searched in an effective manner to find the relevant extrema (minima or maxima) using the scoring function. An interesting point to note is that the 'docking problem' is closely related to the 'protein folding problem' in the sense that the latter can also be seen as composed of the same three components and many methodologies developed for the protein folding problem may directly or indirectly be applicable to the docking problem.

Different representations lead to different levels of computational complexity. In terms of flexibility, there are several algorithms that represent both the protein and the ligand molecule as completely rigid (Rigid Ligand Rigid Protein: RLRP). That is to say that the search for the structure of the ligand-protein complex is contained in a six-dimensional space of the relative translation and rotation coordinates of one molecule with respect to the other. The next level of complexity in the representation is to include ligand flexibility (Flexible Ligand Rigid Protein: FLRP), by allowing dihedral angle rotations around ligand single bonds or performing RLRP simulations on different conformers of the ligand. Further on, one needs to consider side chain flexibility as well as backbone movements, either keeping the ligand rigid (RLFP) or flexible (FLFP). The most accurate representation would be one that allows unrestricted side chain, backbone and ligand flexibility.

There are numerous docking algorithms such as Flex (and its derivatives, FlexX, FlexE), DOCK, AutoDOCK, GOLD and many others (for reviews see Halperin et al., 2002; McConkey et al., 2002; Taylor et al., 2002). Most are designed to treat the RLRP and FLRP cases.

The inclusion of flexibility of the receptor is still an area of research in its infancy largely due to the computational complexity of including so many degrees of freedom in the simulations. Some works on docking predictions have been published in recent years which include side chain flexibility as well as backbone movements (for reviews see Carlson, 2002; Halperin et al., 2002). The majority of side chains do not undergo conformational changes upon ligand binding (Najmanovich et al., 2000b), nonetheless, the effect known as induced fit, where either side chains obstruct a binding site and need to be moved to accommodate the ligand or the binding site is too loose and the ligand induces a tightening of the pocket, does occur. Although protein flexibility is important for binding in cases of induced fit, it is my view that docking algorithms with flexible side chains (and backbone loops), will really be of utmost relevance in application to docking of small ligands to low resolution protein structures, as well as structures obtained through homology modelling. Such low resolution and homology modelling docking targets are likely to require in many cases a global search for the binding site that is likely to be unknown as well as the rearrangement of side chains in the binding site to accommodate the ligand.

One of the first attempts to include side chain flexibility in a docking algorithm was that of Leach *et. al.* (Leach, 1994). The author restricts the analysis to the binding site area. A 1.0-1.5 Å grid is used to define possible anchorage points for the ligand. A total of 22 different putative positions for the ligand are generated for each grid point. In the case of a flexible ligand, the same procedure is used for each ligand conformation separately. Receptor flexibility is introduced at the final step using the Dead End Elimination Theorem (DEE) (Desmet et al., 1992) to probe the different rotamers for all side chains in the putative binding site and select those combinations of rotamers that yield low energy complexes with the ligand. Apart from the use of the DEE theorem, the search performed in the docking procedure is not a directed search; rather, it is some sort of enumeration scheme (not an exact enumeration). The author reports that the lowest energy conformations found have small RMSD values with respect to the crystallographic structure (between 0.7-2.5 Å). This study was performed for two ligand-protein complexes using the AMBER force field as the scoring function.

Totrov and Abagyan (Totrov & Abagyan, 1997) developed the ICM method for ligand-protein docking that describe the ligand and protein in terms of a generalized internal coordinates system (this description is equal to that applied in our own docking algorithm). Any set of torsional angles including backbone angles (defining loop movements) can be set flexible. The six rotation and translation coordinates together with the protein internal coordinates set flexible are simulated using the Monte Carlo Method. The dynamics used is such that random rotations are performed on the optimisation variables (angles) followed with a gradient descent energy minimization. The energy of the final structure obtained after rotation and energy minimization is used in the metropolis criterion to determine whether or not to accept the new structure. The authors utilize the ECEPP/3 energy function (Nemethy et al., 1992).

Mangoni *et. al.* (Mangoni et al., 1999) use a modified molecular dynamics procedure to perform the docking procedure. The authors assign different temperatures to the different components of the system (ligand, surface residues, receptor centre of mass) to sample more effectively the conformational space. Flexibility is introduced in the form of reducing force constants associated to dihedral angles such that rotations that were prohibited at the given temperature are permitted. The algorithm is tested on the phophocholine-immunoglobulin McPC603 complex.

Ota and Agard (Ota & Agard, 2001) use molecular dynamics and simulated annealing to produce an ensemble of putative conformations which are used to generate a pseudo-crystallographic electron density map that is used to create a final solution using standard crystallography refinement tools. The authors tested the method on a single complex, that of a viral peptide (VSV8) to the major histocompatibility complex (MHC). Other approaches using either molecular dynamics or Monte Carlo simulations are capable of simulating receptor flexibility (Apostolakis et al., 1998; Trosset & Scheraga, 1999; Broughton, 2000).

Knegtel et. al. (Knegtel et al., 1997) use interaction grids defined with weighted averages with respect to energy and geometry over ensembles of proteins structures to describe the flexibility of proteins keeping the ligand rigid. The protein structures used to create the ensemble may include X-ray crystallography as well as NMR determined structures.

Claussen *et. al.* (Claussen et al., 2001) have expanded the FlexX (Kramer et al., 1999a; Kramer et al., 1999b) algorithm to include protein flexibility. The main idea is to describe the protein structure variations with a set of protein structures representing the flexibility, mutations or alternative models of a protein using a united protein description created from the superimposed structures of the ensemble. Similar parts of the structures are merged whereas dissimilar areas are treated as separate alternatives. Like FlexX, the ligand is broken down into building blocks that are sequentially linked on an initial anchoring block using a set of permitted torsional angles specific for each bond being recreated. The rotational and translational search for the ligand with respect to the protein is performed using the anchoring building block of the ligand. Interactions are calculated with all atoms in the united protein description. Graph theory as well clustering methods are used to make sure that the interaction taken in consideration on any given energy evaluation include atoms that create one single coherent conformation of the protein structure. Taking in account the top ten solutions, FlexE finds a ligand position with an RMSD below the 2.0 Å mark to the reference structure in 67% percent of the cases. Alternatively, docking the ligand using FlexX separately on each protein an overall 63% of cases are found to lie bellow the 2.0 Å mark.

Jones *et. al.* (Jones et al., 1995) developed a genetic algorithm to perform ligand-protein docking simulations. The authors use a scoring function comprising a weighted sum of a hydrogen-bonds component and a van der Waals energy component. A given chromosome is composed of four strings, two integer-coded and two grey-binary coded. The two grey-binary coded strings define values of angles. Each angle is coded using one byte (8 bits) corresponding to 256 different values between 0 and $2\pi$. The second pair of chromosomes is cleverly used to propose putative hydrogen bonds. Each gene is assigned for a possible hydrogen bond donor or acceptor in the ligand molecule and encodes a prospective acceptor or donor in the protein molecule. In order to evaluate the interaction energy (fitness) of a solution (individual), the coded rotatable bond angles are set and a least-square fit algorithm is used to try to accommodate, as much as possible, pairs of atoms interacting through the suggested hydrogen bonds. Although the authors explicitly mention that their algorithm allows for the inclusion of receptor flexibility they don't apply it in the cases shown and do not suggest also how to deal with the problem of properly sampling the considerably larger number of variables involved in such a case.

The present study of side chain flexibility and docking predictions is divided in four parts: a database analysis of side chain flexibility upon ligand binding, the creation of a classifier system based on support vector machines (SVM) to predict which side chains on a docking target are likely to be flexible, the generation of an atomic pairwise scoring function and, finally, the creation of a docking program implementing a genetic-algorithm-based search procedure incorporating side chain flexibility on local (binding site known) and global simulations (whole protein surface is searched) whose input of which side chains are to be set flexible may be manually chosen or utilizing the SVM classifier system.

# 2. Database Analysis

A significant number of protein structures have been determined by X-ray crystallography in both complexed and uncomplexed forms and are available from the macromolecular protein databank (Bernstein et al., 1977). Comparison of these structures is valuable for revealing general features of ligand binding.

Ligand binding may induce rearrangements of the protein structure, in particular, side chain movements. Knowledge of the extent with which side chain rearrangements occur is therefore important to developing improved docking prediction algorithms. Predicting the structure of the complex of a small ligand with a protein (molecular docking) is still a complex task, the two major problems being the definition of an appropriate scoring function of the candidate structures and the size of search space. Assuming that one knows the correct scoring function (as discussed in Petrella et al., 1998), a successful searching procedure should consider the three factors that give rise to the size of search space: the relative positions of ligand and receptor, the different ligand conformations and protein flexibility.

Ligand binding may induce large structural changes in the receptor protein, such as the movement of loops (Fraser et al., 1992; Hecht et al., 1992) or even large domains (Lesk & Chothia, 1988; Ikura et al., 1992). Nevertheless, in most cases changes in backbone structure are negligible and only side chain reorientation (if any) occur upon ligand binding (Katzin et al., 1991; Xu et al., 1992). Consequently, combinatorial approaches making use of side chain rotamer libraries are considered very efficient. Accounting for side chain reorientation during docking procedures is a similar task to that of predicting side chain conformations in homology modeling (Bower et al., 1997; Huang et al., 1998; Samudrala & Moult, 1998).

A fast method for finding the global minimum (or maximum) of the scoring function in side chain rotamer space uses the dead-end elimination theorem (Desmet et al., 1992; Goldstein, 1994; Lasters et al., 1995; De Maeyer et al., 1997). Later, it was shown that this method could miss the global minimum during the searching procedure. Its modification, based on the fuzzy-end elimination theorem (Lasters & Desmet, 1993; Keller et al., 1995), corrects this problem but the size of the search space becomes huge. Thus, additional information that allows the restriction of search space can be very valuable.

In this paper we analyze to what extent amino acid side chains belonging to the binding pocket undergo conformational changes upon binding of a small ligand. We constructed a complete database (3,827 entries) of protein structures in complexed (holo-protein) and uncomplexed (apo-protein) forms from the PDB macromolecular structural databank of March 1999 from which two different non-redundant databases (980 and 353 entries) were defined. The number and type of binding pocket undergoing side chain conformational changes in holo- and apo-proteins were then analyzed.

## *2.1 Database Creation*

An entry in our database is composed of a pair of PDB files representing the same protein, and a ligand ID which is present in at least one of the files and defines the binding site for that entry (e.g., 3pcb 3HB 551 O 2pcd; where the ligand – defined by its 3-letter code, residue number and chain ID, if applicable – is bracketed by the pair of PDB file names).

The first step in creating our database was the selection of ligands. In the PDB, a ligand is described by its three-letter code name and listed as HETATM or ATOM in the coordinates section of an entry. Our analysis is restricted to ligands listed as HETATM (thus excluding nucleic acids and peptides that are listed as ATOM). Furthermore, we excluded from our analysis ligands that are covalently bound to protein atoms as well as $PO_4^-$ and $SO_4^-$ molecules. If different parts of a single ligand appear with different codes they are considered as separate ligands. No distinction was made between ligands diffused into the protein crystal or co-crystallized.

The binding pocket is defined as consisting of those amino acids in contact with ligand atoms (Sobolev et al., 1999). "Interatomic contact" refers to the contact surface between atoms (Sobolev & Edelman, 1995). Two binding pockets are considered different if the list of residues in contact with the ligand differs by one or more residues.

Our goal is to use the present analysis to improve docking algorithms. At the atomic level, numerous locations can be found at the protein surface having surface complementarity for very small ligands. Similarly, non-specific locations can be found for ligands that form only a small number of contacts. We therefore imposed a minimum number of five heavy atoms (viz. non-hydrogen) to the ligand as well as a minimum number of five contact residues between the ligand and the protein. The minimum number of atoms required for the ligand excludes atomic ions and water molecules.

A PDB entry that contains a ligand (as defined above) is termed holo-protein with respect to the ligand in consideration. To be considered as an apo-protein (of a given holo-protein), a PDB entry must have an identical amino acid sequence as that of the holo-protein, and none of its binding pocket residues can be in contact with another ligand that is not also present in the given holo-protein. The last requirement ensures that the only difference between the two proteins in the region of the binding pocket is due to the ligand being considered. Only PDB entries determined by X-ray crystallography to a resolution equal or better than 2.5 Å were used in the present study.

## *2.2 Analysis of Side Chain Rearrangements*

For the purpose of the present analysis, we compared the value of side chain dihedral angles for binding pockets residues in both holo- and apo-protein entries. We define a conformational change to have occurred if a difference larger than a threshold value exists for at least one dihedral angle.

Several studies stress that in known protein structures a particular dihedral angle could differ from the angle determined by the torsional energy minima or from the statistical average by more than ±40° and still belong to the same rotamer (De Maeyer et al., 1997; Petrella et al., 1998). It then follows that occasionally a difference of ~80° can be found for a particular angle with the side chains still belonging to the same rotamer. However, in the great majority of instances a difference of ~80° would indicate structures belonging to different rotamers. For completeness, our analyses were performed at three different threshold values: 45°, 60°, and 75°. The trends for all were very similar. As expected, the higher the threshold value the smaller the number of binding pocket residues which undergo side chain conformational change. However, the differences are not pronounced (e.g., the percentage of binding pockets with up to three flexible residues was about 80% at a threshold of 45°, 85% at 60°, and 90% at 75°). Moreover, the probability for a specific amino acid to change side chain conformation upon ligand binding is insensitive to variation in the threshold. Thus, for clarity, we present only the results for the threshold value of 60°. A similar threshold value has been used in recent analysis of amino acid conformational changes in protein association (Betts & Sternberg, 1999).

## *2.3 Databases*

### *2.3.1 Maximal database*

Using the definitions and rules discussed in Methods, we built a database with 3,827 entries. This maximal (MAX) database is the largest one that meets our criteria. It contains 221 different protein sequences, defining a total of 980 binding pockets and 353 different ligands. The MAX database is composed of blocks of entries, with all PDB files in a given block having the same amino acid sequence (i.e., all deriving from the same protein). Every block is further divided into sub-blocks, with all PDB files in a given sub-block being grouped together because they have the same ligand and list of binding-pocket residues. Blocks and sub-blocks may be composed of single entries. A small section of the MAX database is shown in Figure 1 to help visualize how the apo- and holo-structures are paired for analysis in the derived, non-redundant databases.

The MAX database is large due to the combinatorial nature of the process used to create it: 1. A given protein (apo- and/or holo-form) might have been crystallized several times with the same ligand(s) but under different conditions, giving rise to different PDB entries; or 2. The same PDB entry might be considered as a holo-protein for a given ligand but as apo-protein for another ligand, thus increasing the total number of entries in the database.

The MAX database is clearly redundant yet useful as a starting point for studies related to ligand binding that require structural comparisons between apo- and holo-forms. In this study, we have derived two subsets from the MAX database, each with different minimal redundancy criteria.

```
MAX DATABASE         BPK DATABASE         LIG DATABASE
----------------     ----------------     ----------------
2IZF BTN 01 2IZD     2IZF BTN 01 2IZD
2IZF BTN 01 2IZE
2IZG BTN 01 2IZD
2IZG BTN 01 2IZE
2IZH BTN 01 2IZD
2IZH BTN 01 2IZE

2IZL IMI 01 2IZD     2IZL IMI 01 2IZD
2IZL IMI 01 2IZE                          2IZL IMI 01 2IZE
----------------     ----------------     ----------------
1SWD BTN 01 1SWA     1SWD BTN 01 1SWA     1SWD BTN 01 1SWA
1SWD BTN 01 1SWB
1SWE BTN 01 1SWA
1SWE BTN 01 1SWB

1SWE BTN 02 1SWC     1SWE BTN 02 1SWC

1SWE BTN 04 1SWC     1SWE BTN 04 1SWC

1SWD BTN 02 1SWC     1SWD BTN 02 1SWC
----------------     ----------------     ----------------
```

**Figure 1.** Database structure. An entry in each of the three databases is composed of a holo-protein PDB ID (e.g., 2IZF), followed by a modified ligand ID (e.g., BTN 01) and, finally, an apo-protein PDB ID (e.g., 2IZD). The number in the modified ligand ID numbers the ligand in the order of its appearance in the PDB file. All MAX database entries between two consecutive dashed lines form a block which represents the same protein sequence, while empty lines are used to separate entries, forming sub-blocks, that differ either in the ligand or the set of residues in contact with the ligand. The BPK database was created by choosing one entry from each sub-block (in the example shown, the first entry in each case). The LIG database was created by choosing one entry for each three-letter ligand ID.

### 2.3.2 *Binding pocket database*

The binding pocket database (BPK) consists of one entry for each different binding pocket found in the MAX database. Any two entries in this database must differ in at least one of three factors: the protein sequence, the ligand and the binding pocket. In terms of the structure of the database, applying the above criteria is equivalent to choosing one entry from each sub-block in the MAX database (see Figure 1). In the particular analysis presented, we used the first entry appearing in each sub block. However, the results were essentially the same when we used different choices. The BPK database consists of 980 entries.

### 2.3.3 *Ligand database*

The ligand database (LIG) was created by randomly selecting a single entry for each different ligand present in the MAX database. Having only a single entry for each ligand, stringently avoids the possibility that a set of binding-pocket contacts will be counted more than once. Either because the ligand defining the binding pocket does not duplicate all the contacts in two different entries, or two protein sequences differ in a small number of distal amino acids which do not influence the structure of the binding pocket. On the other hand, in this way we may lose cases in which a given ligand binds genuinely to different binding pockets, as well as cases where the same ligand was crystallized with decidedly different proteins. The LIG database consists of 353 entries.

A potential source of redundancy that was not excluded in either of our sub-databases is that of structural similarity of ligands. Two ligands are considered different if their three-letter PDB codes are not the same. However, there might be several such ligands that are structurally or functionally very similar. It is very difficult to judge in advance the effect on binding of even a single atom difference between ligands.

We present further information with respect to the composition of the three databases in Table I. The three databases mentioned in this study can be obtained at the following URL: http://sgedg.weizmann.ac.il/ferafael/ligdb.html.

Table I. Database Characteristics

| Database | Entries[a] | PDB files[b] | Unique protein sequences |
|---|---|---|---|
| MAX | 3827 | 998 | 221 |
| BPK | 980 | 729 | 221 |
| LIG | 353 | 473 | 154 |

[a] Number of holo- and apo-protein pairs present in each database. The difference between the number of entries and the number of unique protein sequences is mainly due to the frequent existence of several independent PDB entries for the same apo- (or holo-) form of a protein (see example in Figure 1).
[b] Number of different PDB files used to build each database irrespective of their role (apo or holo) in each database.

### *2.4 Analyses*

Two questions were addressed: 1. How many binding pocket residues of paired protein structures undergo side chain conformational changes? 2. Which amino acids are more likely to undergo such changes (i.e., which amino acids are more flexible upon ligand binding)?

#### *2.4.1 Binding pocket flexibility*

We determined the frequency of side chain conformational changes occurring in binding-pocket residues upon ligand binding. Only dihedral angles of rotatable side chain bonds were considered, thus excluding alanine and glycine from the present analysis. Side chain conformational changes in proline were not considered since they are invariably accompanied by changes in backbone conformation. The number of rotatable bonds for each amino acid as well as the total number of residues studied is indicated in Table II.

Table II. Distribution of Flexible Side Chains

| Amino acid[a] | Rotatable bonds | BPK database | | LIG database | |
|---|---|---|---|---|---|
| | | Total | Flexible side chains[c] | Total | Flexible side chains[c] |
| Arg[b] | 4 | 1199 | 285 | 418 | 113 |
| Asn | 2 | 686 | 78 | 248 | 24 |
| Asp | 2 | 893 | 51 | 327 | 19 |
| Cys | 1 | 204 | 4 | 108 | 3 |
| Gln | 3 | 474 | 107 | 205 | 51 |
| Glu | 3 | 795 | 109 | 308 | 48 |
| His | 2 | 862 | 45 | 370 | 27 |
| Ile | 2 | 708 | 89 | 273 | 43 |
| Leu | 2 | 932 | 118 | 381 | 52 |
| Lys | 4 | 653 | 247 | 274 | 111 |
| Met | 3 | 196 | 48 | 92 | 15 |
| Phe | 2 | 616 | 9 | 311 | 5 |
| Ser | 1 | 751 | 46 | 366 | 25 |
| Thr | 1 | 875 | 63 | 338 | 30 |
| Trp | 2 | 674 | 14 | 205 | 4 |
| Tyr | 2 | 955 | 71 | 337 | 23 |
| Val | 1 | 731 | 61 | 349 | 27 |

[a] Side chain conformational changes in proline were not considered since they are invariably accompanied by changes in backbone conformation. Glycine and alanine do not have any rotatable bonds.
[b] Rotation of the NE-CZ bond in Arg was not considered because the CD, NE, CZ, NH1, NH2 atoms form a structure close to planar that practically does not change in shape under various conditions (atom names follow PDB nomenclature).
[c] The probability to undergo side chain conformational change for each amino acid is obtained by dividing the number of flexible side chains by the total number of observed side chains. We assume that this probability is a quantitative measure of flexibility.

The distributions of the number of amino acid residues that undergo side chain conformational change for the BPK and LIG databases are shown in Figure 2. The probability of a pocket to have N residues undergoing conformational change decreases asymptotically with N in such a way that changes in up to three residues account for ~85% of the cases (inset, Fig. 2). This result supports restriction of side chain flexibility to a small number of residues in docking predictions. In addition, 94.4% of $\chi_1$ angles and 95.7% of $\chi_2$ angles (irrespective of the amino acid) do not undergo conformational change. These results are somewhat higher than those presented by Betts & Sternberg, 1999, (83.1% and 87.9% respectively for $\chi_1$ and $\chi_2$ for surface exposed residues) in the case of protein-protein association. This difference suggests that side chains in binding pockets are more rigid than those in protein interfaces (perhaps due to functional constraints in ligand recognition). Although the averages for $\chi_1$ and $\chi_2$ are similar, detailed analysis for each residue show larger differences. In most cases $\chi_1$ is more flexible than $\chi_2$ however, we see the opposite behavior in Asn (96.6% and 92.0%) and Ile (95.9% and 91.5%).

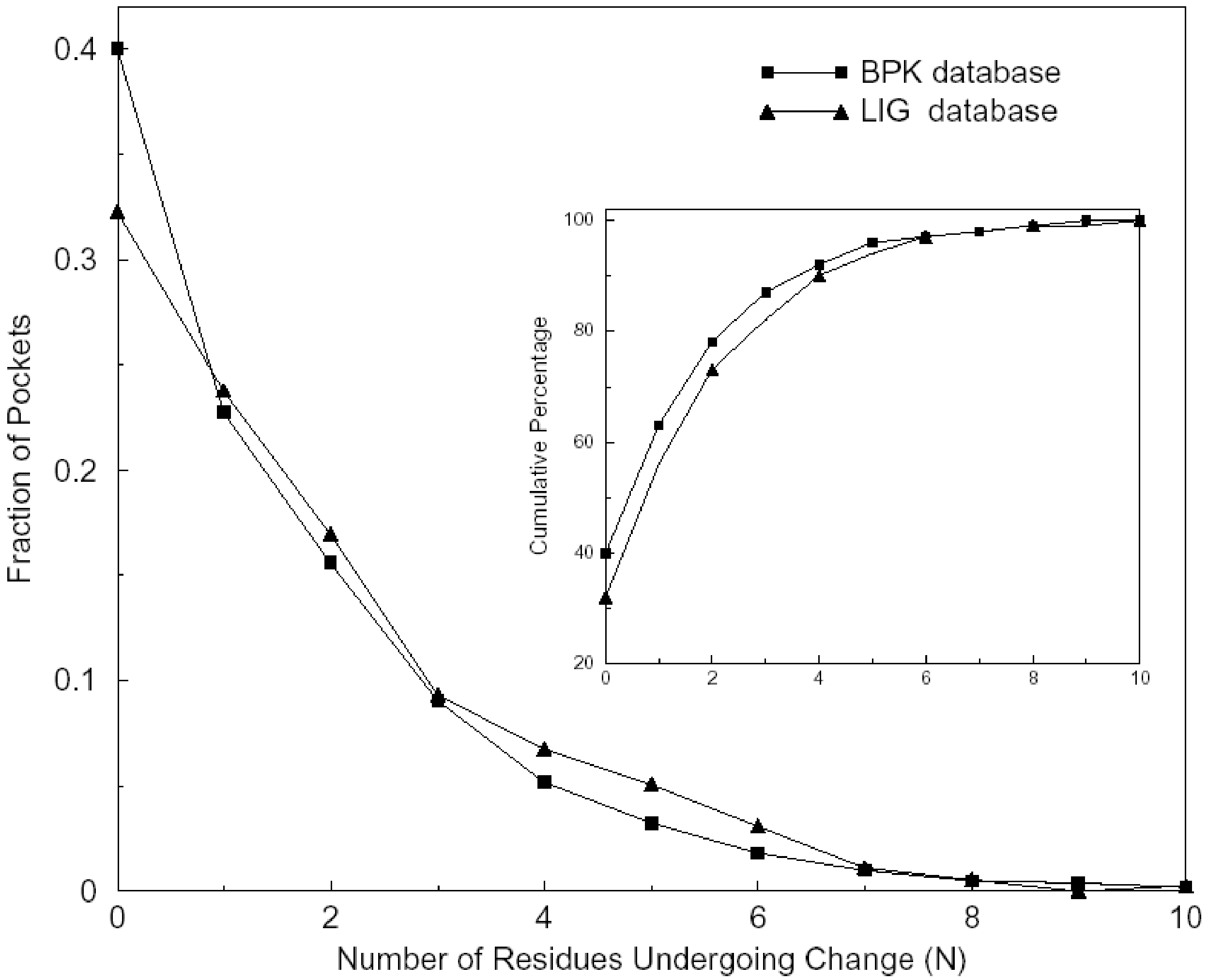

**Figure 2**. Flexibility of binding pockets. Changes in side chain conformations were analyzed in 980 pockets for the BPK database and 353 pockets for the LIG database. The fraction of pockets was plotted versus the number of pocket residues undergoing changes in their side chain conformation upon ligand binding. The inset shows the cumulative percentage of pockets in which not more than the indicated number of residues (co-ordinate axis) undergo conformational change.

### *2.4.2 Amino acid flexibility*

We calculated the probability with which each amino acid undergoes side chain conformational changes as follows:

$$p_i = \frac{N_C^i}{N_T^i} \pm \frac{\sqrt{N_C^i}}{N_T^i} \tag{2.1}$$

where $N_C^i$ is the total number of amino acids of type $i$ undergoing conformational changes and $N_T^i$ is the total number of amino acids of type $i$ present in all binding pockets, the second term is the error estimation involved in the measurement.

Our purpose is to estimate the probability of an amino acid already present in a binding pocket to undergo conformational changes, therefore, we did not normalize $p_i$ by the probability of occurrence of different amino acids in binding pockets.

The data summarized in Table II is sufficient for statistical analysis of flexibility for individual amino acids, including the least frequently occurring one (Met) with 196 cases in the BPK database and 92 cases in the LIG database. Side chain flexibility upon ligand binding, $p_i$ (Equation 2.1), for the amino acids listed in Table II is presented in Figure 3. The results indicate the following order: Lys > Arg, Gln, Met > Glu, Ile, Leu > Asn, Thr, Val, Tyr, Ser, His, Asp > Cys, Trp, Phe; with a 25-fold difference in the probability to undergo side chain conformational changes between Lys and Phe.

The low flexibility of Cys can only be partially explained by its participation in disulfide bonds since, about 50% of the cysteines are involved in disulfide bonds irrespective of whether their side chains undergo conformational change.

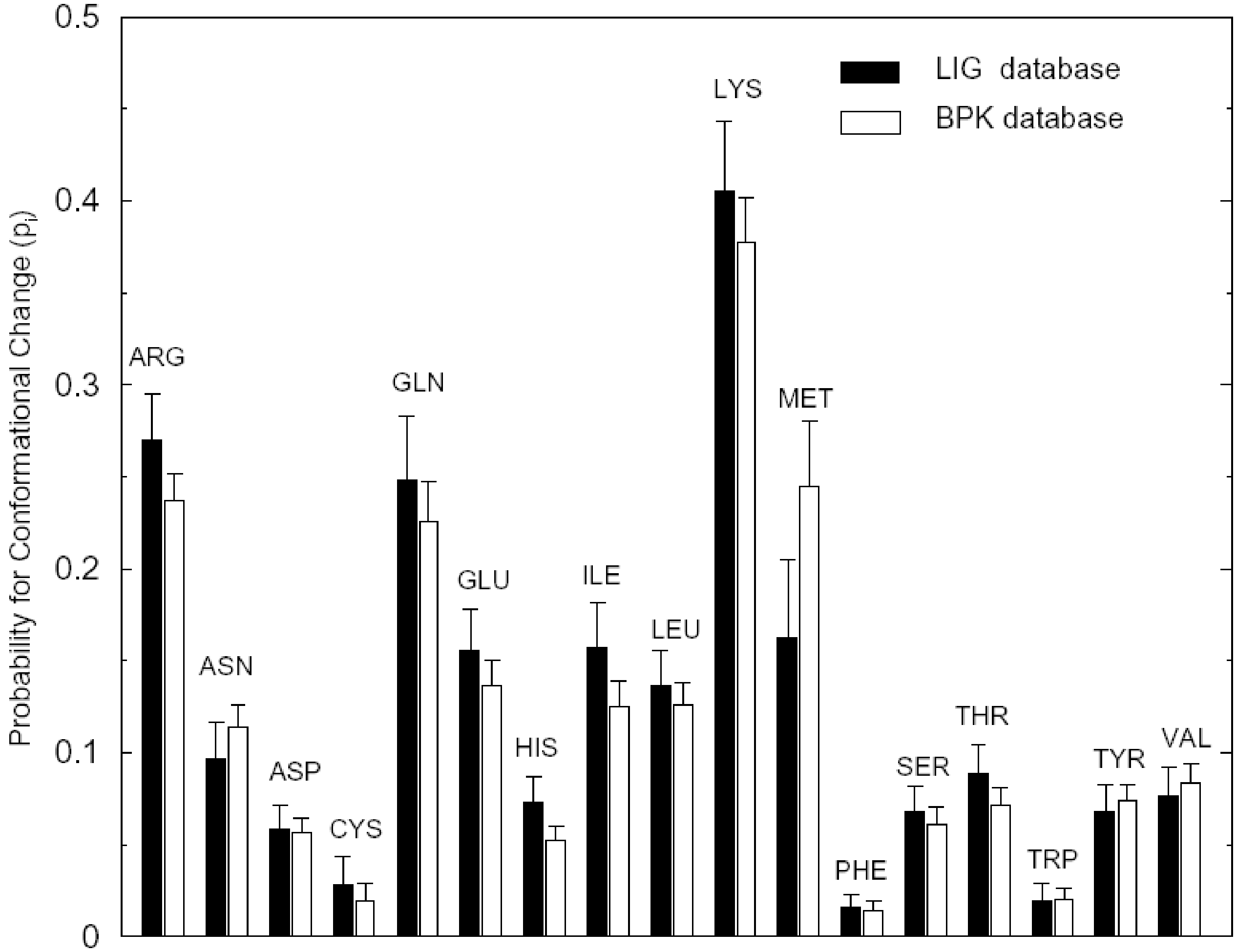

**Figure 3**. Flexibility residue side chains. The probability for side chain conformational change for the different amino acids ( $p_i$ ) is shown for the BPK and LIG databases. Bars represent error estimation according to Equation 2.1.

### 2.4.3 CORRELATION WITH STRUCTURAL AND CHEMICAL FEATURES

We noticed a tendency in the results of Figure 3 for some amino acids with three or four side chain rotatable bonds (such as Arg, Gln, Lys, and Met) to be more flexible, while several amino acids with one or two such bonds (such as Asp, Cys, Phe, and Trp) were more rigid. As a first approximation, one can suppose that the different dihedral angles can rotate independently. We denote p as the probability of a single dihedral angle to undergo conformational changes in amino acid i. In order to estimate p, we denote the probability that a given bond does not undergo conformational change as $1 - p_i$. For an amino acid side chain with $n_d$ flexible dihedral angles, $p_d^i$ is then given by:

$$p_d^i = 1 - \sqrt[n_d]{1 - p_i} \qquad (2.2)$$

The root comes from the fact that there are $n_d$ independent dihedral angles in a given side chain.

Analysing the calculated bond probabilities (Equation 2.2), we still find significant differences in flexibility among the amino acids. Thus, while there is a correlation between the number of flexible dihedral angles and the probability for a side chain to undergo conformational change, in general, differences among amino acids present in Figure 3 are still apparent in Figure 4, although somewhat attenuated. No correlation was observed between side chain flexibility and number of atoms rotated.

Recently, Kawashima and Kanehisa (Kawashima & Kanehisa, 2000) created a database of amino acid indexes containing 437 different sets of values that reflect structural propensities as well as physicochemical properties of the different amino acids. We calculated the correlation between the flexibility scale (Fig. 3 and Table II) and every entry in their database. The highest correlation obtained (0.74) is to an index of average accessible surface area (Janin et al., 1978). Analysis of other high ranking indexes (correlation larger than 0.70) showed clustering in two categories; viz., indexes related to surface accessibility and $\alpha$-helix stability.

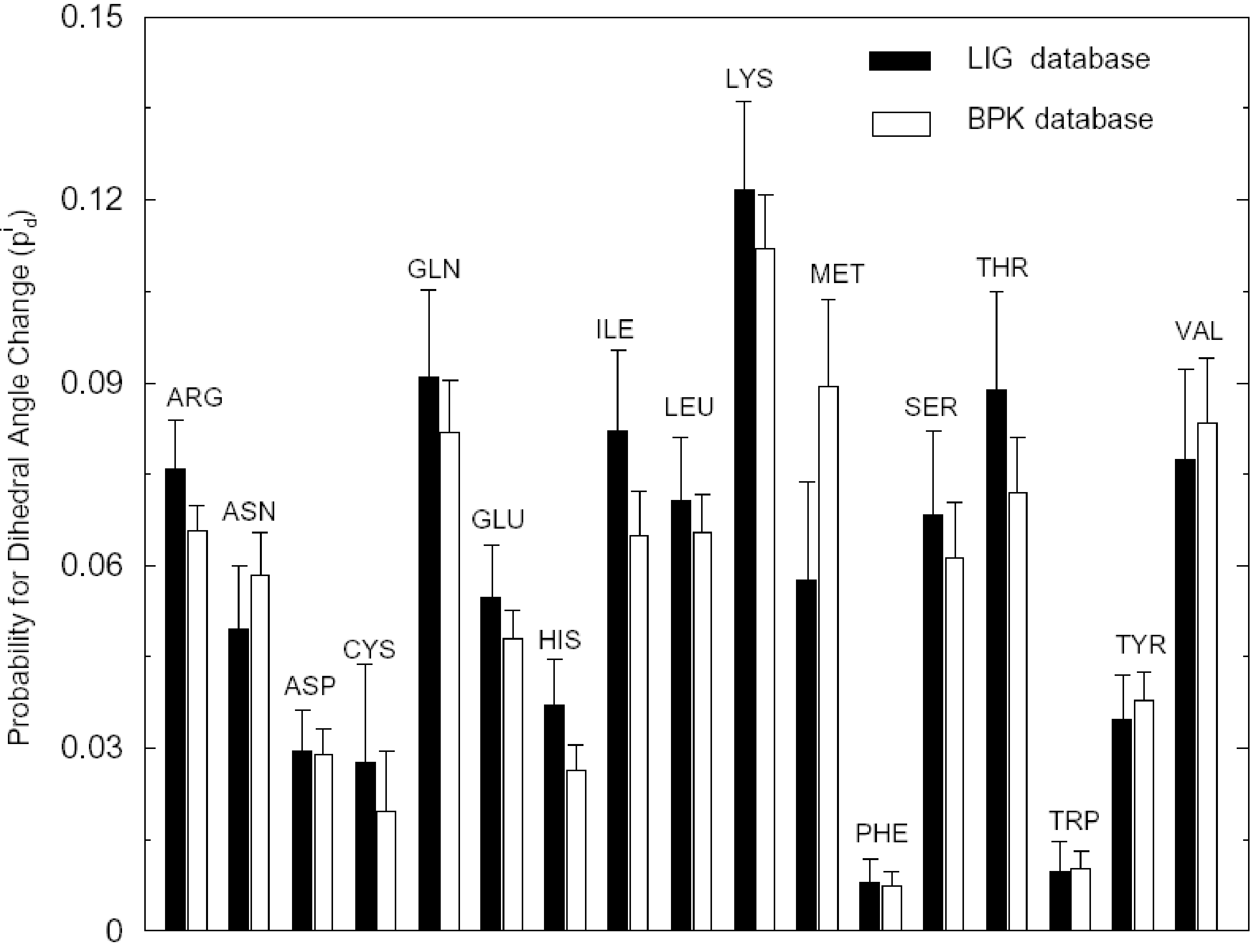

**Figure 4**. Flexibility of dihedral angles. The probability for side chain conformational change for the different amino acids from Figure 3 were used to estimate the probability of a single dihedral angle to undergo change according to Equation 2.2.

### *2.4.4 Backbone versus side chain flexibility*

The parameter we use to estimate the extent of backbone movements of the binding pocket residues is the maximal displacement of $C_\alpha$ atoms ($\Delta d_{\max}$). From the list of residues in contact one calculates all possible pairwise distances, $d_{i,j}$, between $C_\alpha$ atoms in the apo- and holo-protein entries. The value of $\Delta d_{\max}$ is given by:

$$\Delta d_{\max} = \max_{<i,j>} \left| d_{i,j}^{APO} - d_{i,j}^{HOLO} \right| \qquad (2.3)$$

where $i, j$ denotes all pairwise combinations of $C_\alpha$ atoms from residues in contact with the ligand.

Ligand binding can cause large backbone displacements. If side chain flexibility were of minor importance compared to backbone movements, it would be unrealistic to study the former in a database where the latter effect was major. Analysis of the BPK database shows that only 12% of the cases have backbone displacements ($\Delta d_{\max}$) larger than 2 Å; indeed, 75% of the cases show $\Delta d_{\max}$ of less than 1 Å (see Figure 5). Thus, backbone displacements in binding pockets are, on the average, of minor importance when compared to side chain flexibility. Similar conclusions were recently arrived at for protein-protein association (Betts & Sternberg, 1999).

The inset in Figure 5 is an enlargement of the plot of backbone versus side chain movements in the region of 1 to 5 Å of $\Delta d_{\max}$. One can readily see that the data is scattered, suggesting that there is little correlation between backbone displacement and side chain flexibility. We note that in the few cases of very large backbone displacement (i.e., >18 Å) in our database, the fraction of residues undergoing side chain conformational changes is not larger than average. Thus, it is likely that the flexibility of side chains in pockets subject to very large motions does not differ from that of side chains involved in more usual, smaller backbone displacements.

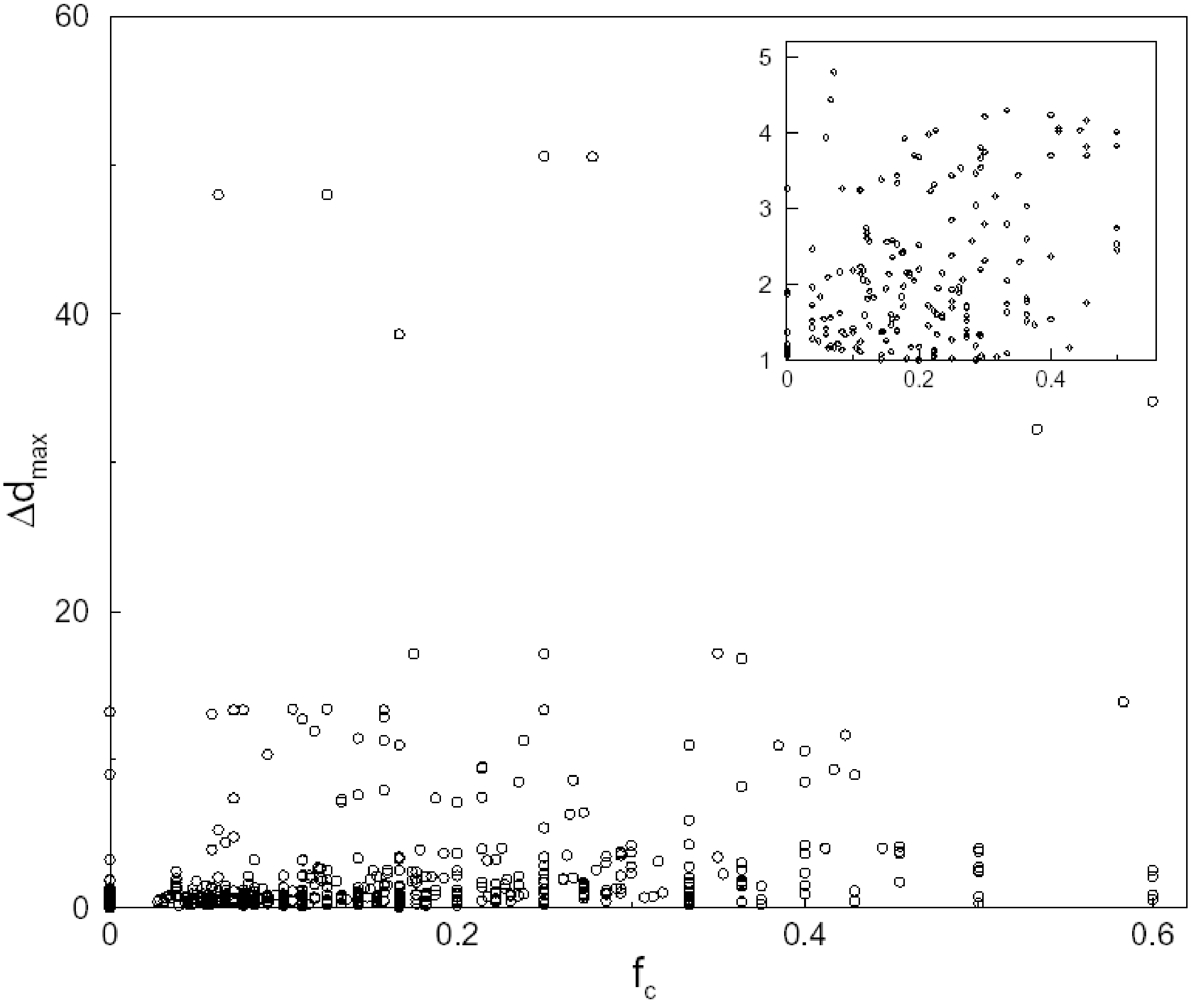

**Figure 5**. Correlation between backbone movements and side chain flexibility. $\Delta d_{\max}$ is the maximal displacement of $C_\alpha$ atoms of the binding pocket (Equation 2.3) while $f_c$ is the fraction of residues undergoing side chain conformational changes. $f_c$ is calculated as the ratio of the number of amino acids undergoing conformational change to the total number of residues in that pocket. Each point represents a single binding pocket. The data shown is for the BPK database. The inset presents a magnified view for the 1-5 Å range of $\Delta d_{\max}$.

### 2.4.5 SIDE CHAIN CONFORMATIONAL CHANGES IN APO PROTEIN ENTRIES

We analysed pairs of apo protein entries belonging to the same sub block on the MAX database to determine to what extent side chain conformational changes are due to ligand binding or to variations among independently determined structures of the same protein. We observed that changes of more than 60° are rarer among apo-apo pairs than among holo-apo ones. However, the flexibility scale for the two cases showed the same trend, suggesting that this is an intrinsic property of the amino acids. This suggestion is further corroborated by the analysis of the side chain flexibility under different circumstances as described in the next section.

### 2.4.6 ALTERNATIVE SIDE CHAIN FLEXIBILITY SCALES

The methodology developed for the analysis of side chain flexibility was subsequently used to analyze side chain flexibility in the vicinity of point mutations (Eyal et al., 2003a). A database of 393 pairs of PDB files was created, each consisting of PDB files differing in a single mutation. It was found that in 91–95% of cases, two or fewer residues underwent side-chain conformational change. If mutation sites with backbone displacements were excluded, the number increased to 97%. The majority of rearrangements (over 60%) were due to the inherent flexibility of side-chains, as derived from analysis of a control set of protein subunits whose crystal structures were determined more than once. Different amino acids demonstrated different degrees of flexibility near mutation sites. Large polar or charged residues, and serine, are more flexible, while the aromatic amino acids, and cysteine, are less so. The probability for conformational change was correlated with B-factor frequency of the side-chain conformation in proteins and solvent accessibility. The last trend was stronger for aromatic and hydrophilic residues than for hydrophobic ones. The relative patterns of side chain flexibility for the three cases studied (upon ligand binding, around mutations and the inherent flexibility) were found to be consistent with each other thus leading to the idea that this pattern is a fundamental characteristic of amino acid side chains (Figure 6). My contribution to this study (Eyal et al., 2003a) was that of creating some of the basic software used in the analysis as well as in the methodology and discussions.

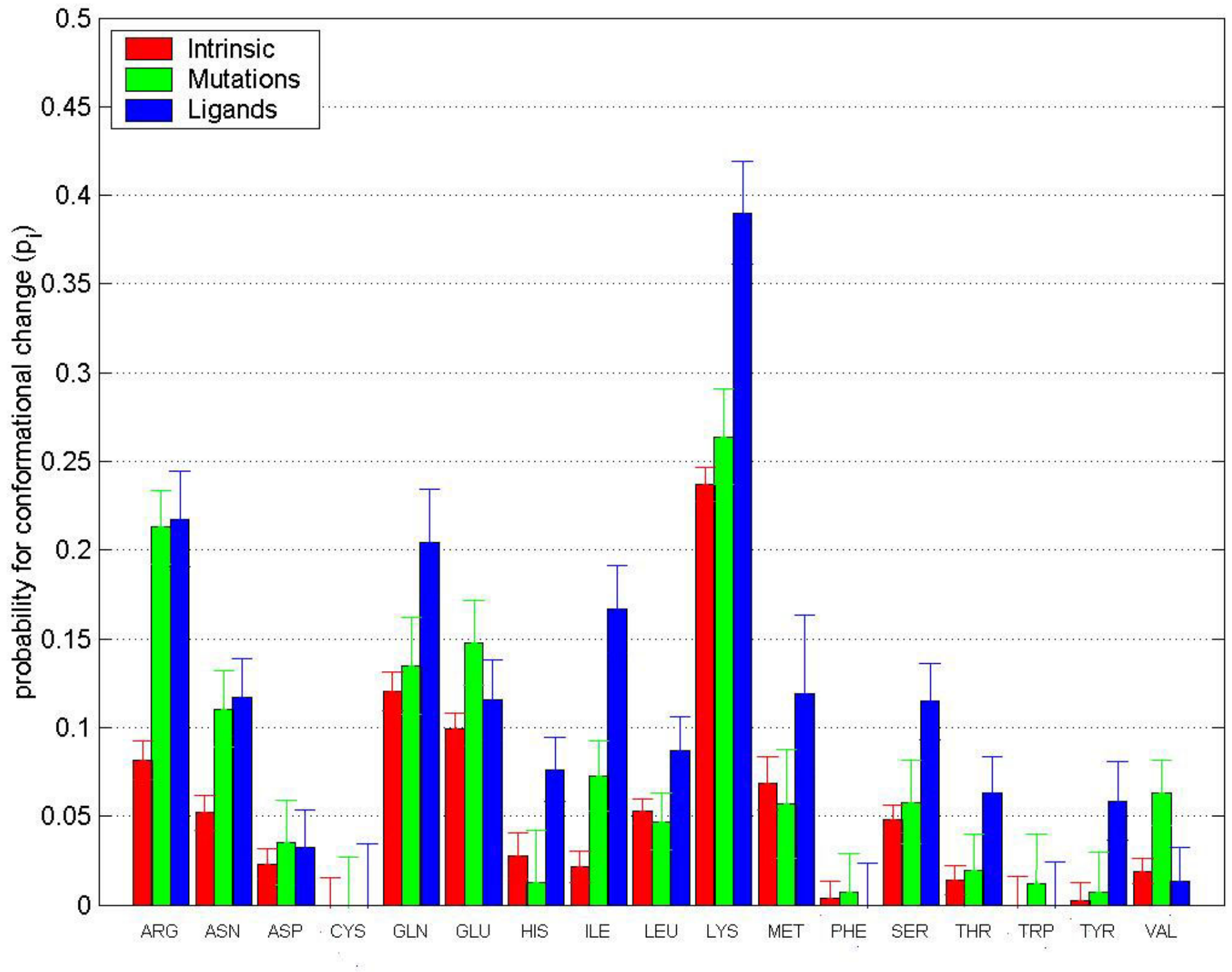

**Figure 6**. Comparison of three scales of side chain flexibility: Inherent flexibility (red) derived from the comparison of pairs of PDB files representing the same protein and cofactors, Flexibility in the vicinity of point mutations (green), derived from the comparison of amino acids in the vicinity of point mutations and Flexibility of binding site side chains. The methodology used is described in Section 2.4.2 while the datasets are described in (Eyal et al., 2003a).

## *2.5 LigProt: A Database and visualization System for the analysis of binding site structural changes*

Our previous study of side chain flexibility upon ligand binding (Najmanovich et al., 2000b) included the development of a database of pairs of PDB files and associated ligand defining holo-apo proteins pairs that were used to study the effect of ligand binding on side chain flexibility. This database was expanded by loosening several constraints necessary in that study and made available on the WWW through a web-interface that allows the user to query the database. LigProt complements other databases related to ligand-protein complexes such as RELIBASE (Hendlich, 1998) and LigBASE (Stuart et al., 2002) in that LigProt allows the retrieval and direct analysis of pairs of PDB files emphasizing differences in binding site occupancy.

### *2.5.1 LigProt Implementation*

The database is implemented in mySQL and the web-interface uses a CGI script to query and parse the database search results. Once an entry is selected for further analysis, the user is presented with links to several other resources containing information about the PDB files, the ligand and its interactions, as well as a structural superimposition of the binding site residues on both PDB files. A LigProt ligand is one whose atoms are listed as HETATM (hetero atom) in the PDB record, excluding water molecules. In the present version, nucleic acids and short peptides are not considered. Entries are generated by selecting a ligand and its associated PDB file (i.e., the holo form) and combining these with all other complementary PDB files which describe the same protein (as defined by sequence comparison, allowing up to two amino acid substitutions). The binding site is defined as the set of residues in contact with the ligand under consideration. In complementary files, it can be occupied by the same ligand (in which case, the two PDB files may differ in other factors such as crystallization conditions, amino acid substitutions, etc.), a different ligand or be empty (apo form, comprising about one third of all entries). Residues in contact with the ligand are determined by the LPC program (Sobolev et al., 1999). Classification with respect to binding site occupancy is crucial when comparing the structural effects due to ligand binding vis-à-vis the different ligands present in the binding site. Only structures determined by x-ray crystallography are utilized in LigProt. Of a total of 3,445 different ligands appearing in the PDB that pass the selection criteria described above, 75% appear in LigProt, the remainder lack a suitable PDB file to serve as the complementary form.

A given PDB file can appear in more than one LigProt entry. For example, a PDB file containing more that one ligand may generate one or more LigProt entries for each of its ligands depending on the number of suitable PDB files which can serve as the complementary form for the given holo protein. Furthermore, the same PDB file that appears as holo protein in one case may serve as complementary form in another. Due to the combinatorial nature of its entries, LigProt currently encompasses 7,800 PDB files (6770 appearing both as holo and complementary forms) and contains over 458,000 entries (about half of which contain one or two amino acid substitutions).

### *2.5.2 Searching*

The LigProt database contains ancillary information such as ligand name, chemical formula, binding mode (covalent or non-covalent, according to a 2 Å distance cut-off), number of ligand atoms, protein name and function, PDB submission date, authors and resolution. All this information can be combined and used to search LigProt. Depending on the complexity of a search query, the results may take several minutes. There is an option to receive the results through email as an attachment in HTML format with all necessary links to LigProt resources.

### *2.5.3 Linked Resources*

Once an entry is chosen for further analysis, the user is presented with links to related resources, such as Pubmed bibliographic search results, retrieval of the PDB records, the PDBsum (Laskowski, 2001) page for each of the PDB records, LPC (Sobolev et al., 1999) analysis of the contacts between the ligand and the holo-protein form, LigPlot (Wallace et al., 1995) graphical representation of ligand protein contacts as well as a 3D and sequence representation of the ligand-protein contacts obtained by LPC and CSU (Sobolev et al., 1999) in the first and second spheres of interacting residues. The holo and complementary forms are superimposed and the user can download the superimposed structure in PDB format, view it using RasMol (Bernstein, 2000) or Chime (MDL information systems, Inc.) and follow a link to a more detailed visualization of the superimposed binding site area. This offers the possibility for direct inspection of the structural differences between the two PDB records present in the LigProt entry. An outline of the LigProt search interface, linked resources and visualization capabilities are given in Figure 7. LigProt is under development, some of the more advanced capabilities are not yet fully operational. LigProt can be found at http://www.weizmann.ac.il/ligprot.

A similar web tool, MutaProt, was previously developed to analyse and visualize regions of mutations (Eyal et al., 2001). I have contributed in this work helping in the development and implementation of the database. MutaProt is accessible through a web interface (http://bioinformatics.weizmann.ac.il/MutaProt) and contains additional data not used in the statistical analysis study.

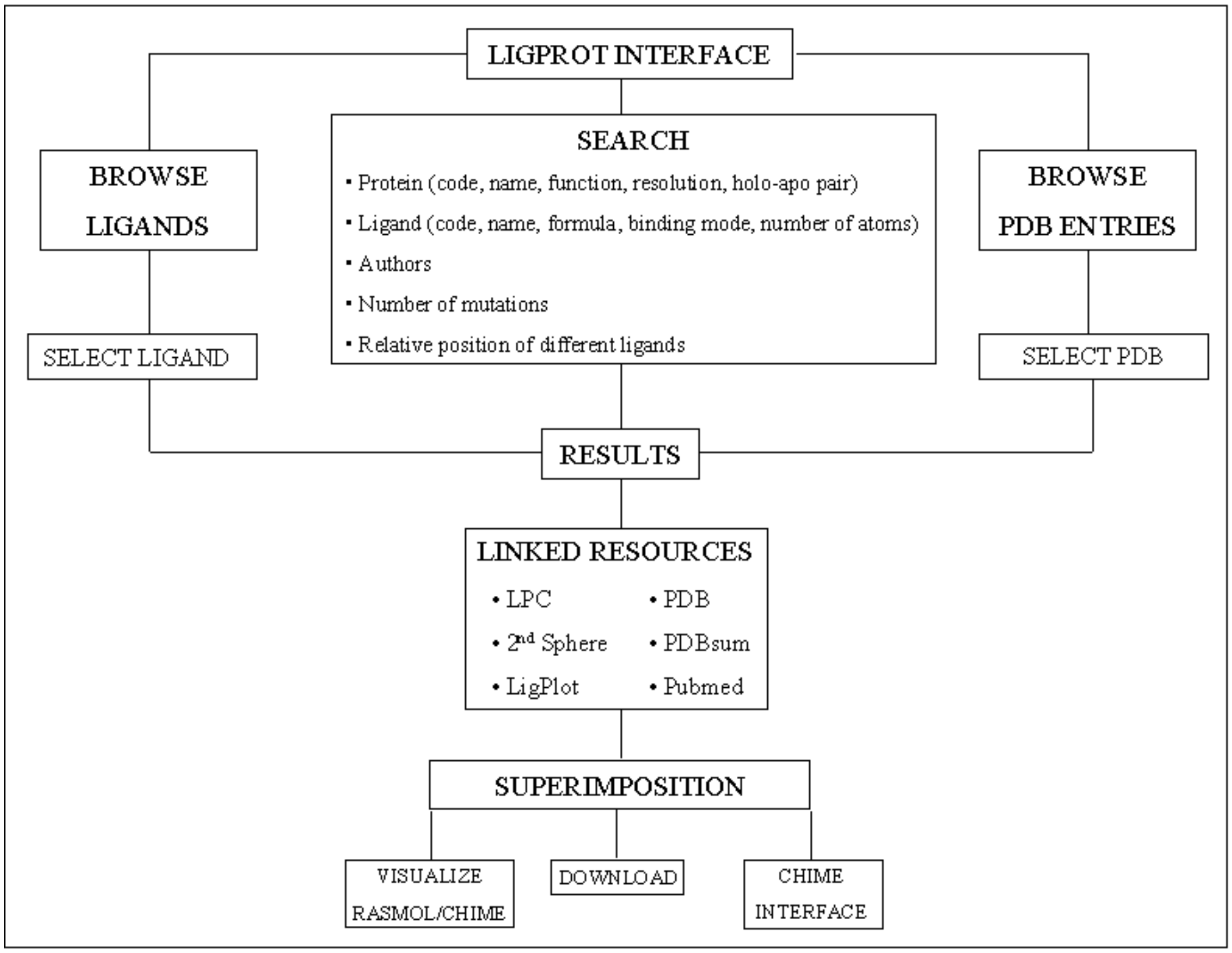

**Figure 7.** LigProt outline. The LigProt database web-interface (http://www.weizmann.ac.il/ligprot) contains ample browsing and searching capabilities for selecting the holo and complementary forms as well as the ligands present in each. Search results are linked to analysis and visualization resources related to the pair of PDB entries, the ligand(s) present and their interactions. The superimposed structure of the two PDB files can be downloaded or visualized using a Chime interface.

# 3. Classification of Side Chain Flexibility using Support Vector Machines

We determined (Najmanovich et al., 2000b) that side chain flexibility is restricted to a small portion of binding site residues, such that consideration of side chain flexibility on up to three binding pocket residues accounts for 85% of the observed cases. Furthermore, we found that different residues have considerably different intrinsic side chain flexibility propensities (Najmanovich et al., 2000b; Eyal et al., 2003a). These two findings suggest that it is possible in principle to include side chain flexibility in a docking algorithm without unduly expanding the time requirements of such a computationally complex task. However, knowing that only a small number of side chains may undergo conformational changes upon binding and even guessing which side chains on a given pocket might be good candidates in terms of side chain flexibility propensities does not help much in the general case to identify as accurately as possible which are in fact the side chains that will undergo conformational changes. So far, all our efforts to rationalize the causes behind side chain flexibility (Najmanovich et al., 2000b; Eyal et al., 2003a) did not yield correlations strong enough to serve as guides in predicting such flexibility. I therefore turned to a machine learning approach to predict side chain flexibility based on examples drawn from our LigProt database.

Machine learning approaches include among others, decision trees, neural networks and support vector machines (Duda et al., 2000). Machine learning algorithms use examples to build an internal model of the phenomena at hand that can be used to classify previously unseen examples. Examples are N-dimensional vectors whose components represent characteristics of interest thought to be relevant for the classification task. Each example used in training contains also a label describing to which class it belongs (supervised learning). In principle, the internal model built by the classifier system is not available to the user and does not help to understand correlations between the different characteristics. Classifier systems are essentially black boxes that once trained can be used to classify new cases (assign labels) without shedding any light on explaining the phenomena. The success of machine learning approaches rests in great part on the appropriate choice of the characteristics used to describe the phenomena (and build the examples). This choice can be seen as a hypothesis. It is up to the researchers to come up with a good hypothesis since machine learning cannot be used for hypothesis generation.

I choose to use support vector machines for the reason that a robust, well-tested implementation is readily available, the $SVM^{light}$ implementation (Joachim, 1999). $SVM^{light}$ has been recently used in different areas of bioinformatics such as classification of protein classes (Cai et al., 2002a; Zavaljevski et al., 2002), prediction of solvent accessibility of amino acids (Yuan et al., 2002) and prediction of protein cellular location (Cai et al., 2002b).

## *3.1 Support Vector Machines*

In what follows I will delineate very briefly the theory behind support vector machines (for a more complete explanation refer to Burges, 1998; Cristianini & Shawe-Taylor, 2000; Duda et al., 2000). Support Vector Machine (SVM) learning theory can be more easily introduced for the separable case, where there is a boundary that perfectly separates the examples into the two classes. The classification task of a SVM is that of finding the largest margin decision boundary that separates the examples such that all examples of a given class are on the same side of the decision boundary (see Figure 8).

The maximal margin hyperplane is defined by the vector $\vec{w}$ normal to the hyperplanes $H_1$ and $H_2$. The maximal margin decision boundary is the one that has the smallest generalization error, thus being the most effective in classifying new, previously unseen examples. The margin itself is equal to $2/|\vec{w}|$ so that in order to maximize the margin, one has to minimize $\|\vec{w}\|^2$ subjected to the constrains of keeping all examples on their appropriate side of the decision boundary:

$$y_i(\vec{x}_i \cdot \vec{w} + b) - 1 \geq 0 \quad \forall i \tag{3.1}$$

where $y_i = \pm 1$ represents the labeling of each example $i$ represented by the vector $\vec{x}_i$.

Positive slack variables $\xi_i$ are introduced for the non-separable case (see Figure 9) to account for misclassified examples on the constrain equations which become:

$$y_i(\vec{x}_i \cdot \vec{w} + b) - 1 + \xi_i \geq 0 \quad \forall i \tag{3.2}$$

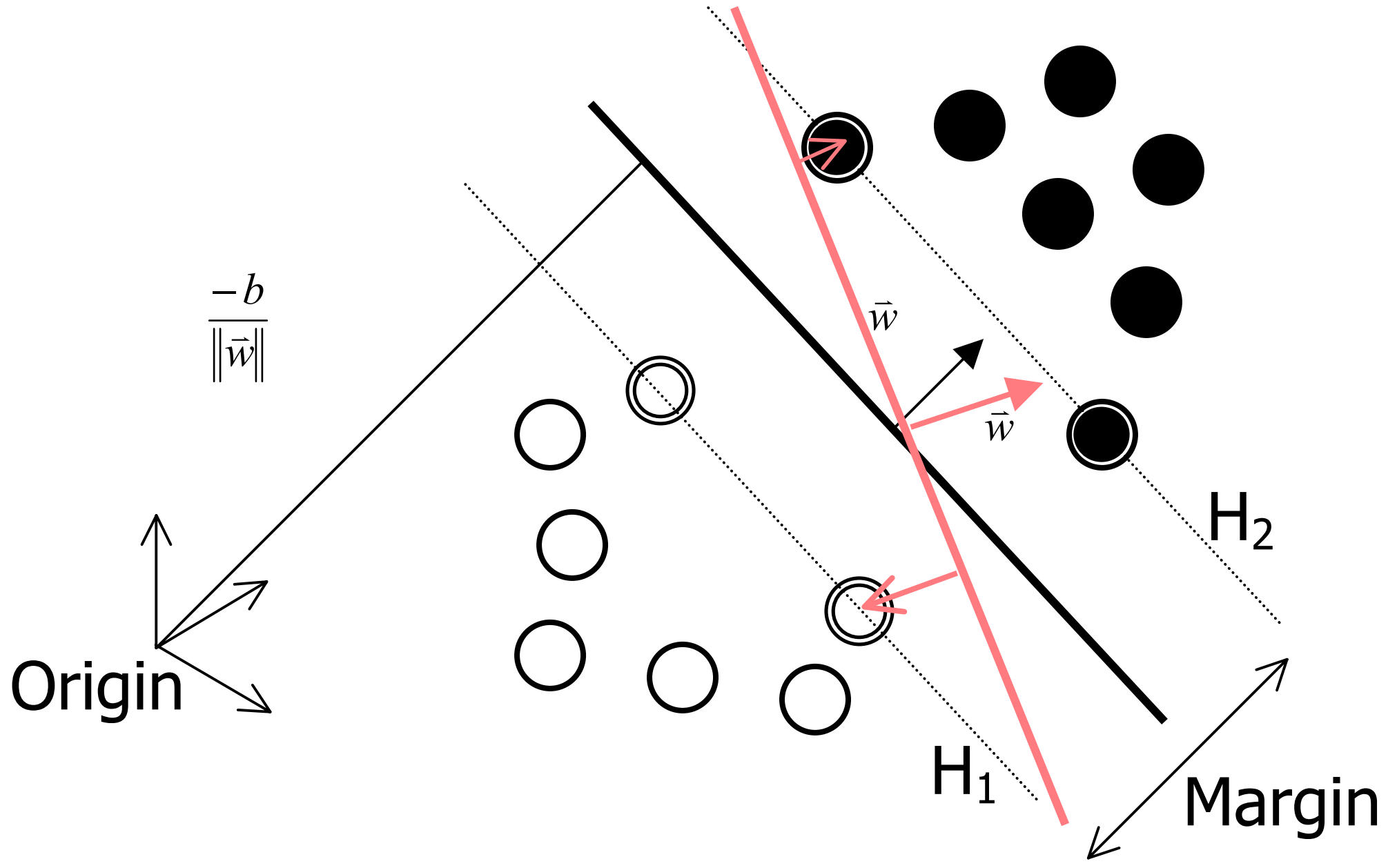

**Figure 8**. Two dimensional representation of a classification task showing the largest margin decision boundary (bold line), a second possible decision boundary (red) with smaller generalization capability as well as the support vectors (double circles) that define the $H_1$ and $H_2$ hyperplanes used to determine the margin for the bold decision boundary line.

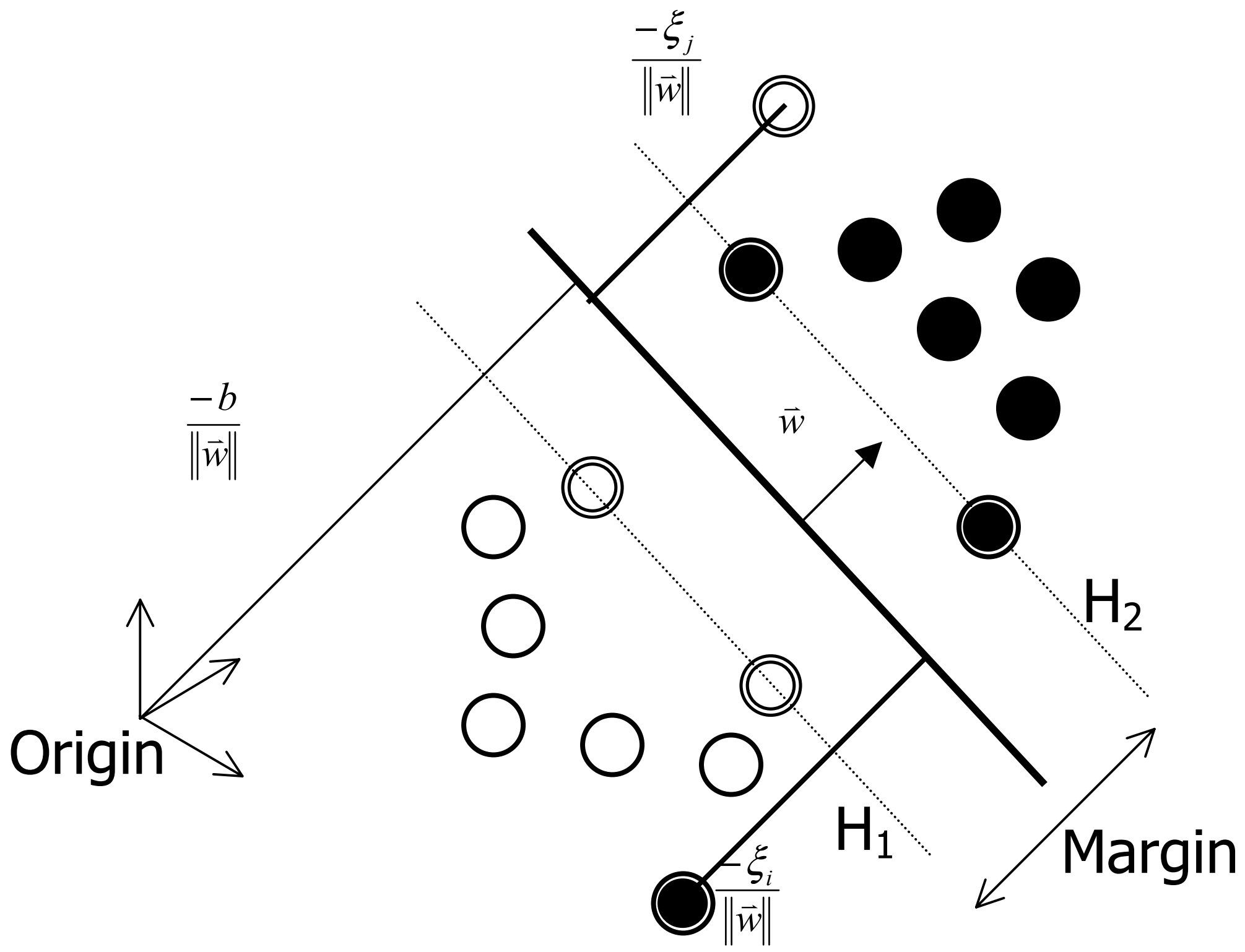

**Figure 9**. A non-separable task. The misclassified examples become support vectors as well, with a weight proportional to their distance from the boundary (proportional to $\xi_i$).

Again, the solution is found by minimizing $\|\vec{w}\|^2$ subject to the constrains in equation 3.1 as well as the positivity of the slack variables $\xi_i$.

This constrained minimization can be rewritten in lagrangian form:

$$L_p = \frac{1}{2}\|\vec{w}\|^2 + C\sum_i \xi_i - \sum_{i=1}^{l} \alpha_i \{y_i(\vec{x}_i \cdot \vec{w} + b) - 1 + \xi_i\} + \sum_{i=1}^{l} \mu_i \xi_i \qquad (3.3)$$

where $C$ is a term to penalize the misclassification of examples, $0 \leq \alpha_i \leq C$ are the Lagrange multipliers introduced to enforce the constrains introduced in equation 3.2 and $\mu_i$ are the Lagrange multipliers introduced to enforce the positivity of the slack variables $\xi_i$. The larger the value of $C$ the longer it takes to train the support vector machine. However, the larger accuracy thus obtained in classifying training set examples does not necessarily guarantee a higher success in classifying independent test set examples. When such scenario occurs, the SVM is said to be over-fitting the data.

The decision boundary, $\vec{w}$, can be expressed as a linear combination of those example vectors $\vec{x}_i$ for which $\alpha_i \neq 0$, which are called support vectors:

$$\vec{w} = \sum_i \alpha_i y_i \vec{x}_i \tag{3.4}$$

Substituting the above equation on the primal lagrangian $L_p$, one can note that the example vectors appear solely as scalar products, in this form the rewritten lagrangian is called dual lagrangian:

$$L_D = F(\vec{x}_i \cdot \vec{x}_j) \tag{3.5}$$

In general a linear decision function (boundary) might not be sufficient to describe the complex nature of the data and thus be unable to properly classify it. For example, the XOR function (Figure 10) cannot be classified with a linear decision function, but it can be classified using other functions.

The manner in which support vector machines are generalized for the case of non-linear decision functions is by introducing a mapping $\Phi$ of the data to some other (possibly infinite dimensional) space $H$, $\Phi : \Re^d \mapsto H$, prior to the classification (see Figure 11). Since the classification depends only on scalar products of example vectors, in the new space, the classification depends only on scalar products of the form $\Phi(\vec{x}_i).\Phi(\vec{x}_j)$. Any function of the form $K(\vec{x}_i, \vec{x}_j) = \Phi(\vec{x}_i).\Phi(\vec{x}_j)$ can be used to create the mapping $\Phi$ and can be used directly in the classification task with no need to ever know explicitly what the mapping is, such functions are called Kernel Functions (this is a necessary but not sufficient condition for a function to be a kernel function but this in not relevant at this point). The dual lagrangian can be written as:

$$L_D = F(\phi(\vec{x}_i)\cdot\phi(\vec{x}_j)) = F(K(\vec{x}_i,\vec{x}_j)) \tag{3.6}$$

Examples of Kernel functions are:

- Polynomial kernel (POL):

$$K(\vec{x}_i\cdot\vec{x}_j) = (s+\vec{x}_i\cdot\vec{x}_j)^p \tag{3.7}$$

- Radial Basis Function (RBF):

$$K(\vec{x}_i\cdot\vec{x}_j) = e^{-\left\|\vec{x}_i-\vec{x}_j\right\|^2/2\sigma^2} \tag{3.8}$$

- Hyperbolic Kernel (HYP):

$$K(\vec{x}_i\cdot\vec{x}_j) = \tanh(k\vec{x}_i\cdot\vec{x}_j-\delta) \tag{3.9}$$

Unfortunately, only in very special cases it is possible to decide in advance which kernel function will be best suited for the given classification task. In general, the only possibility is to try several different kernels and choose the one that offers the best accuracy.

Once the set of support vectors were determined for a given kernel function and set of examples, the classification of a new, previously unseen example, $\vec{x}$, is given by:

$$f(\vec{x}) = \sum_{i=1}^{N_s}\alpha_i y_i \phi(\vec{s}_i)\phi(\vec{x}) + b = \sum_{i=1}^{N_s}\alpha_i y_i K(\vec{s}_i\cdot\vec{x}) + b \tag{3.10}$$

where $\vec{S}_i$ are the support vectors, which are specific for the kernel function being used since they determine the decision boundary after mapping.

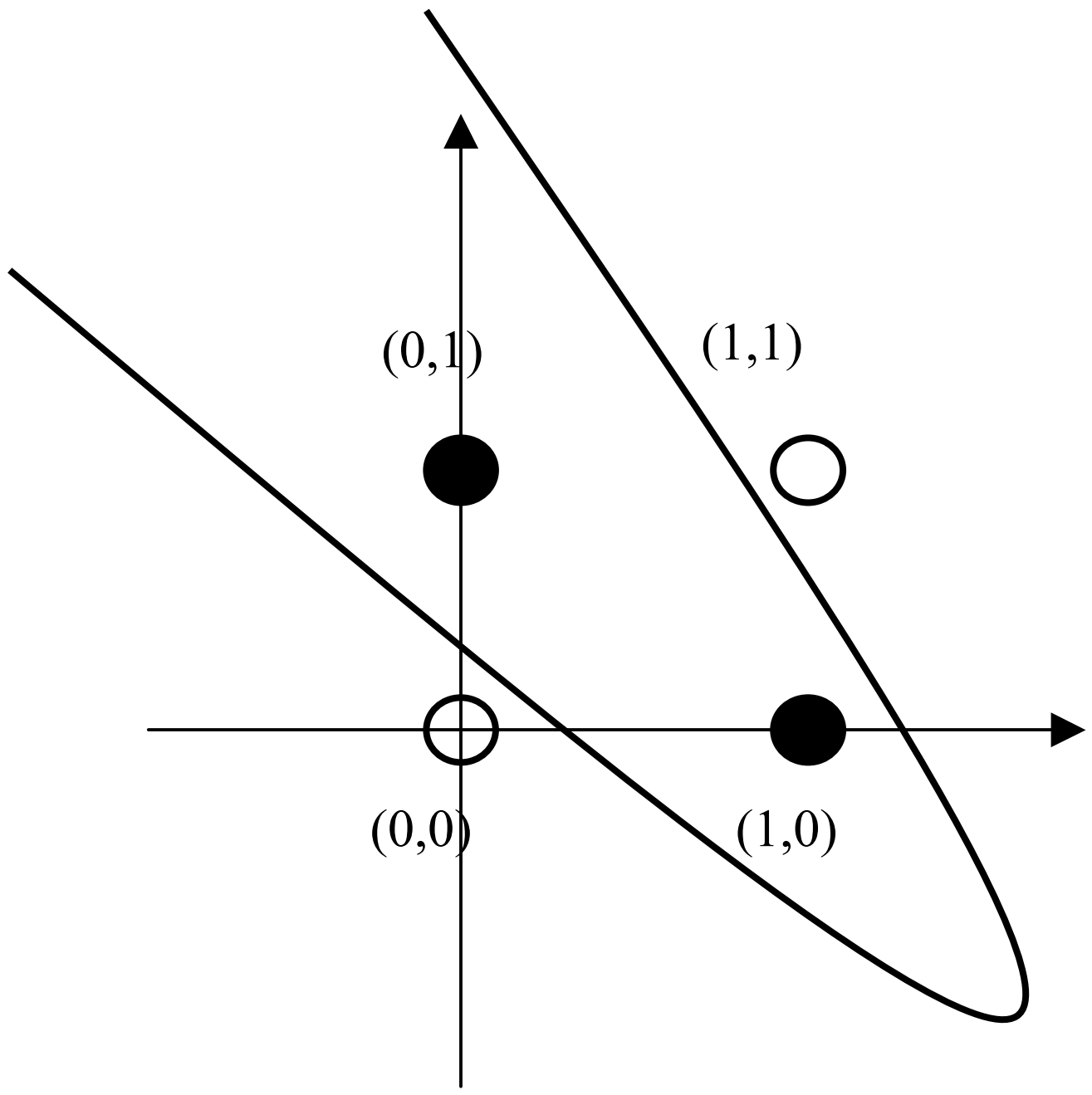

**Figure 10**. XOR function. This function cannot be correctly classified using a linear boundary function. Other functions like the quadratic function shown can classify the data properly.

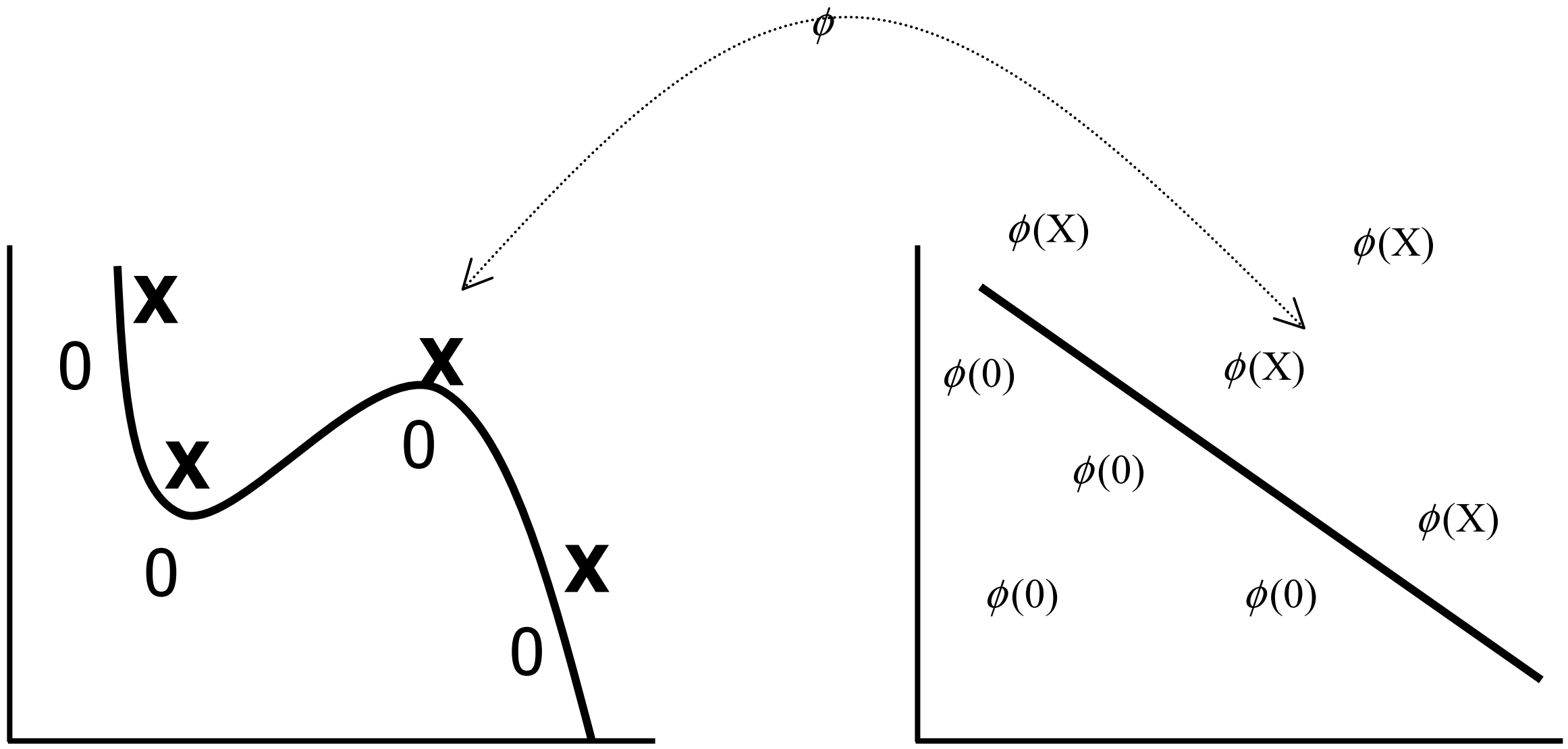

**Figure 11**. Pictorial representation of the mapping of complex data to a possibly higher dimensional space where the decision function becomes linear.

## *3.2 Generation of Examples*

My goal is to classify side chains as to whether they will undergo a conformational change upon binding. Thus, each example represents a specific binding-site side-chain in terms of its properties and those of its neighbors while the label designs whether at least one dihedral angle in the side chain under consideration undergoes a rotation larger than a threshold of $60^{o}$ upon ligand binding. Thus, an example has the following form:

<class> <feature>:<value> <feature>:<value> … <feature>:<value>

where <class> stands for a value $\pm 1$, for flexible and rigid side chains respectively, <feature> is an index characteristic for each feature and <value> is the value associated with the given feature in the specific example.

Three different characteristics are being tested to describe side chains to be classified and their neighbors:

1. Solvent accessibility.
2. Flexibility scale value.
3. Hydrophobicity scale value.

*Sovent Accessibility.* Solvent accessibility for a given amino acid is calculated using the LPC software (Sobolev et al., 1999). The larger the solvent accessible area of an amino acid, the larger its chance to be able to undergo a conformational change in order to accommodate changes in its environment.

*Flexibility Scale*. Using the flexibility scale value, one is at the same token making use of the statistical knowledge of the propensities of the different side chains to undergo conformational change and labeling the different amino acids in terms of a meaningful scheme rather than some artificial labeling scheme (e.g., using the integers 1 to 20 to label each of the 20 naturally occurring amino acids present in proteins, Table I).

*Hydrophobicity Scale*. Different amino acids have different hydrophobic propensities. The Hydrophobic effect is in general one of the most important determinants of packing in protein structures. The consideration of the hydrophobic

'nature' of a given amino acid and its neighbors might be important in altering the propensity of a side chain to undergo conformational changes. I use the values determined by (Cid et al., 1992), listed in Table III.

TABLE III. Amino acid properties used on SVM learning

| Amino acid | Flexibility | Hydrophobicity |
|---|---|---|
| **ALA** | 0.00 | 0.02 |
| **ARG** | 0.26 | -0.42 |
| **ASN** | 0.12 | -0.77 |
| **ASP** | 0.12 | -1.04 |
| **CYS** | 0.04 | 0.77 |
| **GLN** | 0.26 | -1.10 |
| **GLU** | 0.19 | -1.14 |
| **GLY** | 0.00 | -0.80 |
| **HIS** | 0.03 | 0.26 |
| **ILE** | 0.09 | 1.81 |
| **LEU** | 0.07 | 1.14 |
| **LYS** | 0.44 | -0.41 |
| **MET** | 0.20 | 1.00 |
| **PHE** | 0.13 | 1.35 |
| **PRO** | 0.00 | -0.09 |
| **SER** | 0.05 | -0.97 |
| **THR** | 0.06 | -0.77 |
| **TRP** | 0.01 | 1.71 |
| **TYR** | 0.05 | 1.11 |
| **VAL** | 0.03 | 1.13 |

Different sets of features can be combined and tested to determine which generates the best generalization. In total we can therefore have

$$\text{Number of Features} = N_c(N+1) \tag{3.11}$$

where $N_c$ respresents the number of characteristics used and $N$ represents the number of neighboring side chains taken in consideration. Neighboring side chains are those that are in contact with the side chain being classified using the CSU software (Sobolev et al., 1999) and are selected according to their solvent accessible area, in decreasing order. Since different side chain may have different numbers of neighbors, each side chain needs to be classified using the appropriate SVM; i.e., a SVM trained on a database specific for the required number of neighbors.

I used a database containing 3510 holo-apo protein pairs (Najmanovich et al., 2000b), each defining a binding site with an average of 10 residues. Overall, a total of approximately 33000 individual side chain examples are available, of which only approximately 10% undergo side chain conformational change upon binding. The effective number of independent (non-redundant) examples for a given choice of features is considerable smaller. For example, utilizing two features, e.g. the flexibility scale (Table III) of an amino acid and one neighbor, a maximum of 400 different examples can be created, which can be assigned as either flexible or rigid such that in total there might be a maximum of 800 independent examples, the remaining examples in the database, despite representing different entries in the original database are effectively copies of the 800 base-set examples. The same is true for other choices of characteristics and number of features (see equation 3.11 and Tables A.I-A.VII in appendix A). In particular, on can see in Tables A.I-A.VII that there is a number of overlapping examples (OE), i.e. examples that differ solely in their labeling (flexible or rigid) specific for each combination of characteristics and number of neighbors.

To test the accuracy of a given SVM (i.e., choice of kernel type and associated parameters) I calculate the average number of misclassified examples over consecutive learning and testing rounds. A learning and testing round comprises the selection of two independent sets of examples drawn from the available pool, subsequently using one set to train the support vector machine and the other to determine its accuracy rate (number of misclassified examples, NME). The choice of the ratio of rigid to flexible examples in learning and testing sets is crucial. Examples sets maintaining the original ratio of rigid to flexible examples of approximately 9:1, are not useful as the support vector machine makes the simple choice of classifying all examples as rigid obtaining a ~90% accuracy level. I chose to create example sets with a fixed 1:1 ratio of rigid to flexible examples such that classification as rigid or flexible is made solely based on the features that compose the example and do not consider the overall frequency of flexible and rigid examples. Alternatively, I could utilize the original 9:1 ratio but associate a penalty for misclassifying flexible examples. The net effect of choosing an optimal value for the penalty of misclassifying flexible examples would the same as that of using a 1:1 ratio.

One major consequence of the redundancy analysis is that utilizing the original database to draw 2 sets one for training and one for testing, the sets will not be truly independent despite having been drawn from different entries in the original database, artificially increasing the accuracy since several examples used in training the support vector machine are subsequently used during testing.

Table IV shows the percentage of misclassified examples for different kernels functions and penalties for misclassification of examples ($C$) for the redundant and non-redundant databases. The features used were flexibility and solvent accessibility for the amino acid under classification and its 5 first neighbors (in order of largest solvent accessibility area), a total of 12 features. The larger the value of $C$, the smaller the percentage of misclassified examples on training; however, this forced improvement performance is not followed, as expected, by an equal classification improvement on the test set (as seen most dramatically for the case of the RBF kernel). The best performance ($\mathrm{NME} \cong 0.30$) is obtained with a linear kernel (POL kernel with $s = 0$ and $p = 1$). The accuracy of the redundant datasets are approximately 5% higher than those of non-redundant datasets.

Figure 12 shows the effect of different combinations of the three characteristics (surface accessible area – SAA; flexibility – FLX; hydrophobicity - FOB) as well as number of neighbors on accuracy utilizing a linear kernel (POL kernel with $s = 0$ and $p = 1$) and training error penalty $C = 1$ for the non-redundant datasets (the data is presented in full in Tables A.I-A.VII). The Figure shows that for larger number of neighbors, FLX as the sole characteristic or in combination with any of the other two characteristics obtains the best accuracy (lowest NME value) while SAA and FOB together or on their own perform considerably worst. For smaller numbers of neighbors FLX on its own or together with FOB performs poorly while in combination with SAA or SAA and FOB maintain the same levels of accuracy. The data suggests that there is no advantage in using FOB if one is already using FLX and SAA for any number of neighbors. The trend seen for large number of neighbors is still seen utilizing the redundant dataset (Figure 13).

Table IV. Fraction of misclassified examples[a]

| Kernel | | C | NME | | | |
|---|---|---|---|---|---|---|
| | | | Training Set | | Testing Set | |
| | | | Original set | NR set | Original set | NR set |
| POL | S=0 P=1 | 0 | 0.38 (0.02) | 0.36 (0.01) | 0.40 (0.02) | 0.39 (0.03) |
| | | 1 | 0.26 (0.01) | 0.29 (0.04) | 0.26 (0.02) | 0.31 (0.02) |
| | | 10 | 0.25 (0.01) | 0.41 (0.03) | 0.26 (0.01) | 0.34 (0.04) |
| | S=0 p=2 | 0.2 | 0.31 (0.02) | 0.33 (0.05) | 0.31 (0.02) | 0.37 (0.04) |
| | | 0.5 | 0.31 (0.01) | 0.31 (0.03) | 0.31 (0.02) | 0.34 (0.03) |
| | | 1 | 0.31 (0.01) | 0.35 (0.06) | 0.32 (0.02) | 0.40 (0.06) |
| RBF | $\sigma^2$=1/2 | 0.01 | 0.37 (0.04) | 0.00 (0.00) | 0.44 (0.01) | 0.49 (0.01) |
| | | 0.3 | 0.36 (0.01) | 0.00 (0.00) | 0.43 (0.01) | 0.48 (0.01) |
| | | 1 | 0.07 (0.02) | 0.00 (0.00) | 0.37 (0.02) | 0.50 (0.01) |

[a] Each support vector machine was trained on a training set and subsequently tested on an independent test set. Train and test sets were composed of 1000 examples in the Original set (redundant) and 200 examples in the non-redundant set (NR set), with a fraction of rigid:flexible examples of 1:1 in all cases. The Percentage of misclassified examples was averaged on 5 rounds. An example is composed of two characteristics, Flexibility and Surface Accessible Area for the amino acid under classification and 5 neighbors, a total of 12 features.

As a control in the SVM approach I have created three different randomized controls and compared NME values to those obtained with the non-randomized set. A set of examples utilizing flexibility and surface accessible area (linear kernel, POL kernel with $s = 0$ and $p = 1$, five neighbor and $C = 1$) was generated (FLX_SAA_FOB set). The first random control consists of randomly shuffling the labels of the examples in the FLX_SAA_FOB set (RLS set). The second set is created by discarding the original labels and randomly assigning new labels to the examples in a way to create 50% flexible and 50% rigid examples (RLA). The difference between the RLS and RLA is that the original ratio between flexible and rigid examples is lost, considerably increasing the number of flexible examples. The third set is obtained by keeping the original ratio of labels but randomly assigning the values of the features (RVA set). The NME values obtained for the three randomized controls (Figure 14), as expected are 50% with slight differences in the classification error of flexible and rigid examples for the RLS and RLA sets.

The classification of examples as flexible or rigid is a necessary pre-condition to create training sets. This classification depends on the difference between the values of dihedral angles before and after ligand docking. If the difference is higher than a threshold value the side chain is labeled as flexible. The choice of the threshold value is based on the knowledge that a rotation beyond a certain threshold is likely to signify a change in conformation. Thus although not completely arbitrary, the choice of threshold is, for lack of a better word, rather flexible. Figure 15 presents the effect of different choices of threshold on the number of misclassified examples in training and testing. One can see that the differences for different threshold values are very small.

The results obtained show that it is possible to classify side chains as rigid or flexible utilizing support vector machines. A level of approximately 70% classification accuracy is obtained. I am not aware of another study of this kind and it is not clear if other current techniques could reach a higher classification accuracy rate.

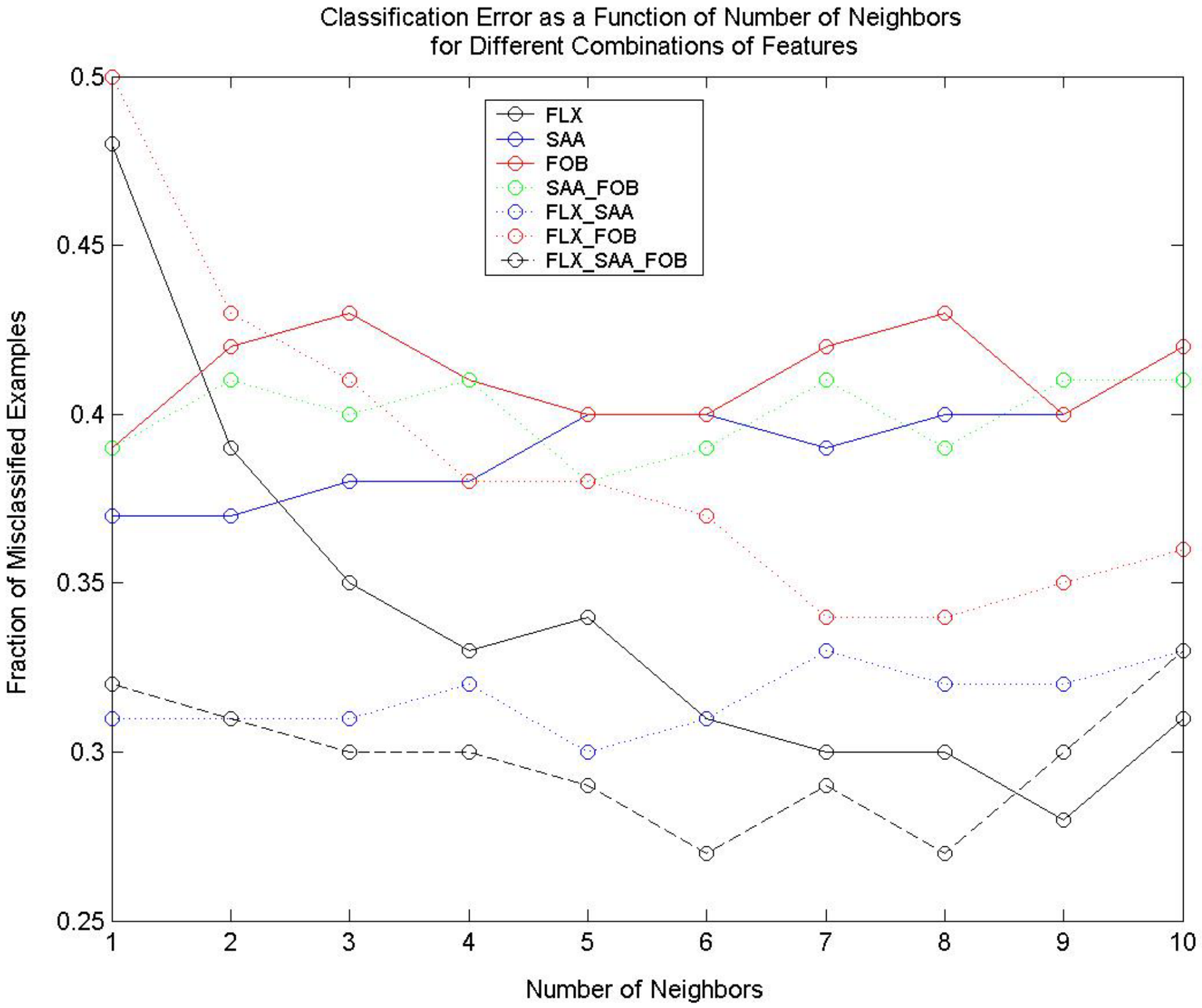

**Figure 12**. SVM learning accuracy for different combinations of characteristics and number of neighbors. SVM learning was performed using the linear kernel (POL kernel with $s=0$, $p=1$ and $C=1$). The features used were flexibility (FLX), surface accessible area (SAA) and hydrophobicity (FOB). The error bars are not shown for clarity (see Tables A.I-A.VII). For larger number of neighbors, FLX as the sole characteristic or in combination with any of the other two characteristics obtains the best accuracy (lowest NME value) while SAA and FOB together or on their own perform considerably worst. For smaller numbers of neighbors FLX on its own or together with FOB performs poorly while in combination with SAA or SAA and FOB maintain the same levels of accuracy.

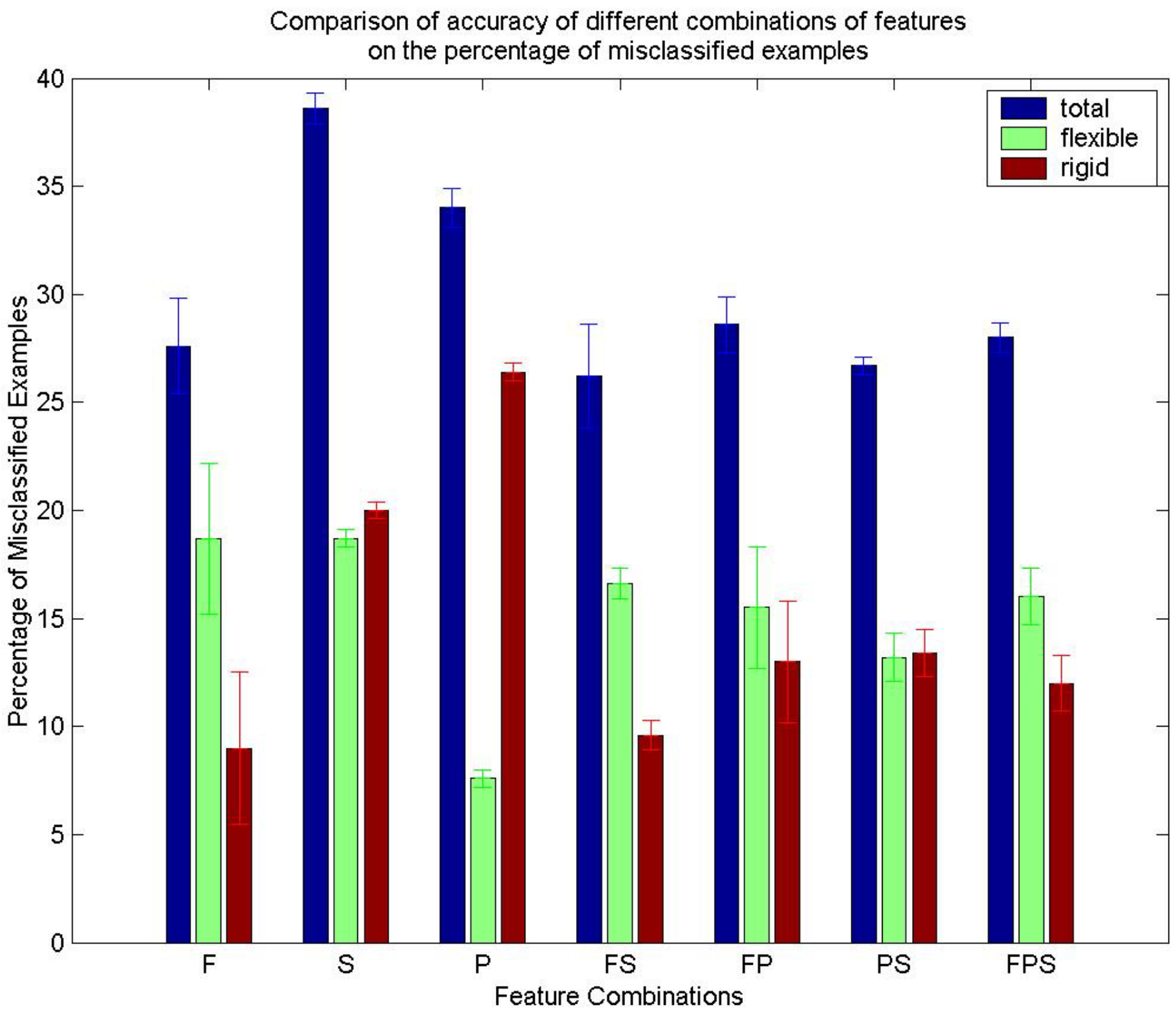

**Figure 13**. Comparison of different combinations of the three characteristics being used to describe side chains: flexibility (F), surface accessible area (S) and hydrophobicity (P). The number of misclassified examples is shown for all examples irrespective of the original label of the example (blue) as well as separated in flexible (green) and rigid (red) examples.

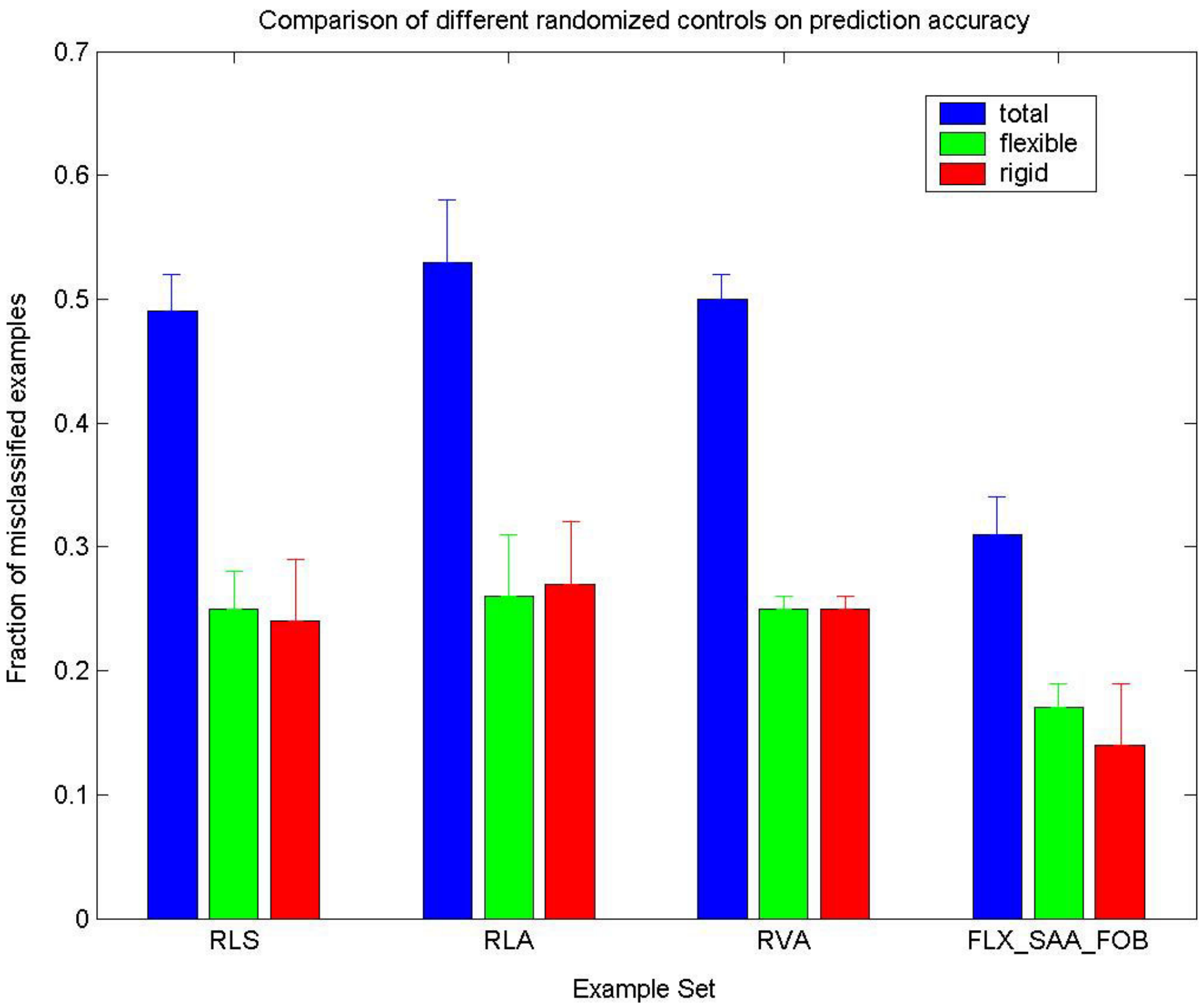

**Figure 14**. Comparison of different randomized controls on prediction accuracy for all examples, flexible examples and rigid examples. A set of examples utilizing flexibility, surface accessible area and hydrophobicity (linear kernel, POL kernel with $s=0$ and $p=1$, five neighbor and $C=1$) was generated (FLX_SAA_FOB set). The first random control consists of randomly shuffling the labels of the examples in the FLX_SAA_FOB set (RLS set). The second set is created by discarding the original labels and randomly assigning new labels to the examples in a way to create 50% flexible and 50% rigid examples (RLA). The difference between the RLS and RLA is that the original ratio between flexible and rigid examples is lost, considerably increasing the number of flexible examples. The third set is obtained by keeping the original ratio of labels but randomly assigning the values of the features (RVA set).

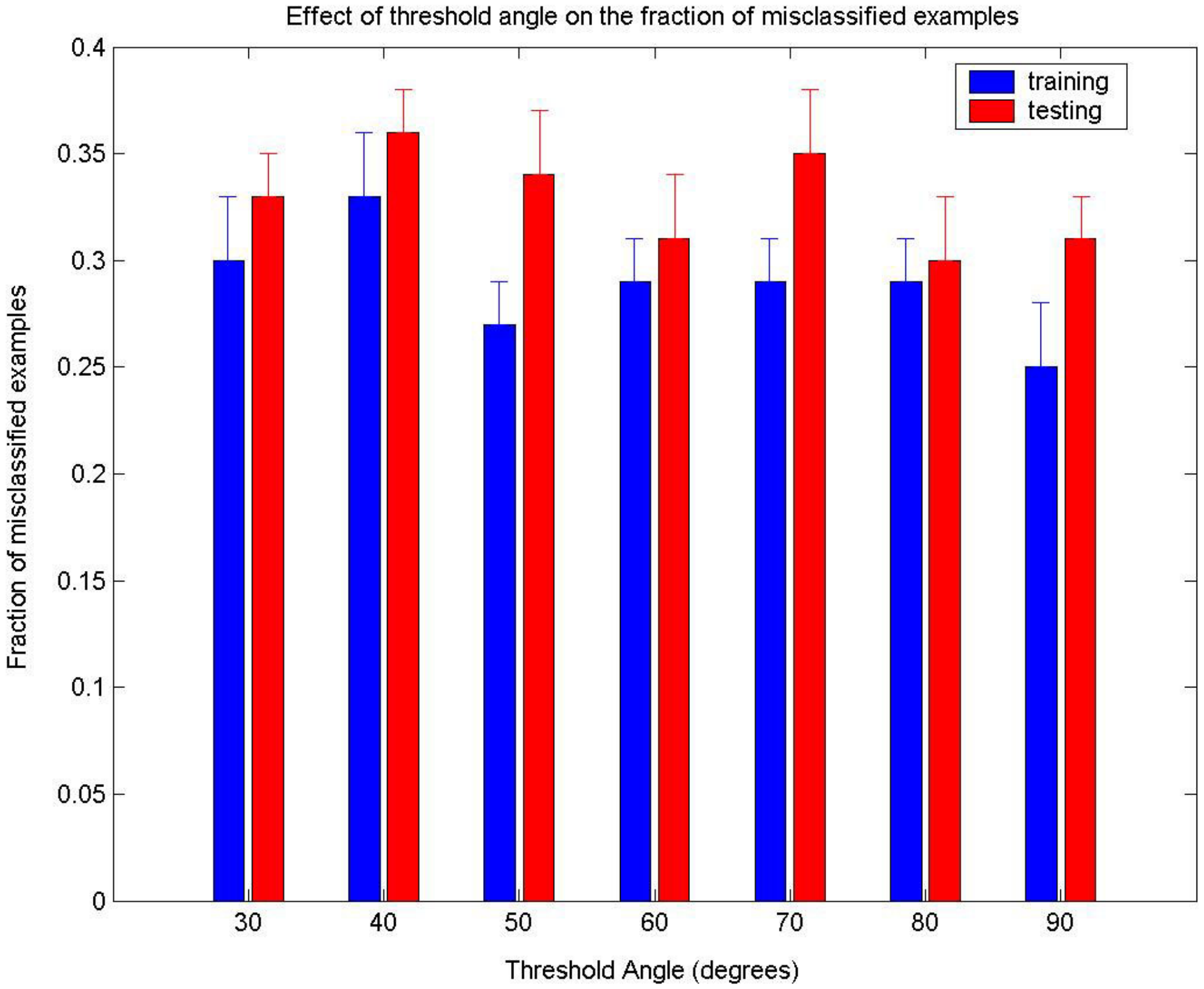

**Figure 15**. Effect of different choices of threshold on the fraction of misclassified examples in training and testing. One can see that the differences for different threshold values are very small. Training and testing were performed using flexibility, surface accessible area, hydrophobicity and five neighbors with the linear kernel (POL kernel with $s=0$ and $p=1$, five neighbor and $C=1$).

# 4. SCORING FUNCTIONS

Scoring functions are used to quantify the relative quality of different solutions of an optimization problem. Scoring functions are also called energy functions or force fields although they are not required to bear any resemblance to the real (i.e., physical) energy landscape of the system. The energy landscape produced by a scoring function represents the surface utilized in the search procedure to find the desired global extreme (either global minima or maxima). In general, it is difficult to assert whether or not an optimization procedure reached the global extreme since the latter is not generally known. In the realm of biomolecular structure simulations, it is hoped that the parameters of a scoring function are such that experimentally determined solutions are given high enough scores (low energy) as to discriminate them from other extrema that do not bear resemblance to experimentally determined solutions.

There are several approaches for the generation of scoring functions (Jernigan & Bahar, 1996;Halperin et al., 2002). I will describe my results with two knowledge-based approaches, namely, the utilization of a database to derive parameter values from contact probabilities for ligand-protein interactions and the utilization of neural networks to derive interaction parameters for a pairwise amino acid protein folding scoring function. The latter method is devised explicitly to look for parameter sets that assign the experimentally determined solution as the global minima. First I introduce the scoring function utilized in our docking simulations, the complementarity function.

## *4.1 COMPLEMENTARITY FUNCTION*

The complementarity function (CF, Sobolev & Edelman, 1995; Sobolev et al., 1997) accounts for the area in contact between atoms participating in favourable and unfavourable interactions:

$$CF = S_f - S_u - E \tag{4.1}$$

Where $S_f$ is the surface area of favourable contacts, $S_u$ is the surface area of unfavourable contacts (a contact is deemed favourable if the pairwise parameter for the involved atoms is positive, see Table V) and E is an energy term to penalize the placement of atoms at distances closer than the sum of their van der Waals radii effectively prohibiting the placement of atoms at distances significantly closer than the sum of their van der Waals radii:

$$E = \frac{1}{2}\sum_{a}\sum_{b} E_{ab} \tag{4.2}$$

where

$$E_{ab} = \begin{cases} 0 & \text{if } R_{ab} \geq R_o \\ k(1/R_{ab}^{12} - 1/R_o^{12}) & \text{if } R_{ab} < R_o \end{cases} \tag{4.3}$$

$R_0 = 0.9(R_a + R_b)$ and $K = 10^6$ Å$^{14}$. $R_a$ and $R_b$ are the van der Waals radii of the contacting atoms (Bondi, 1964; Lee & Richards, 1971) and $R_{ab}$ is the distance between them.

TABLE V. Pairwise contact energy parameters.

| LigIN ATOM TYPES | I | II | III | IV | V | VI | VII | VIII |
|---|---|---|---|---|---|---|---|---|
| I. Hydrophilic | +1 | | | | | | | |
| II. Acceptor | +1 | -1 | | | | | | |
| III. Donor | +1 | +1 | -1 | | | | | |
| IV. Hydrophobic | -1 | -1 | -1 | +1 | | | | |
| V. Aromatic | +1 | +1 | +1 | +1 | +1 | | | |
| VI. Neutral | +1 | +1 | +1 | +1 | +1 | +1 | | |
| VII. Neutral-Donor | +1 | +1 | -1 | +1 | +1 | +1 | -1 | |
| VIII. Neutral Acceptor | +1 | -1 | +1 | +1 | +1 | +1 | +1 | -1 |

The contact surface area between atom *a* and *b* is defined as the surface area of a sphere of radius $(R_a + R_w)$ centred at $a$ which penetrates the sphere of radius $(R_b + R_w)$ defined by atom $b$, $R_w$ is the van der Waals radii of the solvent molecule, 1.4 Å.

When a patch of surface is in contact with more than one atom, it is assigned to be in contact to the closest atom.

In a more general sense, the CF function (equation 4.1) can be seen as a pairwise potential as follows:

$$CF = \sum_{i} \sum_{\substack{j \\ j \neq i}} \varepsilon_{ij} S_{ij} - E \qquad (4.4)$$

Where the summation over $i$ runs through all atoms in the molecule whose CF is to be evaluated and the summation over $j$ runs through all atoms in the system; $\varepsilon_{ij}$ is the contact energy of a contact between atoms $i$ and $j$ according to their atom types (Table V), while $S_{ij}$ is the surface in contact in square angstroms.

The pairwise energy parameters used in LigIn were chosen according to general chemical principles, they are not supposed to be a quantitatively correct representation of the relative strength of the interactions among different atom types.

## *4.2 Knowledge Based Potential Generation*

The method is based on the assumption that the probability of the different states of the (ligand-protein) system follows a Boltzmann distribution, one therefore calculates the probability with which each of the 36 different pairwise parameters among the 8 atom types appear in a database, $p_{ij}$:

$$p_{ij} = e^{-\frac{\varepsilon_{ij}}{k_b T}} \qquad (4.5)$$

Where $i,j=\{I,II,III,...,VIII\}$ represent the atom classes, $k_b$ is the Boltzmann constant, $T$ is the temperature and $\varepsilon_{ij}$ is the pairwise contact energy between atom types $i$ and $j$. To derive the pairwise energy parameters one has to use the following relation:

$$\varepsilon_{ij} = -k_b T \ln p_{ij} \tag{4.6}$$

In order to calculate the $p_{ij}$ values we created a non-redundant database of 962 PDB entries. All entries contain one or more ligands. A PDB file needs to fulfil the following criteria to be included:

- Resolution better or equal to 2.5 Å.
- No DNA or RNA ligand(s).
- Ligands with at least 5 non-hydrogen atoms.
- No covalent bonds between ligand and protein atoms.

These criteria are the same used in the generation of the databases in the study of side chain flexibility. In order to reduce the redundancy, we chose to select one single entry for each ligand so that we do not over represent the interactions present in common ligands in deterrence of rare ones.

This method developed to derive amino acid pairwise contact potentials (Godzik et al., 1995) can be readily applied to derive atomic contact potentials as well. The interaction probability $p_{ij}$ of atom types $i$ and $j$ can be calculated according to:

$$p_{ij} = \frac{N_{ij}^{observed}}{N_{ij}^{expected}} \tag{4.7}$$

While $N_{ij}^{observed}$ is simple a count of the number of times a contact of a given type was observed, the calculation of $N_{ij}^{expected}$ is more cumbersome.

In order to derive an expression for $N^{\text{expected}}$ between atom types $i,j$ in a database with M binding pockets, lets assume that the $k^{th}$ binding-pocket consists of $L(k)$ atoms, the number of atoms of type $i$ in pocket $k$ is $number_i^k$ and the total number of interactions is $N_{total}^{k}$. Atom $n$ in pocket $k$ has $ncont_{n}^{k}$ interactions and its atom type is given by $typ(n)$. The derivation follows four steps:

1. We calculate for every atom type $i$, the total number of interactions $S_j$ in the whole database.

$$S_i = \sum_{k=1}^{M} \sum_{\substack{n=1 \\ typ^k(n)=i}}^{L(k)} ncont_n^k \tag{4.8}$$

the first summation run over all pockets in the database, the second over all atoms belonging to the pocket under the condition that an atom at position $n$ is of type $i$. $S_i$ in turn is used to calculate the mean number of interactions for every atom type.

$$q_i = \frac{S_i}{\sum_{k=1}^{M} number_i^k} \tag{4.9}$$

2. A contact fraction for every atom type $i$, is calculated for every pocket $k$.

$$X_i^k = \frac{q_i number_i^k}{\sum_{j=1}^{8} q_j number_j^k} \tag{4.10}$$

3. The expected number of interactions between atom types $i$ and $j$ is calculated as the product of the contact fractions of the respective atom types in a given protein binding pocket and the number of interactions present on it.

$$n_{ij}^k = X_i^k X_j^k N_{total}^k \tag{4.11}$$

Steps 1 to 3 are repeated for each protein binding-pocket in the database.

4. Equation 4.11 is used to calculate the expected number of interactions between atom types $i$ and $j$ simply by summing the values obtained for each protein.

$$N_{ij}^{\exp ected} = \sum_{k=1}^{M} n_{ij}^k \tag{4.12}$$

The new parameters are derived using equations 4.12, 4.7 and 4.6. Although not necessary, the parameters obtained were normalised such that:

$$< \varepsilon > = \frac{1}{36} \sum_{k=1}^{36} \varepsilon_k = 0 \tag{4.13}$$

$$\sigma_\varepsilon = \sqrt{< \varepsilon_{ij}^2 > - < \varepsilon >^2} = 1 \tag{4.14}$$

The resulting values are presented in Table VI where attractive interactions are represented by negative values. Note that this convention is the opposite of that used in Table V.

TABLE VI. Knowledge based pairwise contact energy parameters.

| ATOM TYPES | I | II | III | IV | V | VI | VII | VIII |
|---|---|---|---|---|---|---|---|---|
| I. Hydrophilic | 0.7194 | | | | | | | |
| II. Acceptor | -1.0257 | 1.3608 | | | | | | |
| III. Donor | -1.9224 | -0.9237 | 1.7833 | | | | | |
| IV. Hydrophobic | 0.0155 | 0.2405 | 1.6505 | -0.1682 | | | | |
| V. Aromatic | 0.7689 | -0.1116 | -0.0826 | -1.0392 | 0.2383 | | | |
| VI. Neutral | -0.5468 | -0.3272 | -0.3300 | 0.2284 | -0.3560 | 0.1758 | | |
| VII. Neutral-Donor | -1.9143 | -0.5031 | 1.7833 | 1.0032 | -0.6013 | 0.2422 | 1.7833 | |
| VIII. Neutral-Acceptor | 0.7812 | -0.5771 | -1.3905 | -0.5540 | -0.0972 | -0.2134 | -1.5649 | 1.4747 |

### *4.2.1 Correlation between Potentials*

There is agreement among 27 out of 36 parameters with respect to attractive or repulsive interactions on both data sets. From the remaining 9 parameters only one, that between II-VIII (acceptor-neutral acceptor) atom types, is attractive in the knowledge-based case but repulsive in the binary parameter set. The remaining 8 pairs are attractive interactions in the binary case but are found to be repulsive in the knowledge based case: I-I (hydrophilic-hydrophilic), I-V (hydrophilic-aromatic), I-VIII (hydrophilic-neutral-acceptor), IV-VI (hydrophobic-neutral), IV-VII (hydrophobic-neutral-donor), V-V (aromatic-aromatic), VI-VI (neutral-neutral), VI-VII (neutral-neutral-donor).

One possible explanation for the latter 8 pairs being repulsive, while common wisdom would say them to be attractive, lays in the nature of the process used to generate the knowledge-based parameters. Pairs that are less common are deemed repulsive. However, a pair can appear to be less common due to the automatic process of atom type assignment used for the ligand atoms.

Let us, for example, think of the case of neutral-neutral interactions. If in most cases an atom will for one reason or another be assigned as either neutral-donor or neutral-acceptor instead of neutral, the number of interactions of type neutral will be misrepresented causing problems in the statistics used to calculate the parameter value.

One last point to be mentioned is that the average of the parameter set (Equation 4.13) can be set to any arbitrary value, since that would be cancelled out on any energy difference calculation. By choosing a more appropriate value for the average, we can fit more pairs into agreement with the binary values.

### *4.2.2 CF function Correlation with Experimental Binding Energies*

My first aim from deriving a new set of parameters is that of trying to fit experimental values of binding affinities. At the time these results were generated, one of the most serious problems was that of obtaining compiled data of experimental binding affinity measurements with associated PDB codes as well as clear descriptions of the ligand in consideration. The majority of this data were compiled by authors that have moved to pharmaceutical industries and are not willing to share it. Eventually I managed to obtain a list of 36 complexes (Head et al., 1996). The list of affinity values ($-\log K_i$) is presented in Table VII. I calculated the CF value (Equation 4.4) for each complex presented in Table VII with either the LigIn set of parameters (referred to as binary) or the database derived parameters. In the case of the later, I performed a Monte Carlo simulation (Binder, 1984; Binder & Heermann, 1988) to try and optimise the correlation value by changing the origin of energy ($<\varepsilon>$). The correlation coefficients obtained were 0.52 for the binary set and 0.56 for the database set.

There are several approaches for the prediction of binding affinities (for a review, see Gohlke & Klebe, 2002 and references therein). In the realm of knowledge-based scoring functions, Muegge and Martin (Muegge & Martin, 1999) have developed a distance dependent atomic pairwise scoring function based containing 34 ligand atom types and 16 protein atom types. The authors report a correlation coefficient of 0.61 between the predicted and experimental binding constants for a set of 77 protein ligand complexes. Other approaches (Bohm, 1998; Stahl & Bohm, 1998; Gohlke et al., 2000; Muegge, 2000) are more successful in predicting experimental binding constants from structural data. To this point, studies applying scoring functions used for the prediction of experimental binding constants in docking simulations have not been performed.

TABLE VII. Experimental Affinity values

| Protein inhibitor Name | PDB code | Affinity |
|---|---|---|
| HIV MVT101 | 4hvp | 6.12 |
| HIV JG365 | 7hvp | 9.60 |
| HIV acetylpepstatin | 5hvp | 5.60 |
| HIVA74704 | 9hvp | 8.50 |
| HIV hydroxyethylene | 1aaq | 5.50 |
| HIV L-700,417 | 4phv | 9.15 |
| thermolysin-phosphoramidon | 1tlp | 7.55 |
| Thermolysin-N-(1-carboxy-3-phenyl)-L-LeuTrp | 1tmn | 7.47 |
| thermolysin-N-phosphoryl-L-leucinamide | 2tmn | 4.10 |
| thermolysin-ValTryp | 3tmn | 5.90 |
| thermolysin-Leu-NHOH | 4tln | 3.72 |
| thermolysin-ZFPLA | 4tmn | 10.19 |
| thermolysin-ZGp(NH)LL | 5tmn | 8.04 |
| thermolysin-ZGp(O)LL | 6tmn | 5.05 |
| thermolysin-CH2CO-Leu-OCH3 | 7tln | 2.47 |
| endothiapepsin-L-364,099 | 2er0 | 6.40 |
| endothiapepsin-H-256 | 2er6 | 7.20 |
| endothiapepsin-H-261 | 2er7 | 9.00 |
| endothiapepsin-L-363,564 | 2er9 | 7.40 |
| endothiapepsin-CP-71,362 | 3er3 | 7.10 |
| endothiapepsin-PD 125967 | 4er1 | 6.60 |
| endothiapepsin-H 142 | 4er4 | 6.80 |
| endothiapepsin-CP-69,799 | 5er2 | 6.60 |
| L-arabinose-bind-prot-L-arabinose | 1abe | 6.50 |
| L-arabinose-bind-prot-D-fucose | 1abf | 5.20 |
| L-arabinose-bind-prot-P254G-L-arabinose | 1bap | 6.90 |
| L-arabinose-bind-prot-P254G-D-galactose | 9abp | 8.00 |
| L-arabinose-bind-prot-M108L-D-fucose | 7abp | 5.40 |
| L-arabinose-bind-prot-M108L-D-galactose | 8abp | 6.60 |

### 4.3 ARTIFICIAL INTELLIGENCE SCORING FUNCTION FOR PROTEIN FOLDING

A different approach for the generation of a pairwise scoring function is that of using a method to carefully select parameter values in such a way that, by definition, the global extreme of the scoring function will correspond to a chosen structure, for example the experimentally determined structure (native state). Such methodology was used to generate an amino acid pairwise potential for protein folding utilizing a type of neural network known as perceptron (Hertz et al., 1991; Watkin et al., 1993). The parameters of a pairwise scoring function are tuned in such a way that the native structure of any protein present in a database is assigned the highest score vis-à-vis a set of decoy structures obtained by gapless threading for each given protein (Vendruscolo et al., 1999; Vendruscolo et al., 2000).

The proteins used were selected from the list of 312 proteins (R-factor, 0.21 and Resolution, 2.0; list created from the PDB on July 23, 1997) as obtained by WHATCHECK which adopts the following criteria: (a) The keyword MUTANT does not appear in the COMPOUND name; (b) The structure is solved by X-ray crystallography; (c) The resolution is better than 2.0Å; (d) The *R* factor is lower than 0.21; (e) Chains with "abnormal features" were excluded; (f) No more than 749 or less than 32 amino acids; (g) No more than 30% sequence identity; (h) No more than 1 chain break; (i) No more than two $C_\alpha$-only residues. This list was further reduced to 154 proteins, by removing proteins according to the following criteria:

- $C_\alpha$ distance between consecutive residues outside the interval of 4 $\sigma$ (standard deviations) from the mean $d$ ($d$=5 3.81, $\sigma$=5 0.05). In this way we remove the chains with CIS-peptide bonds (including cis-PRO) or backbone chain breaks.
- Any residue (ASX, GLX, UNK, ACE, PCA, etc.) that does not match the 20 standard amino acid names (in the case where the first and/or last residues are undefined, the residues were removed, not the protein).
- Any chain for which the $C_\alpha$ or the backbone-*N* atoms' coordinates are not present.

- Any unexplained mismatch between the sequence of amino acids presented in SEQRES and the actual sequence appearing in the coordinates section.
- In case of multiple locations of amino acids we keep the protein and consider only the location for which altloc= "A".

To keep track of the degree of structural similarity, we made a note of the CATH classification of protein domains. Sixty one of the 126 chains in the set have a CATH code. Comparison of the first domain shows that the proteins in the set belong to one of 21 groups, with at least two proteins in each group, for which the CATH classification codes are identical. These 21 groups comprise a total of 49 chains. Hence, 28 chains could in principle be removed, leaving only one representative from each group. The 21 groups are the following: [1abe, 1gca, 1gd1 O, 1tad A, 2dri]; [1flp, 1mbd, 2gdm, 2hbg, 3sdh A]; [1pot, 1sbp, 2abh]; [1arb, 1hyl A]; [1aru, 2cyp]; [1cot, 1ctj]; [1ifc, 1lid]; [1mba, 1thb]; [1tgs I, 9wga A]; [1wad, 2cy3]; [1atl A, 1iae]; [1bbh A, 2ccy A]; [1bdo, 2bbk L]; [1cka A, 1ptx]; [1cpc A, 1cpc B]; [1cyo, 2ltn B]; [1iro, 1otf A]; [1prn, 2por]; [1rro, 4icb]; [1ycc, 451c]; [2gdm, 2hbg]. This structural analysis was performed by comparing only the first domain of each protein chain; some proteins may have other domains that were not taken into account.

The three-letter PDB code (with a chain identifier where existent) as well as the CATH classification of the 154 protein chains of the set are presented in Table VIII, this dataset is called $SET_{154}$.

The number of decoys that can be generated by gapless threading is limited by $M_p$, the number of proteins in the database and their lengths $N_i$, $(i = 1,\ldots,M_p)$. For proteins $i$ and $j$, of lengths $N_i$ and $N_j$ respectively $(N_j > N_i)$, the number of possible decoys is $2(N_j - N_i + 1)$. Therefore, having ordered the proteins by increasing length, the total number $P$ of decoys is given by

$$P = \sum_{j>i}^{M_p} 2(N_j - N_i + 1) \qquad (4.15)$$

in total, there are 2,497,334 possible decoys for $SET_{154}$.

Table VIII. Perceptron Scoring Function Database

| CATH class[a] | Total | PDB code[b] |
|---|---|---|
| Mainly α-helix | 42 | 1351, 1531, 1aru, 1axn, 1cmb A, 1cot, 1cpc A, 1cpc B, 1csh, 1ctj, 1flp, 1hyp, 1mba, 1mbd, 1osa, 1pnk A, 1rro, 1thb A, 1xik A, 1ycc, 2abk, 2cyp, 2end, 2gdm, 2hbg, 2wrp R, 3sdh A, 451c, 4icb, 256b A, 2ccy A, 2spc A, 1bbh A, 1cpq, 1lis, 1mzm, 1poa, 1vls, 1htr P, 1lts C, 1rop A, 1lmb 3. |
| Mainly β-sheet | 27 | 1ext A, 3ebx, 1cka A, 1ova A, 1arb, 1hbq, 1hyl A, 1ida A, 1ifc, 1lid, 4fgf, 1kap P, 1lop A, 1prn, 1vqb, 2cpl, 2por, 1amm, 1bdo, 1slt A, 1ten, 1tta A, 2rhe, 1gpr, 1xso A, 2aza A, 2bbk L. |
| α-β | 56 | 1aay A, 1brn L, 1cyo, 1doi, 1fkj, 1frd, 1hpi, 1igd, 1mml, 1onc, 1ubi, 1frb, 1nfp, 1tml, 1tph 1, 2mnr, 1bhp, 1bur S, 1cse I, 1daa A, 1fxd, 1iro, 1kpt A, 1npk, 1otf A, 1ptf, 1ptx, 1puc, 1tgs I, 2bop A, 2chs A, 2nll A, 2phy, 2pii, 3cla, 1pot, 1rcf, 9wga A, 1abe, 1atl A, 1cus, 1dad, 1gca, 1gd1 O, 1iae, 1lkk A, 1mrj, 1pdo, 1sbp, 1tad A, 2abh, 2dri, 5p21, 1reg X, 1wad, 2cy3. |
| Few secondary structure | 2 | 2psp A, 2ltn B. |
| Not classified | 27 | 1beb A, 1cei, 1cyd A, 1dut A, 1dxy, 1edm B, 1ept A, 1ept B, 1fle I, 1gnd, 1hrd A, 1kuh, 1kve A, 1kve B, 1lbu, 1mai, 1nox, 1onr A, 1pmi, 1rie, 1tfe, 1whi, 1yas A, 2arc A, 2cbp, 2mbr, 2tsc A. |

a. Structural classes were assigned using the CATH database.
b. Codes include when appropriate chain identifiers following the 3-letter PDB code.

### 4.3.1 Database Examples

An example vector $\vec{X}$ is created for each decoy structure generated by gapless threading of the sequence of the protein of interest through all other structures with at least the same number of amino acids present in the database.

Given a protein with $N_p$ amino acids in a given configuration $k$, we can associate to it a contact map which is a symmetric $N_p \times N_p$ matrix **S**. Whenever amino acids $i$ and $j$ are closer than a predefined threshold distance, $S_{ij} = 1$(i.e., amino acids $i$ and $j$ are said to be in contact), otherwise $S_{ij} = 0$. We can count how many contact are there in **S** between amino acids of type $\mu$ and $\nu$. The resulting number defines a symmetric $20 \times 20$ matrix. Amino acids $i$ and $j$ are said to be in contact whenever the closest distance between two heavy atom belonging to different residues are closer than a threshold distance $R_c$ (in the present case $R_c = 4.5$ Å).

From the 400 elements composing this resulting matrix, the 210 non-symmetric elements can be mapped into a vector with 210 components, call this vector $\vec{N}$. For any given configuration $k$ we can define the vector $\vec{N}^k$ of the number of contacts of each type. In particular we can define $\vec{N}$ for the native state of a protein (which is known *a priori*), call it $\vec{N}^0$.

The vector of examples $\vec{X}^k$ is defined as the difference between the number of contacts (of each type) in a given conformation $k$ and the correspondent number of contacts in the native state:

$$\vec{X}^k = \vec{N}^k - \vec{N}^0 \tag{4.16}$$

The pairwise energy of configuration $k$, $E^k_{pair}$, can be written as the sum of the number of contacts of each type multiplied by the corresponding energy parameter:

$$E^k_{pair} = \sum_i w_i \vec{N}^k_i = \vec{w} \cdot \vec{N}^k \tag{4.17}$$

The interest on defining the examples in that way comes from the fact that for a Given vector of weights (or pairwise energy parameters) $\vec{w}$,

$$h_k \equiv \vec{w} \cdot \vec{X}^k = E^k_{pair} - E^0_{pair} \tag{4.18}$$

and hence, if $h_k > 0$, configuration $k$ has larger energy than the native state.

The purpose is to find a vector $\vec{w}$ with which $h_k > 0$ for any configuration $k$ from a set of configurations of the given protein and eventually for any native state from a given database.

One way to find such $\vec{w}$ is to use a perceptron with 210 inputs (the vector $\vec{X}$ of examples). The synaptic weights are associated to the parameters of the pairwise energy ($\vec{w}$). The vector $\vec{X}$ of examples is not normalized but that of the pairwise energy parameters is normalized, $|\vec{w}| = 1$.

### *4.3.2 The Perceptron Algorithm*

I implemented the boolean perceptron using the algorithm developed by Nabutovsky and Domany (Nabutovsky & Domany, 1991). I proceed presenting it briefly.

- Initialize $\vec{w} = \vec{X}_1 / |\vec{w}|$ and $d = 1/|\vec{w}|$.
- Present the examples cyclically and whenever $h_\mu = \vec{w} \cdot \vec{X}_\mu \leq 0$, a learning step is taken.

The Learning step consists on updating $\vec{w}$ and $d$ in the following way:

$$\vec{w}_{new} = \frac{\vec{w} + \eta \vec{X}_\mu}{\left| \vec{w} + \eta \vec{X}_\mu \right|} \qquad (4.19)$$

$$d_{new} = \frac{d + \eta}{\sqrt{1 + 2\eta h_\mu + \eta^2 N}} \qquad (4.20)$$

where

$$\eta = \frac{-h_\mu + 1/d}{N - h_\mu / d} \qquad (4.21)$$

and $N$ is the number of inputs, in our case $N = 210$. The learning process stops whenever one of three conditions is met:

1. $h_\mu > 0$ for all $\mu$.
2. The quantity $d$ called "despair" reaches a critical value $d_c$ given by
$$d_c = \frac{N^{(N+1)/2}}{2^{N-1}} \qquad (4.22)$$
3. $h \le 0$ for at least one example on each presentation of the set during 10000 consecutive presentations of the whole set of examples.

Whenever the perceptron successfully learns all the presented examples the learning process stops due to the first condition. In Practical terms in the case of an unlernable set of examples, the CPU time needed for $d$ to reach $d_c$ would be prohibitively high, for such cases the third condition is useful.

Using the Nabutovski-Domany perceptron algorithm it was found that $SET_{154}$ is learnable, i.e. a set of 210 parameters that properly classify all 154 native structures as having lower energy than any of their corresponding decoys was found. It is however wrong to assume that it is possible to learn any database (for example all PDB files) or utilizing different contact definitions. As it turns out (Vendruscolo et al., 2000), it is possible to define a region of learnability in the phase space of $M_p$, number of proteins in the learning set, and $R_c$, the contact threshold distance (Figure 16). Furthermore, the solution obtained learning one set of proteins cannot properly classify all proteins contained in a different set. This is exemplified by the fact that learning a subset of proteins (mainly-$\alpha$, mainly-$\beta$ or $\alpha$-$\beta$) cannot be used to properly classify those proteins belonging to the other classes (Table IX).

The results presented above are my contribution to a larger study which proved utilizing the same methodology that even for a single protein one cannot find a set of 210 pairwise contact parameters able to discriminate the native state against a set of challenging decoys (Vendruscolo et al., 1999).

Table IX. Results of Gapless Threading Fold Recognition

| Class potential | No. of Proteins in test Set/ No. of decoys in test set | Misclassified proteins/ misclassified decoys | Percentage of misclassification (%) |
|---|---|---|---|
| $\alpha$-potential | 82<br>713,182 | 19<br>12,617 | 23<br>2 |
| $\beta$-potential | 97<br>931,222 | 21<br>27,660 | 22<br>3 |
| $\alpha$-$\beta$-potential | 68<br>422,576 | 9<br>11,390 | 13<br>3 |

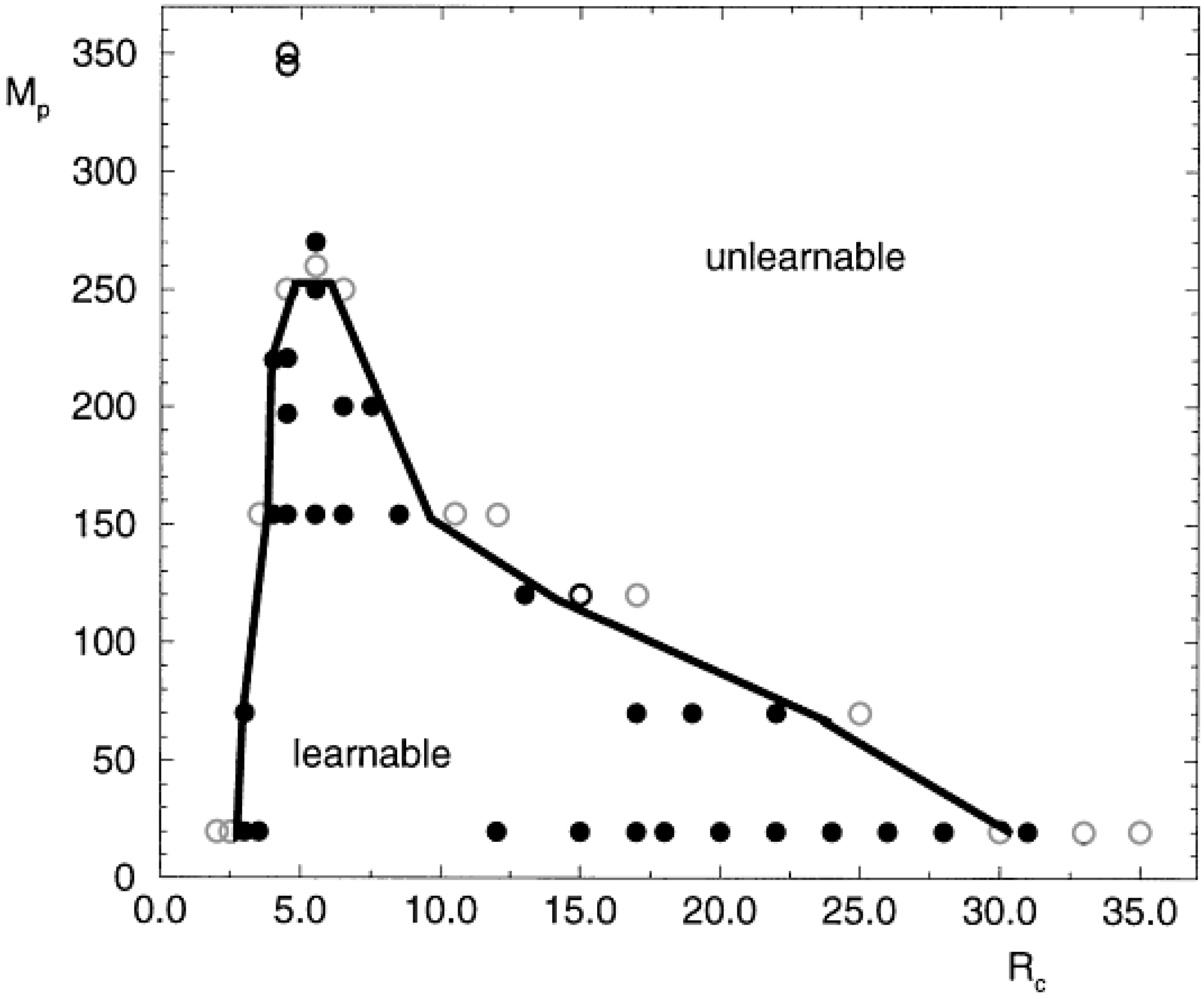

**Figure 16**. Region of learnability for the all atoms definition of contact. Several sets of $M_p$ proteins were generated for each value of $R_c$. Full circles indicate that all the sets considered for a particular value of ($M_p, R_c$) were learnable; otherwise, we use open circles.

# 5. FLEXAID: FLEXIBLE ARTIFICIAL INTELLIGENCE DOCKING APPROACH

A central goal of this study is the creation of a docking approach that takes in account side chain flexibility. As seen in Chapter 2, the inclusion of side chain flexibility is a feasible task due to the fact that in general only a small fraction of binding site residues undergoes conformational changes. Such an algorithm has been created and is called FlexAID for its ability to perform genetic-algorithm based docking of small ligands permitting side chain flexibility on a pre-defined set of residues. FlexAID is described in detail in the following sections.

## *5.1 PROTEIN AND LIGAND REPRESENTATION*

A "full heavy-atom" representation is utilized for both, the protein and ligand molecules, that is, all atoms are taken in consideration except for hydrogen atoms. The chemical identity of an atom is important: First, for defining its atom class which is necessary for calculation of the score of an interaction using the Complementarity Function (section 4.1 and Tables V or VI) and, second, in defining the atomic radius which is required by the Complementarity Function. Covalent bonds are set in advance by means of lists defining which atoms are covalently bound to every atom in the system. Thus, even if a putative ligand position brings one of its atoms closer than say 2.0 Å, a covalent bond will not be assumed to exist.

The atomic coordinates of the protein are read as part of the input; water molecules are ignored. Upon encountering alternative locations for a given atom, a relatively common situation PDB files, the first alternative, which is the most probable by PDB format rules, is utilized. The covalent bonds within the protein and atom types are defined in advance. This has the implication that disulfide bonds are ignored. At least in principle, the chemical nature of the sulfur atom in terms of its interactions with other atoms could be different in the oxidized state (as part of the disulfide bond) and the reduced state (as part of a cysteine not making a disulfide bond). Furthermore, ignoring disulfide bonds has another consequence of more practical importance. Were such a cysteine inadvertently set flexible, and a docking solution found containing a different conformation for the residue without checking the stability of the protein as a whole, one could not claim the effect on stability of the rotamer change to be of minor importance.

The ligand to be docked is represented in terms of internal coordinates. The internal coordinates as well as their values are defined in input files. Ligand atom types are defined in advance. However, the user has the possibility to accept an automatic atom type assignment generated by an auxiliary program created to simplify the process of generating the ligand input files.

Every atom in the system is labeled as rigid or flexible. Rigid atoms are all atoms belonging to the protein backbone, side chains of residues not allowed to be flexible and ligand atoms other than those of the ligand to be docked. The atoms of the ligand to be docked are set as flexible.

## *5.2 Scoring of Putative ligand-protein complexes*

The driving force behind the search for the optimal conformation of a ligand-protein complex is the score given to each putative ligand-protein complex along the optimization. The scoring function used for that purpose is the Complementarity Function, described in detail in section 4.1. It is important to stress that the scoring function does not consider intra-molecular interactions (i.e., interactions between atoms of the same “unit” which may be a ligand or an amino acid). The consequences of this will be discussed at a latter section.

## 5.3 SEARCH PROCEDURE

In three-dimensional space, a total of six variables (one translation and one rotation around each axis) are required to position a rigid ligand with respect to a rigid protein. Alternatively, using three auxiliary fixed points in space and three ligand atoms, here called global positioning atoms, one can define a set of generalized internal coordinates that can be used to position a ligand unequivocally in space (Figure 17). The three fixed space points, $A = (B_x + 1, B_y, B_z)$, $B = (B_x, B_y, B_z)$ and $C = (B_x, B_y + 1, B_z)$ are defined with respect to point $B$ fixed at the center of geometry of the protein. Three atoms of the ligand, say $(O, N, M)$ used together with the points $(A, B, C)$ define a set of six generalized internal coordinates, namely:

1. Distance $MA$.
2. Angle $N\hat{M}A$.
3. Angle $M\hat{A}B$.
4. Dihedral angle around vector $\overline{NM}$ defined by atoms $(O, N, M, A)$.
5. Dihedral angle around vector $\overline{MA}$ defined by atoms $(N, M, A, B)$.
6. Dihedral angle around vector $\overline{AB}$ defined by atoms $(M, A, B, C)$.

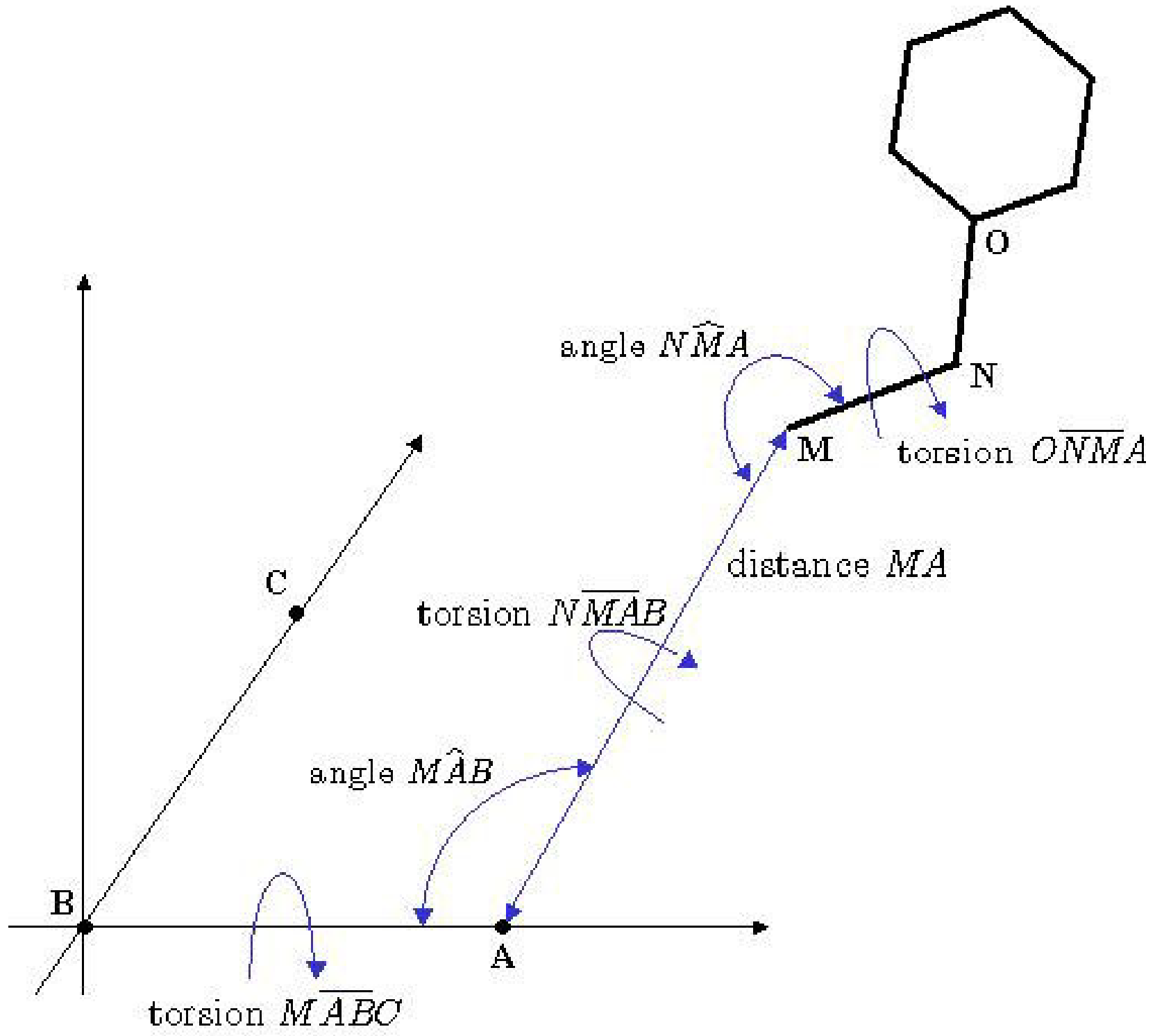

**Figure 17**. Generalized internal coordinates. Three fixed points in space $(A, B, C)$ and three ligand atoms, $(O, N, M)$, are used to unequivocally define the ligand position in space.

The six generalized internal coordinates described above are the variables that are optimized by FlexAID during the search procedure. The three ligand atoms utilized to define the generalized coordinates, atoms $(O, N, M)$ in Figure 17, called global positioning atoms, are defined as part of FlexAID input.

Apart from optimizing the six generalized coordinates, which is equivalent to performing RLRP docking (rigid-ligand, rigid-protein), FlexAID has the possibility of setting flexible any number of dihedral angles belonging to either the ligand or amino acid side chains. However, aside from time considerations, due to the fact that the Complementarity Function does not consider intra-molecular interactions, it would not be appropriate to simulate ligand or side chain flexibility in that manner. Such simulations would be permissible when the Complementarity Function is altered to take into consideration intra-molecular interactions.

The search technique utilized in FlexAID is a genetic algorithm. The next section gives a brief introduction to the theory of genetic algorithms in order to put in context the genetic algorithm specifications implemented in FlexAID.

## *5.4 Genetic Algorithms*

The basic idea of a genetic algorithm is to map (represent) possible solutions of the (docking) problem into strings of characters (of a finite alphabet), such that one can have a population of strings (each representing a possible solution) that can be classified using a fitness function (based on a scoring function). At each generation, the fittest strings have the higher probability to reproduce (and therefore survive), propagating to the next generation the present best solutions. Variability is introduced by using various operators such as the crossover and mutation operators among the population. After a pre-defined number of generations, the population is likely to be dominated by the strings that code for the best solutions to the problem, this convergence has been proven and is known as the Schemata theorem (Goldberg, 1989; Davis, 1991; Spears, 1998).

The implementation of a genetic algorithm implies the choice of several methods as well as parameters associated with them. The choices described below are the result of studying the theory and methods presented in the books by Goldberg (Goldberg, 1989) and Davis (Davis, 1991). In what follows I will present shortly the mathematical foundation of GAs.

As mentioned before, a Genetic Algorithm (GA) is a function optimisation technique. A random initial population will evolve throughout the generations towards a solution of the problem at hand and most likely this solution will be the global solution, i.e., the global extreme (either a minima or maxima depending how the problem is posed). The underlying principle behind the success of GAs is understood using the idea of schemata.

A schema is a similarity template of a string. For example, given a binary alphabet **K**=[0,1] a schema is represented by using the wildcard operator as follows:

$$H=0*111*=[001110, 001111, 011110, 011111] \tag{5.1}$$

The schema presented above requires that all positions denoted by 0 or 1 be fixed as specified while those represented by * can take any value. The number of fixed positions in a schema is called 'schema order' and is denoted as O(H). The distance (in bits) between the two most-distant-defined bits is called 'defining length' and is denoted as δ (H). In the case presented above we have O(H)=4 and δ(H)=4. John Holland (Holland, 1975) proved a theorem (Schemata Theorem) involving the concepts presented above that is recognised as the mathematical foundation of Genetic Algorithms. Without further explanations I will present the Schemata Theorem:

$$m(H,t+1) \geq m(H,t)\frac{f(H)}{\bar{f}}\left(1 - p_c \frac{\delta(H)}{l-1}\right)\left(1 - O(H)p_m\right) \tag{5.2}$$

What the Schemata Theorem says is that the number of strings representing schema H in the new generation $m(H,t+1)$ is larger or equal to the present number $m(H,t)$ multiplied by the ratio of the average fitness of schema H, $f(H)$, and the average fitness of the whole population, $\bar{f}$. This functional form leads to an exponential change in the number of schemata proportional to the amount its average fitness differs from that of the whole population. The growth in the number of above average fitness schemata is solely due to reproduction. The inequality sign is used since the destruction of other schema might create new instances of the one in consideration.

The two terms in parenthesis in the right hand side of the theorem represent respectively the effect of one-point crossover and mutation on the GA evolution. The larger the defining length of a schema the smaller the probability it will survive intact a one-point crossover operation. Likewise, the larger the schema order, the bigger the chance the schema will be destroyed by the mutation operator. In one sentence the Schemata Theorem says that *Schemata of short defining length, low order and above average fitness receive exponentially increasing trials in future generations.*

Although seemingly detrimental, the effect of crossover and mutation is of utmost importance, without them no new points in parameter space would be searched and the GA evolution would simple amplify the best individual in the initial population. By applying crossover and mutation one creates new individuals that will be amplified or not according to their own merit with respect to the rest of the population.

In the previous discussion I used three operators: Reproduction, One-point crossover and mutation. The reproduction method that led to the schemata theorem is called generational replacement where each string produces a number of strings (equal to itself) in the next population directly proportional to its present fitness (using Roulette-Wheel parent selection). It may happen that a very well fit individual is eventually destroyed upon crossover and/or mutation. This is obviously an unwanted effect that might cause the maximal fitness to decrease in consecutive generations:

$$\max(f_i(t+1)) \leq \max(f_i(t)) \tag{5.3}$$

Other reproduction techniques that correct this are:

- Elitism
- Steady State Reproduction
- Steady State Reproduction without duplicates

### *5.4.1 Reproduction Operators*

ELITISM The elitism strategy proposes that a new population is created using Roulette-Wheel parent selection, the crossover and mutation operators are applied as specified by the algorithm and the best fit individual of the previous generation is copied over, while the last fit individual is discarded. Thus, the maximal fitness will always be larger or equal to that of the preceding generation:

$$\max(f_i(t+1)) \geq \max(f_i(t)) \tag{5.4}$$

STEADY STATE REPRODUCTION This technique creates n new individuals and replaces the n last fit individuals of the previous generation to create the present population. When $n = N$ (population size) Steady State Reproduction equals generational replacement, when $n = N - 1$ it corresponds to elitism. In practice a value of $n = 1,2$ is generally used (Davis, 1991).

STEADY STATE REPRODUCTION WITHOUT DUPLICATES One challenge to the GAs performance is that very often, when using Roulette-Wheel parent selection, after a certain number of generations, when best fit individuals start to take the lead, many copies of those individuals will be created and possibly maintained undisturbed by the crossover and mutation operators. This is a heavy toll on the already scarce computational resources available. To deal with this problem it seems advantageous (in terms of performance) to spend the time checking whether or not a new individual already exists in the population.

Steady State without Duplicates is the same as the previous one with the difference that the n new individuals are all different from themselves as well as from those present in the previous generation's population.

### *5.4.2 CROSSOVER OPERATOR*

One of the shortcomings of a one-point crossover operator is that is decreases the survival probability of above-average fitness, large defining-length schemata. To remedy this fact there are at least two alternative crossover operator candidates:

- N-point crossover
- Uniform Crossover

N-POINT CROSSOVER In N-point crossover, N crossover points are randomly chosen. The exchange of substrings proceeds as follows:

- Label the crossing points from left to right as 1 to m.

- If m is odd, treat the string as a circle, e.g., $m+1=1$.
- Exchange between the parents the material between point 1 and 2.
- Skip to point 3
- Exchange between the parents the material between point 3 and 4.
- Proceed until all points have been accounted for.

The most applied operator of this class is the 2-point crossover operator. On which one randomly chooses 2 crossing points and the material between them is exchanged between the parents.

The use of N-point crossover is a mixed blessing in the sense that by choosing more than one point of crossover the bias towards small defining length is decreased. But, considering each possible choice of defining length(s) of the operator, i.e. the distance between two consecutive crossover points, as a different operator, the increase in the number of operators is actually diminishing the chance that each of these operators is chosen in the course of the GA evolution. For this reason 2-point seems to be a reasonable compromise between decreasing the bias for small $\delta(H)$ while not diminishing the operator performance and is more frequently chosen (Davis, 1991; Spears, 1998).

UNIFORM CROSSOVER Uniform crossover is an operator that at each position in the string decides with a certain probability whether or not to exchange the value of that particular bit between the parents.

### *5.4.3 SHARING*

Sharing is a niche formation technique that mimics nature, where the individuals sharing a specific geographical area need to share the resources present in it with the consequence that well fit individuals in an over populated region reproduce less frequently than (perhaps lesser-fit) individuals in yet under-populated regions.

Sharing has the advantage of increasing the reproduction probability of chromosomes in isolated regions of parameter space, improving the sampling of that area at the expense of slowing down the sampling of super-populated areas. Sharing re-scales the fitness of individuals as follows:

$$F_i = \frac{f_i}{\sum_{j=0}^{N} S(d_{ij})} \tag{5.5}$$

with

$$S(d_{ij}) = \begin{cases} 1 - \left(\frac{d_{ij}}{\sigma}\right)^{\alpha} & \text{if } d_{ij} < \sigma \\ 0 & \text{otherwise} \end{cases} \tag{5.6}$$

where $\alpha$ and $\sigma$ are user defined parameters and $d_{ij}$ is the hamming distance between two chromosomes, i.e. the number of differing bits between the two chromosomes.

### *5.5 POPULATION BOOM: A NOVEL REPRODUCTION TECHNIQUE*

In this section I develop a new reproduction strategy that presents some advantages with respect to the existing techniques described in sub-section 5.4.1. The reproduction techniques described ensure that the maximal fitness does not decrease (Equation 5.4). However, I can go one step further and devise a reproduction technique such that not only the maximal fitness will not decrease, its average value will not decrease as well:

$$\bar{f}(t+1) \geq \bar{f}(t) \tag{5.7}$$

To implement this new requirement:

- Create $n$, $(0 < n \leq N)$, new strings with Roulette-Wheel parent selection.
- Apply the Crossover and Mutation operators as required.
- Add this new sub-population to the old one.

- Select from it the $N$ best-fit individuals.

In the worst case when all newly created individuals are worst that the worst individual already present in the population we will have the old population copied to the new generation. On the other hand, in the best case, when $N = n$, we will have $\min(f_i(t+1)) > \max(f_i(t))$.

POPULATION BOOM WITHOUT DUPLICATES The idea is the same as that of Population Boom and like previously explained the $n$ new individuals are required to be different from each other as well as from the previous generation's individuals. The steps in this case are the following:

1. Copy the old population to the new one.
2. Select 2 parents using Roulette-Wheel parent selection.
3. Apply: with probability $p$ the mutation operator otherwise apply the crossover operator.
4. Check if any of the children is different from those in the population.
5. If so copy them to the population.
6. If the population did not reach $N+n$ individuals return to item 2.
7. Select the $N$ best-fit individuals.

## *5.6 GENETIC ALGORITHM IMPLEMENTATION*

Having discussed the various techniques and operators used in GA implementations I can present now the specifications implemented on FlexAID.

I use binary representation, with $l$ bits per gene that represents a parameter given by:

$$l = \text{int}\left( \frac{\ln\left( \frac{R_{\max} - R_{\min}}{p} + 1 \right)}{\ln(2)} \right) + 1 \quad (5.6)$$

where $p$ is the desired precision with which one wishes to sample the parameter inside the pre-defined interval $\{R_{\min}, R_{\max}\}$.

All reproduction techniques discussed in sections 5.4.1 and 5.4.4 as well as Generational Replacement are implemented in FlexAID. Mutation is implemented via a probability to choose to apply the mutation operator, $p_m$, and a different probability of flipping a bit (mutation rate), $p_r$. The crossover operator is 2-point crossover, applied with probability $p_c = 1 - p_m$. Once two parents were selected using roulette wheel parent selection, the mutation or crossover operators are applied respectively according to the probability $p_m$. Furthermore, $p_m$ can be set variable, as a bounded linear function of the number of generations:

$$p_m(g) = p_m^{initial} + \frac{g}{g_{\max} - 1}\left(p_m^{final} - p_m^{initial}\right) \tag{5.7}$$

where $g = [0, g_{\max} - 1]$.

Using chromosome score (in our case CF value) as fitness presents the problem that the more a population has converged, the smaller the advantage of the better-fit individuals with respect to the rest of the population (see equation 5.2). To keep a constant evolutionary advantage for the fittest individuals we use ranking as fitness. The idea is to sort all chromosomes according to score and assign a fitness value from 1 to number-of-chromosomes to each chromosome according to its position in the sorted list. Besides keeping a constant evolutionary advantage for the fittest individuals, ranking does not allow a super-individual (relative to the rest of the population) to take over in the beginning of the GA evolution (when the rest of the population can have much poorer scores).

Preliminary docking experiments using FlexAID failed to reach a solution sufficiently close (in terms of RMSD of atomic positions) to the experimentally determined structure of the complex (i.e., the one found in the PDB file). Inclusion of Sharing as a niche formation technique drastically improved our ability to reach the experimental solution.

A second improvement introduced in our genetic algorithm implementation is that of having the possibility of initializing the GA population partially or completely with pre-selected solutions, thus making it possible for an initial simulation to be coarse grained, and subsequently utilizing the solutions found as part of the initial population in a more detailed simulation. This is done by reading the desired values of the optimization variables from a file and finding the binary representation for each gene that best approximates that desired real value for each of the variables.

## *5.7 Treatment of side chain flexibility*

As described in section 5.3, FlexAID has in principle the capability to treat any number of side chain dihedral angles chosen in advance as optimization variables besides the six rigid-body variables. However, the Complementarity Function, as presently formulated, is not able to account for intra-molecular interactions. The necessity of a detailed consideration of the intra-molecular interactions, but most important, the minimal requirement of forbidding steric clashes between non-covalently-bonded atoms, makes the inclusion of intra-molecular interactions in the Complementarity Function a pre-condition to treating side chain dihedral angles as optimization variables.

Current practical approaches allowing side chain flexibility on docking simulations have mostly followed two paths (see Chapter 1 for details), that of using different snapshots (e.g., NMR models) of the protein molecule, subsequently using this composite picture as docking target, or by allowing severe steric clashes between ligand and protein atoms. The latter approach has two shortcomings: First, it assumes that having placed the ligand in a given position which makes clashes, the side chains involved would somehow adopt a different conformation and permit the ligand to be placed in the chosen position; Second, it does not take in consideration the interactions between ligand atoms and those atoms with which it makes severe steric clashes, since that would require the protein atoms to be placed in realistic positions.

The approach used in FlexAID for introducing side chain flexibility is that of building alternative side chain conformations and performing an exhaustive search over all alternative conformations of flexible side chains that might make contacts with any ligand atom for a given putative ligand position. In order to be assigned as putatively in contact with the ligand, the distance between the side chain $C_\beta$ atom and any ligand atom needs to be smaller than a threshold of 10 Å.

Alternative side chain conformations are created utilizing a backbone independent rotamer library (Lovell et al., 2000). All rotamers not making severe steric clashes with atoms labeled as rigid are built by FlexAID in advance for each of the chosen flexible residues. A threshold distance of 2.0 Å is used to probe whether two atoms are making a severe steric clash. Furthermore, taking in consideration that rotamer angle values are the mean values of the respective dihedral angle distributions, if a given rotamer is rejected due to steric clashes using mean values for each of the dihedral angles, all combinations of dihedral angle values differing by $\Delta\chi = \pm 15°$ of the mean are considered in turn until one is accepted. Thus, for a given rotamer, every angle can be left unchanged or be shifted by $\Delta\chi = \pm 15°$, a total of three states. Therefore, a side chain containing $n_d$ dihedral angles generates a total of $3^{n_d}$ combinations to be probed, where the first one is that with all angles unchanged. For example, in the case of Lysine, containing 4 dihedral angles, a maximum of $80$ conformations associated to a given rotamer would be probed besides the rotamer mean values. It is important to stress that each flexible residue has a specific number of rotamers with specific $\chi$ angle values. For example, Glutamines have a total of 9 different rotamers in Lovell's rotamer library. Two different Glutamines set flexible may have different numbers of available rotamers (but no more than 10, the additional conformation being that present in the input PDB file) and a given rotamer accepted for both Glutamines might have a different combination out of the 27 possible combinations of $\Delta\chi = 0°, \pm 15°$ for each of its angles.

During the genetic algorithm optimization, the following steps are taken when calculating the score of a given putative ligand position (chromosome):

1. The score of the ligand interacting with the protein using the experimentally determined conformation of all residues is calculated independent of the fact that some of the residues in contact may be flexible. If it is detected that the ligand makes severe steric clashes with protein atoms labeled as rigid, the calculated score is assigned to the given chromosome; otherwise, step 2 is taken. The rational is that there is no point on probing different side chain conformations because the score will in any case be dominated by the unfavorable steric clash term (Equations 4.2 and 4.3).
2. A list is made of all flexible residues that might be in contact with the ligand. If the list is empty, as might be the case when only a few residues are allowed to be flexible during a global optimization, the score calculated in step 1 is assigned; otherwise, step three is taken.
3. An exhaustive search is performed over all possible combinations of alternative conformations of each of the selected flexible residues, to determine the combination with the highest Complementarity Function value. This value is then assigned to the given putative ligand position (chromosome) instead of that calculated in item 1. Furthermore, the information on which combination of alternative conformations was selected is stored for possible future use (in case that putative ligand position is selected as one of the final solutions) so that each flexible side chain conformation can be set appropriately. It might happen that of all possible combinations the one using the original flexible side chain conformations is still the best one. In this case, the resulting CF value is the same as that calculated in item 1.

The algorithm described here does not incur a high computational cost in terms of time for including side chain flexibility into the docking simulation for a number of reasons: 1. Not all side chain rotamers for a given residue are included in the exhaustive search; 2. The permitted alternative side chain conformations are built in advance, thus no extra time is required to probe alternative side chain conformations; and finally, 3. Only a fraction of putative ligand solutions during the genetic algorithm optimization require an exhaustive search to be performed.

The FlexAID approach uses a rotamer library to create alternative side chain conformations but does not take in consideration possible unfavorable interaction between side chain atoms in alternative conformations and other protein atoms other than by checking for the occurrence of severe steric clashes. When comparing different PDB files representing the same protein structure (also containing the same ligands), as described in sub-section 2.4.6, one finds that amino acid side chains display intrinsic flexibility. Thus, one can argue that in most cases the energetic considerations (aside from clashing) of altering the side chain conformations of a small number of residues can be safely ignored.

The method described here has potential advantages over earlier approaches (Mangoni et al., 1999) that treated side chain flexibility by means of permitting severe steric clashes between ligand and protein atoms. Here physically permissible protein structures are actually used during the simulation, thus guaranteeing that a given docking solution is feasible.

The method used by FlexE (Claussen et al., 2001) could be seen as being more general in the sense that by utilizing a set of structures to create a combined protein description one can in fact take in consideration not only different side chain conformations but also point mutations as well as different loop conformations. One could create a set of structures utilizing a rotamer library and in this situation FlexE would be similar to FlexAID. However, FlexE is restricted to utilize a maximum of 16 different conformations. Also, FlexE does not generate the different structures itself and, therefore, a user who whishes to utilize different side chain conformations in the different base set structures, needs to create them with other programs. Last, FlexE is restricted to local docking, that is, to a small area containing the binding site. FlexAID on the other hand, can build any number of side chain rotamers without the need of any auxiliary program and can perform global searches. In other words, any previous knowledge of the location of the binding site is not required, a definite advantage when dealing for example with homology modeling structures, where side chain conformations may not be correct as well as the binding site be unknown.

## *5.8 FLEXAID DOCKING OUTPUT*

The genetic algorithm optimization finishes after a pre-set number of generations returning the final population sorted according to CF values. Each chromosome in the population determines a possible ligand-protein complex structure.

Given the array of chromosomes sorted in descending order of CF values, the solutions are clustered as follows:

1. The solution with highest CF value still not assigned to a cluster is determined. This solution is called a cluster top solution and the cluster is said to be active. Thus, the chromosome with the highest CF value in the population is the cluster top of the first cluster.
2. All other chromosomes that were not yet assigned as members of a cluster are assigned to the active cluster if their RMSD with respect to the ligand atom coordinates of the cluster top solution is lower than 2.0 Å. By definition, all chromosomes belonging to a cluster have CF values lower than that of the cluster top solution.

3. If there are chromosomes that were not yet assigned to a cluster, return to item 1, otherwise end.

The docking solutions generated by FlexAID are based on the cluster top solutions described above. The number of solutions generated by FlexAID is given by:

$$N_{solutions} = \max\left[N_{clusters}, S_{\max}\right] \tag{5.8}$$

where $N_{clusters}$ is the total number of clusters obtained and $S_{\max}$ (maximum allowed number of solutions) is equal to 10% of the number of chromosomes utilized during the simulation. The first top $N_{solutions}$ clusters (according to the CF value of the cluster top chromosome) are utilized in turn to generate separate atomic coordinates files in PDB format. A file is created containing information on the number of chromosomes associated to each cluster as well as the RMSD values between the cluster top solutions. Furthermore, a file is created containing the values of each of the six optimization variables (Section 5.3 and Figure 17) as well as their RMSD with respect to both their cluster top solution as well as to reference structure (if one is provided as part of the input). The utilization of a reference structure is of most interest when comparing the solutions found by FlexAID to that of an experimentally determined complex.

## *5.9 Evaluation of Performance*

We evaluate the performance of FlexAID in two different instances, that of local docking simulations, when the search is restricted to an area encompassing the binding site, and that of global simulations, where no information on the location of the binding site is utilized. These two situations are evaluated with different criteria. At the moment, FlexAID is not envisioned as a docking program to be used in high-throughput applications and, therefore, we are content if at least one among the top 10 solutions satisfies the success criteria. Additional steps are therefore required to judge the solutions obtained by FlexAID to determine which are more appropriate according to other criteria.

Local docking simulations are evaluated in terms of the root mean square distance (RMSD) of the coordinates of the ligand atoms from those obtained experimentally. A threshold value of 2.0 Å is used to judge whether a local docking simulation successfully found the ligand position. A threshold value of 2.0 Å is a commonly used value (see references cited in Chapter 1).

Global docking simulations are evaluated in terms of the minimal distance between any two atoms, one belonging to the simulated ligand, the other to the experimentally determined ligand (MIND). A threshold value of 3.0 Å is used to judge whether a global docking simulation has been successful. This criterion is devised to signal whether the binding site is found, rather than the position of the ligand inside the binding site. This criterion is clearly less stringent than that for local simulations.

The value chosen for the MIND threshold was determined empirically. As an example, Figure 18 shows the different positions of several global rigid docking solutions for the carbonic anhydrase II-PTS complex (PDB code 1cim) with different values of MIND. All solution for which $MIND < 3.0\text{Å}$ are clustered very close to the true position of the ligand (blue) while other solutions with larger values for MIND are not located inside the experimentally determined binding site.

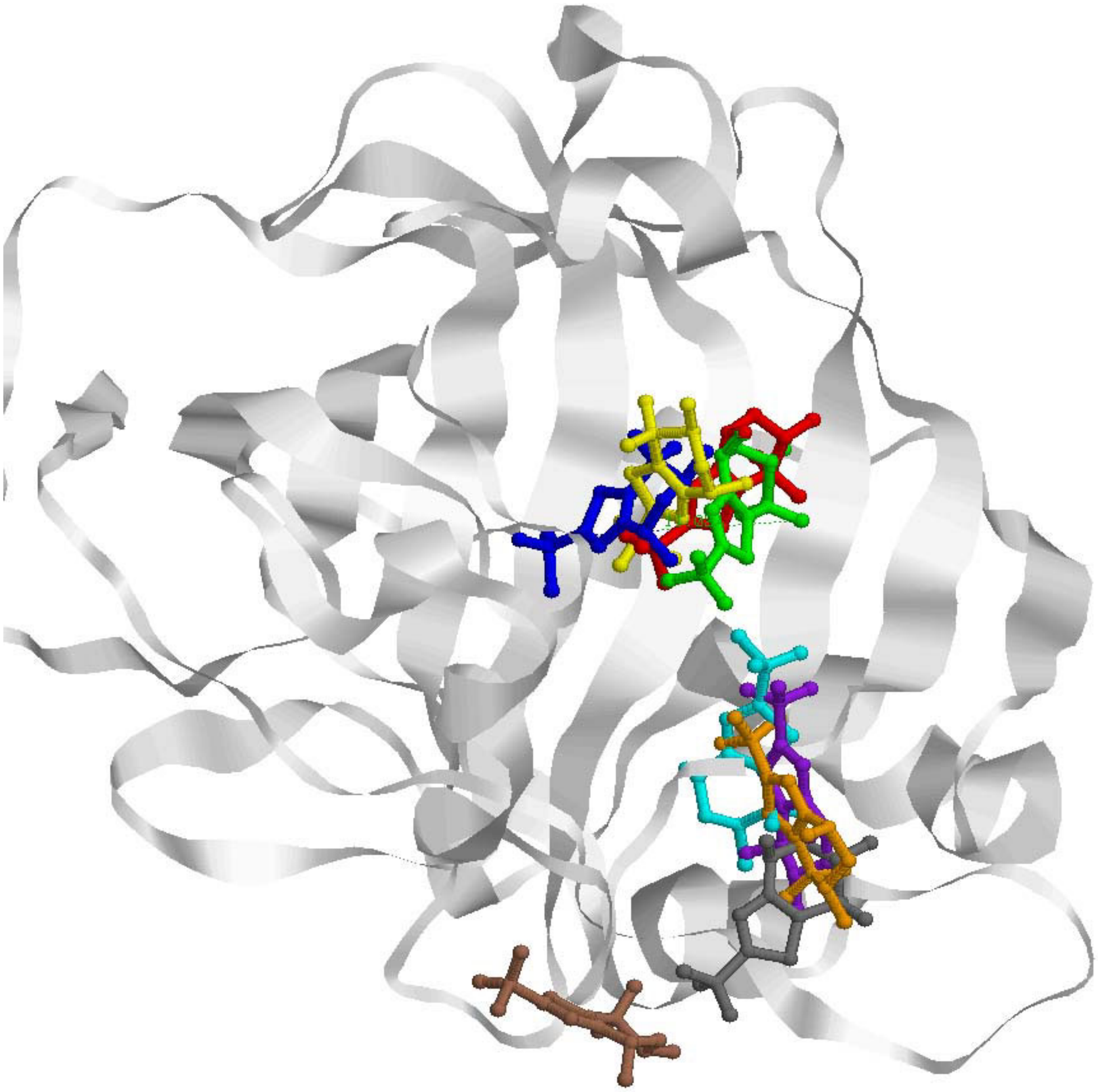

**Figure 18**. Relation between MIND value and ligand positon. Relative positions of different global rigid docking solutions displaying different values of MIND with respect to the experimentally determined ligand position (blue). The complex shown is that of Carbonic Anhydrase II with the ligand PTS. The values of MIND are as follows: blue: 0.00, red: 0.71, yellow: 1.17, green: 2.59, cyan: 6.33, purple: 7.97, orange: 9.21, gray: 13.01 and brown: 19.99.

Although we are satisfied in finding the binding site on a global simulation, in many cases the final solution actually satisfies the criterion of local simulations, that is, not only the binding site is found but also the correct position of the ligand. In principle, the information of the putative binding site locations obtained from the solutions of global simulations can be used to set the location of the binding site for a local docking simulation. This recipe has been successfully tested in one case.

## *5.10 Rigid docking test set*

A test set of eleven protein-ligand complexes (Sobolev et al., 1996) was used to evaluate the performance of FlexAID on local and global rigid docking simulations (Table X). The percentage of success is calculated as the fraction of complexes for which at least one solution among the top $T$ solutions satisfies the relevant success criteria. The performance of FlexAID is analyzed considering the top one ($T = 1$), five ($T = 5$) and ten ($T = 10$) solutions on both local and global simulations.

Given the probabilistic nature of the genetic algorithm optimization, successive docking simulations of a complex may generate different numbers of solutions (Equation 5.8). For this reason, the set of complexes are independently docked 5 different times and the mean percentage of success for the whole set calculated, the error bars correspond to the standard deviation from the mean percentage of success.

Table X. Rigid docking test set

| Protein Receptor | | Ligand[a] | | | |
|---|---|---|---|---|---|
| Name | Code | ID[b] | | | Name |
| Dihydrofolate reductase | 4dfr | MTX | A | 160 | Methotrexate |
| Aconitase | 7acn | ICT | - | 755 | Isocitrate |
| Thermolysin | 1tlp | RHA | I | 1 | Phosphoramidon |
| Thermolysin | 2tmn | PHO | I | 1 | P-*Leu/NH2 |
| Penicillopepsin | 1ppm | CBZ | I | 1 | Cbz-Z-Z-L(P)-(O)Pme |
| Carbonic Anhydrase II | 1cim | PTS | - | 262 | PTS |
| Adipocite lipid binding protein | 1lif | STE | - | 132 | Stearic Acid |
| Retinol binding protein | 1erb | ETR | - | 176 | N-ethyl retinamide |
| Ricin | 1fmp | FMP | - | 301 | Formycin-5'-monophosphate |
| Met-repressor | 1cmc | SAM | A | 105 | S-adenosyl methonine |
| Streptavidin | 1stp | BTN | - | 300 | Biotin |

[a] Ligands used in the rigid docking test set are extracted from the same PDB files of the protein receptors.

[b] Ligands are described by their three-letter PDB code, chain identifier and residue number. The symbol `-` describes the cases for which no chain identifier is associated to the ligand under consideration.

### 5.11 *FlexAID performance on rigid docking*

The performance of FlexAID on rigid docking simulations was evaluated in several situations. The following sub-sections describe each in turn. All simulations are performed using a population of 300 chromosomes, with initial random population, utilizing population boom without duplicates and a fixed mutation probability of $p_m = 0.5$ irrespective of the generation number.

### 5.11.1 EFFECT OF A SOLVENT TERM

The complementarity function (Sobolev et al., 1996) as described in section 4.1 does not take in consideration interactions with the solvent. To determine if there is any advantage in considering interactions with the solvent, we assume a continuum model of the solvent by considering the solvent accessible area of an atom as that area that is not in contact with any other atoms. Interactions with the solvent are introduced by adding a term to the Complementarity Function (Equation 4.1) as follows:

$$CF = S_f - S_u - E - \varepsilon_{sol} S_{free} \tag{5.9}$$

where $S_{free}$ is the solvent accessible area and $\varepsilon_{sol}$ represent the interaction term between an atom and the solvent.

As can be seen, the strength of the interaction between an atom and the solvent is independent of the atom type of the atoms involved in the interaction. Two different values of $\varepsilon_{sol}$ are considered: $\varepsilon_{sol} = [0, -2]$. The rationale of choosing a value of –2 for the interaction between an atom and the solvent is that when using the atom-type pairwise interaction parameters given by Table V, we are actually favoring atom-atom interactions (independent of the atom-types of the atoms involved) with respect to interactions with the solvent. That is, we are introducing a driving force into the optimization procedure whereby it is preferable for a ligand atom to make unfavorable atom-atom contacts rather than being solvent exposed. Such a term has been shown to be favorable when using contact surfaces to predict side chain conformations (Eyal et al., 2003b).

In Figure 19 we present the effect of the two values for the solvent interaction parameter on the percentage of success of local rigid docking simulations as a function of the number of generations for the top solution.

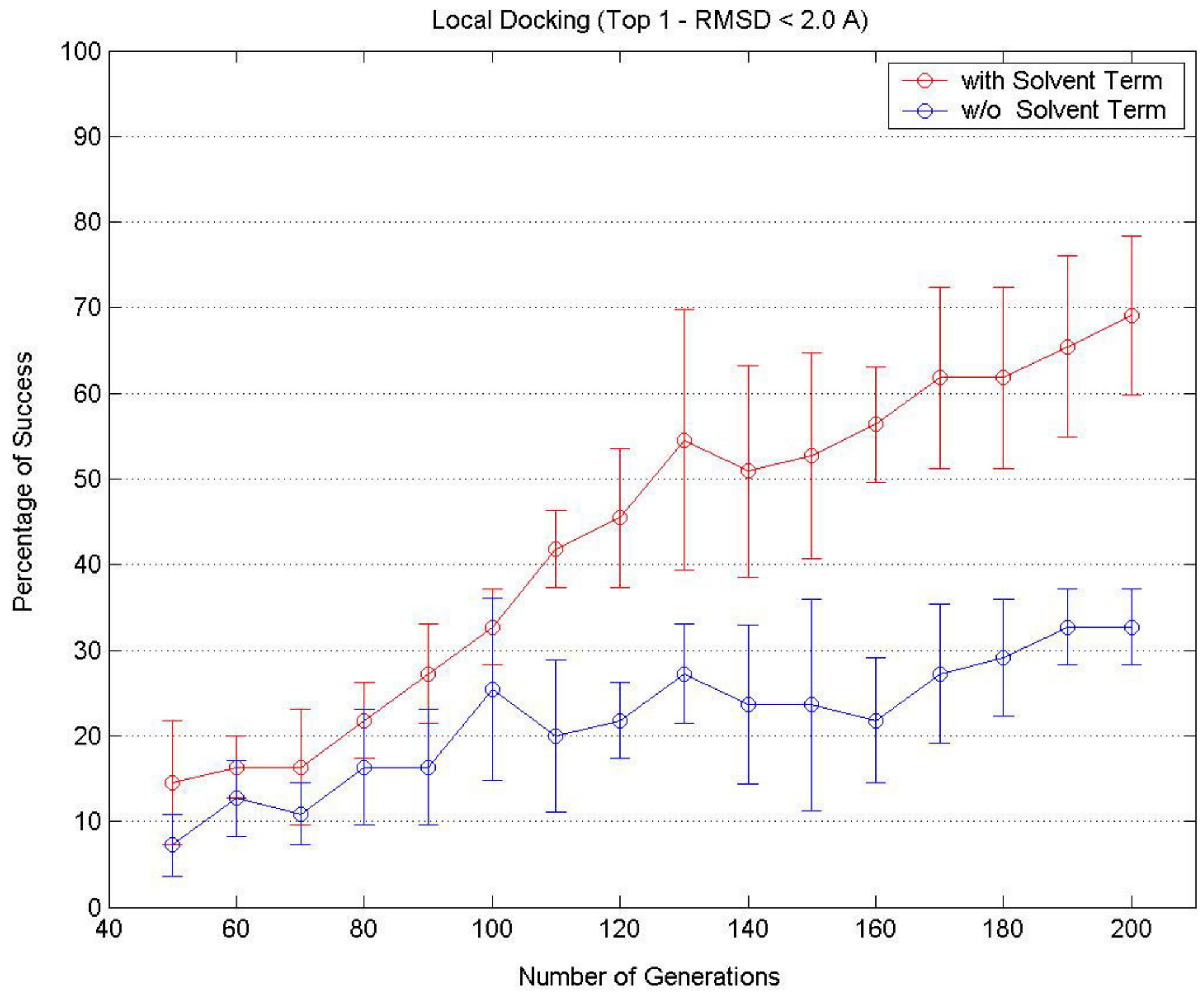

**Figure 19**. Percentage of success as a function of number of generations for two different values of $\varepsilon_{sol}$. In red we show the case where the solvent term is taken in consideration ($\varepsilon_{sol} = -2$) and in blue we show the case where the solvent term is not taken in consideration, $\varepsilon_{sol} = 0$, that is, when Equation 5.9 becomes equivalent to Equation 4.1.

The introduction of the solvent term increases the percentage of success drastically to an average level of approximately 70%. That is to say that in an average 70% of simulations (of 11 distinct complexes), the top solution found corresponds to the experimentally determined solution. If we consider the top 10 solutions instead of only the top solution we obtain a slight increase to an average of 75% percentage of success (Figure 20). This result shows that in most cases, if a solution corresponding to the experimental ligand pose is present among the top 10 solutions, it is most likely to be the top solution.

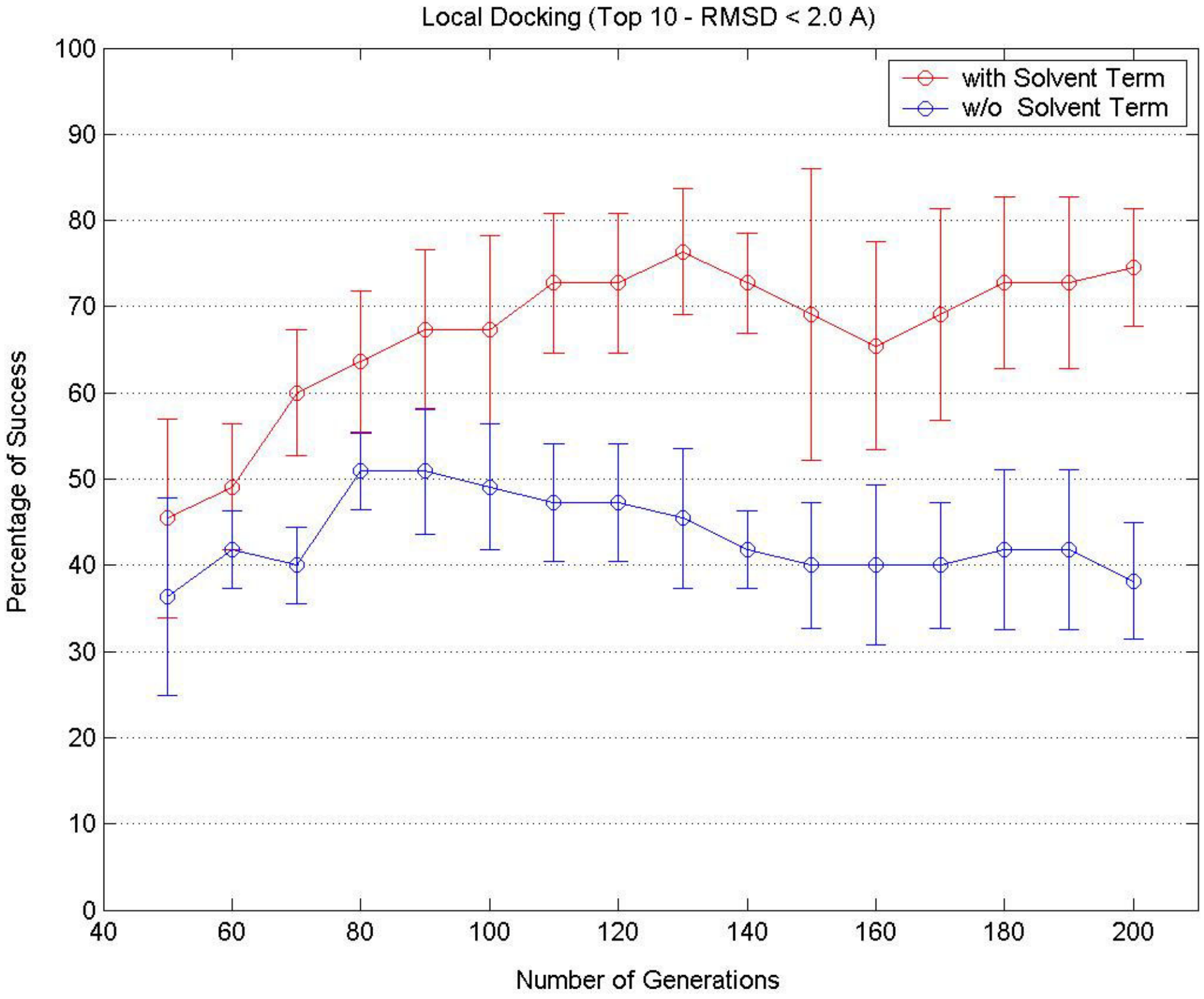

**Figure 20**. Percentage of success as a function of number of generations for two different values of $\varepsilon_{sol}$ considering the top 10 solutions. In red the case where the solvent term is taken in consideration ($\varepsilon_{sol} = -2$) is shown and in blue we show the case where the solvent term is not taken in consideration, $\varepsilon_{sol} = 0$, that is, when Equation 5.9 becomes equivalent to Equation 4.1.

If one relaxes the threshold RMSD value of 2.0 Å to a value of 2.5 Å, one finds that the percentage of success does not improve significantly when considering the top solution (Figure 21). However, when considering the top 10 solution a gain of 10% to a level of 80% success is reached, suggesting that a number of solutions in the top 10, other than the top one itself, are not too far from the experimental ligand position (Figure 22).

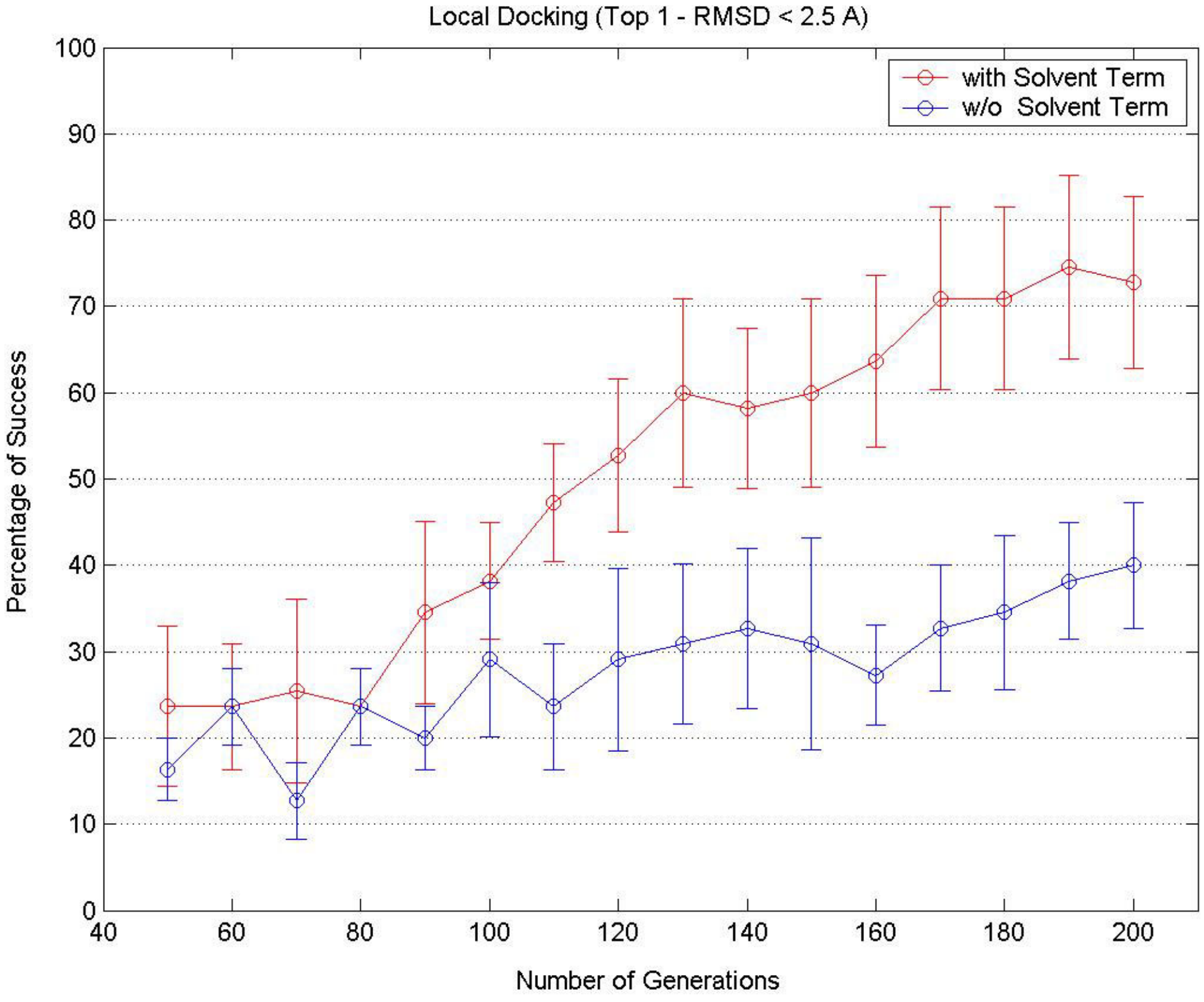

**Figure 21**. Percentage of success as a function of number of generations for two different values of $\varepsilon_{sol}$ for RMSD threshold of 2.5 Å. In red we show the case where the solvent term is taken in consideration ($\varepsilon_{sol} = -2$) and in blue we show the case where the solvent term is not taken in consideration, $\varepsilon_{sol} = 0$, that is, when Equation 5.9 becomes equivalent to Equation 4.1.

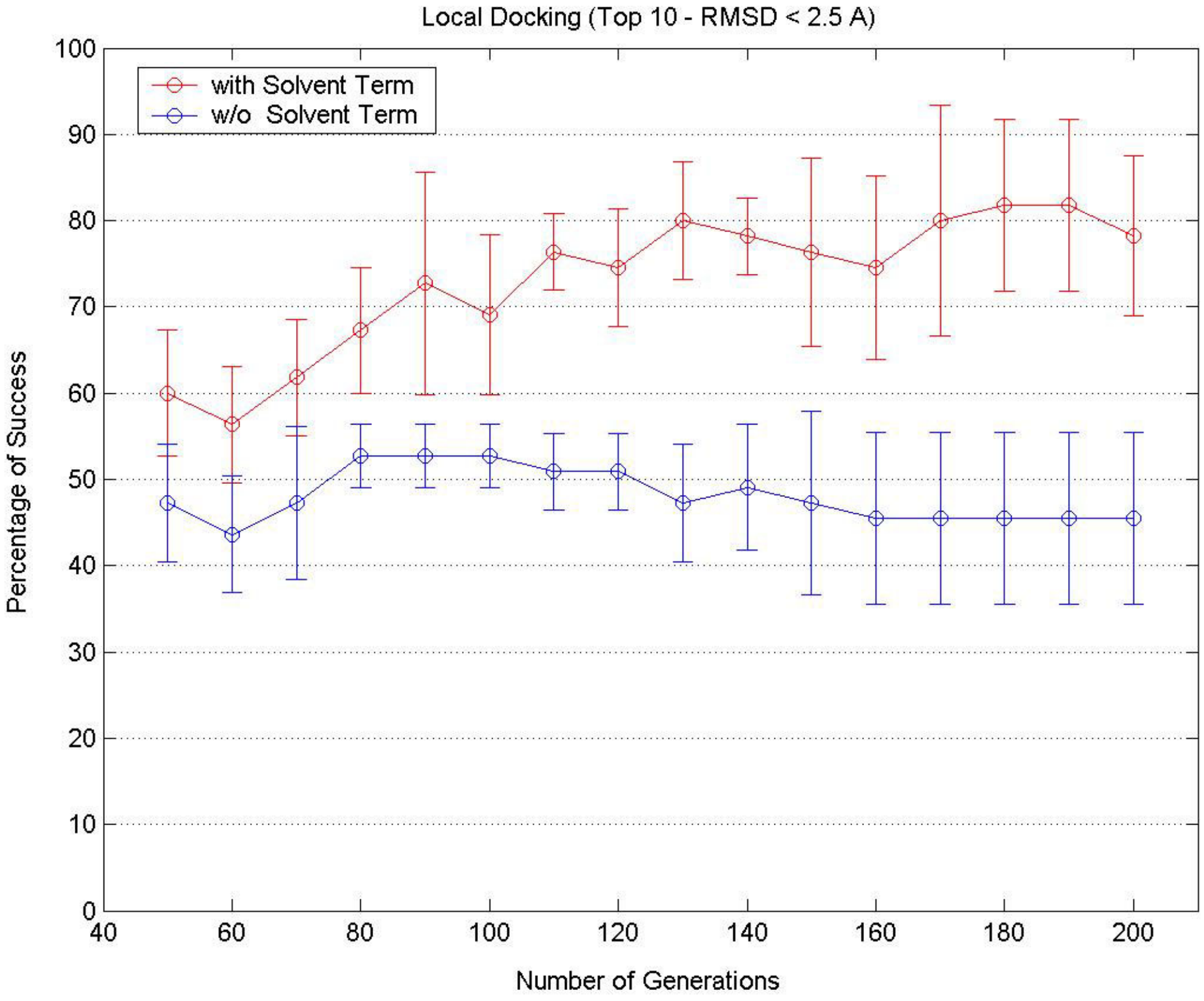

**Figure 22**. Percentage of success as a function of number of generations for two different values of $\varepsilon_{sol}$ for RMSD threshold of 2.5 Å considering the top 10 solutions. In red we show the case where the solvent term is taken in consideration ($\varepsilon_{sol} = -2$) and in blue we show the case where the solvent term is not taken in consideration, $\varepsilon_{sol} = 0$, that is, when Equation 5.9 becomes equivalent to Equation 4.1.

The same analysis is performed for global rigid docking simulations. In Figure 23 we present the percentage of success on global rigid simulations as a function of number of generations for the case where no interactions with the solvent are considered ($\varepsilon_{sol} = 0$) as well as considering unfavorable interactions with the solvent ($\varepsilon_{sol} = -2$).

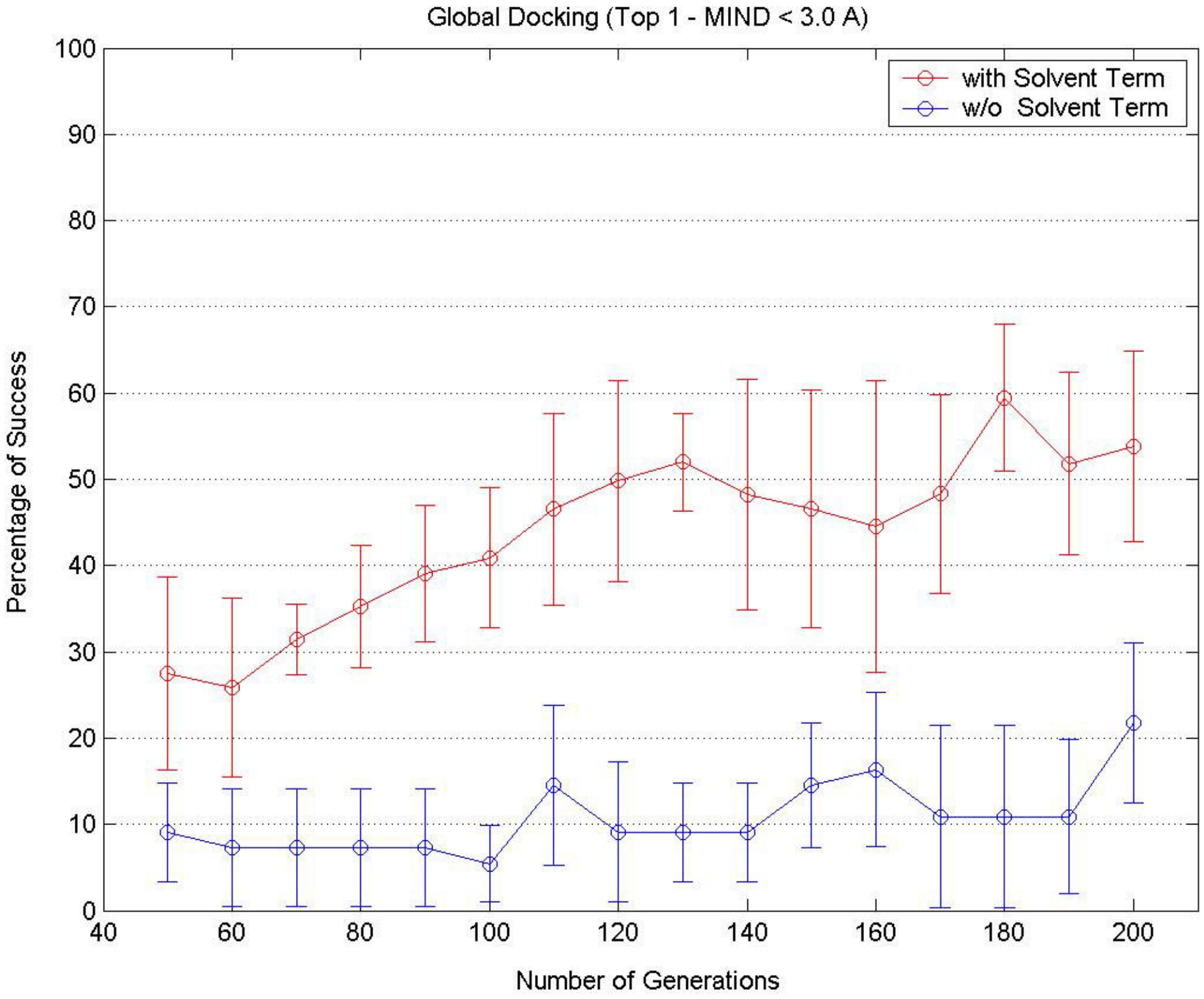

**Figure 23**. Effect of inclusion of solvent interactions on global rigid docking simulations. In red we show the case where the solvent term is taken in consideration ($\varepsilon_{sol} = -2$) and in blue we show the case where the solvent term is not taken in consideration, $\varepsilon_{sol} = 0$. Past approximately 120 generations, an average percentage of success of 50% is obtained irrespective of the number of generations.

When considering different numbers of top solutions in the calculation of the percentage of success one sees that at intermediate stages, there are a number of solutions that satisfy the success criteria but are lost as the simulation proceeds. The explanation for this fact is that those desirable solutions at intermediate stages of the simulation have poor CF values and are eventually lost as the number of generations increases (Figure 24).

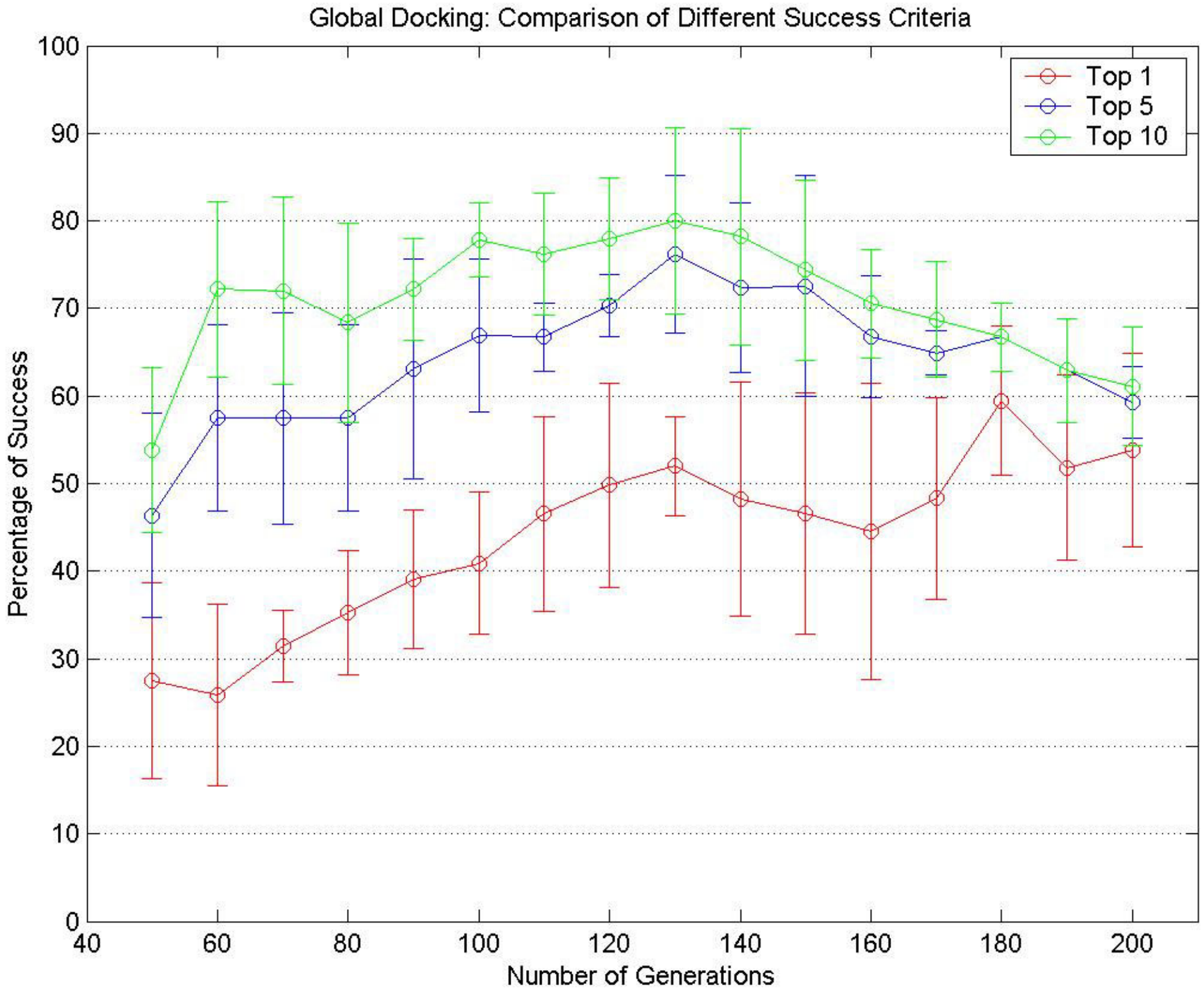

**Figure 24**. Comparison of percentage of success when considering different numbers of top solutions on global rigid docking simulations.

For the purpose of finding the position of candidate binding sites to be used subsequently as input on local simulations, a maximum of approximately 100 generations is sufficient to obtain a percentage of success of approximately 80% considering the top 10 solutions.

### 5.11.2 Relaxing the wall term

As described in Section 4.1, there is a penalty to be paid when non-covalently-bound atoms are placed at too close of a distance (Equations 4.2 and 4.3). In this section we want to probe whether allowing for closer atomic distances before starting to penalize steric clashes would increase the percentage of success. For this matter we rewrite equation 4.3 altering the threshold used to start penalizing steric clashes as follows:

$$E_{ab} = \begin{cases} 0 & \text{if } R_{ab} \geq k_o(R_A + R_B) \\ k\left(1/R_{ab}^{12} - 1/R_o^{12}\right) & \text{if } R_{ab} < k_o(R_A + R_B) \end{cases} \qquad (5.10)$$

For $k_0 = 0.9$ the equation above is similar to equation 4.3. Choosing values for $k_0$ smaller than 0.9 we allow closer interactions but once the threshold is reached, the value of the penalty is the same irrespective of the value of $k_0$.

In Figure 25 we show different combinations of values of $(\varepsilon_{sol}, k_0)$ for local rigid docking simulations. The results show that a decrease of the threshold value is detrimental in local rigid docking simulations.

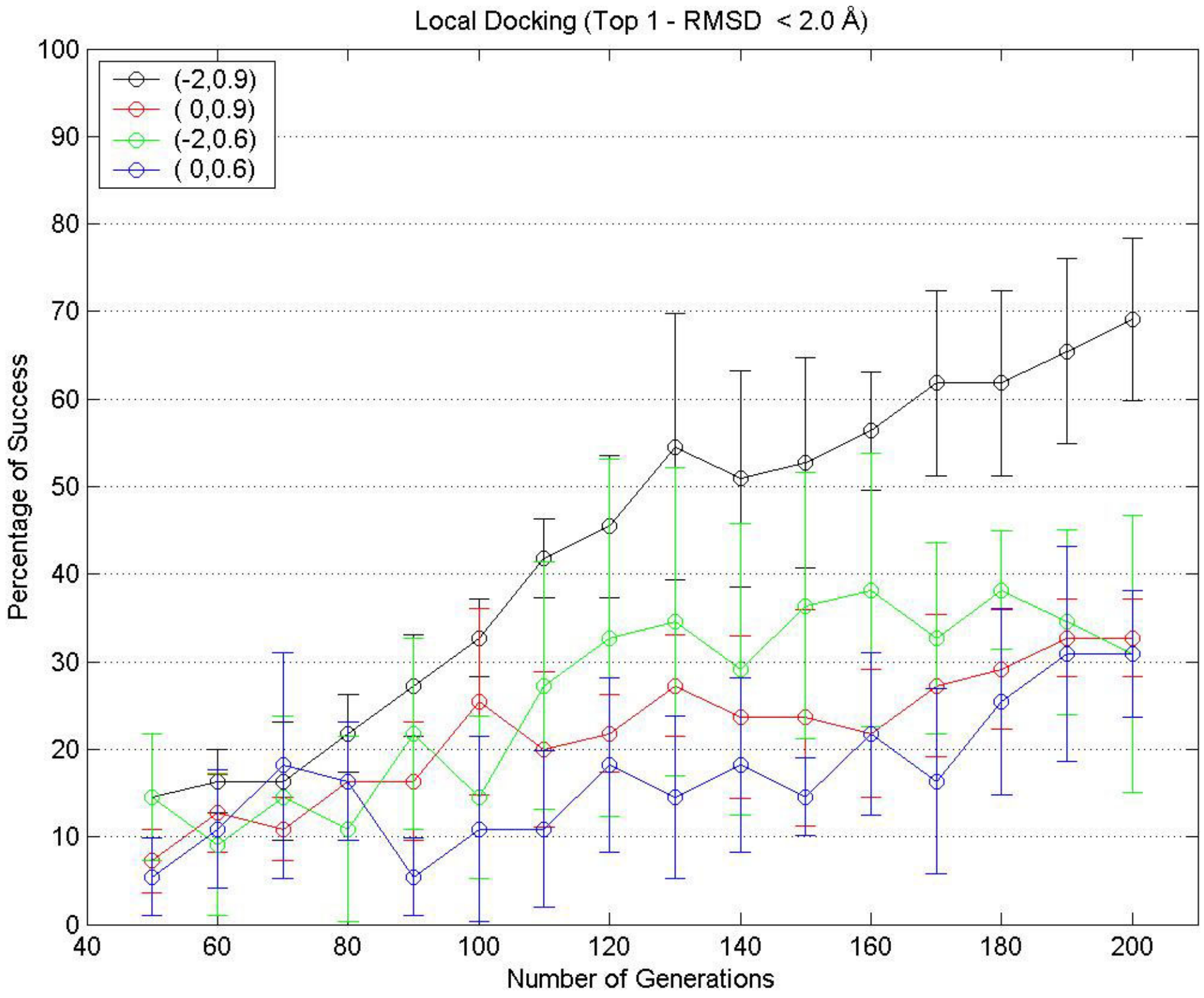

**Figure 25**. Effect of different combinations of $(\varepsilon_{sol}, k_0)$ on local rigid docking simulations. A relaxation of the parameter $k_0$ does not improve the percentage of success.

For global rigid docking simulations, on the other hand, relaxing the parameter $k_0$ actually improves the percentage of success when considering the top solution (Figure 26). However, if one considers the top 10 solutions the improvement is rather sharp, reaching a level of 90% as early as 50 generations through the simulation (Figure 27).

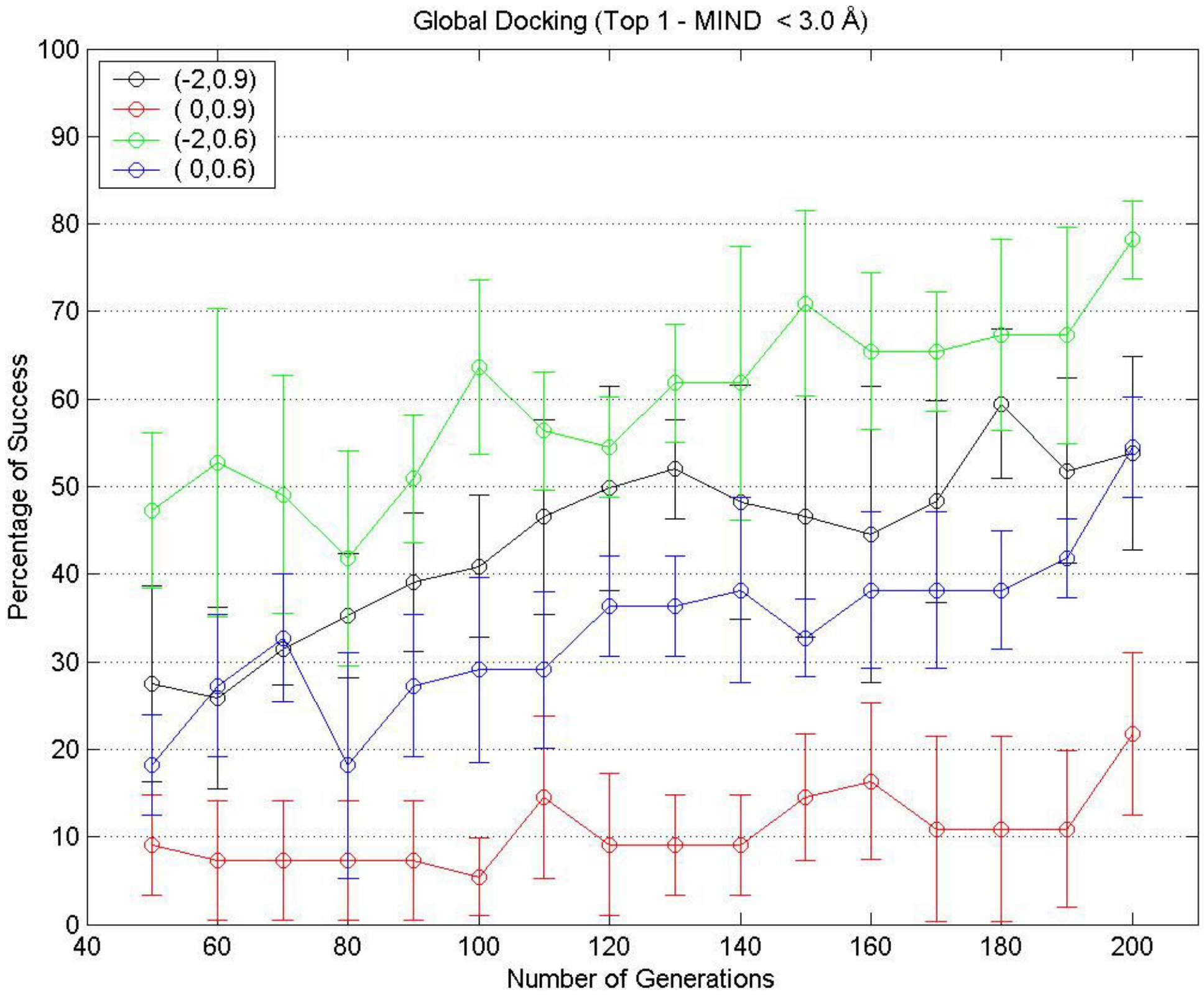

**Figure 26**. Effect of different combinations of $(\varepsilon_{sol}, k_0)$ on global rigid docking simulations. A relaxation of the parameter $k_0$ improves the percentage of success by approximately 10% with respect to the case where $k_0 = 0.9$.

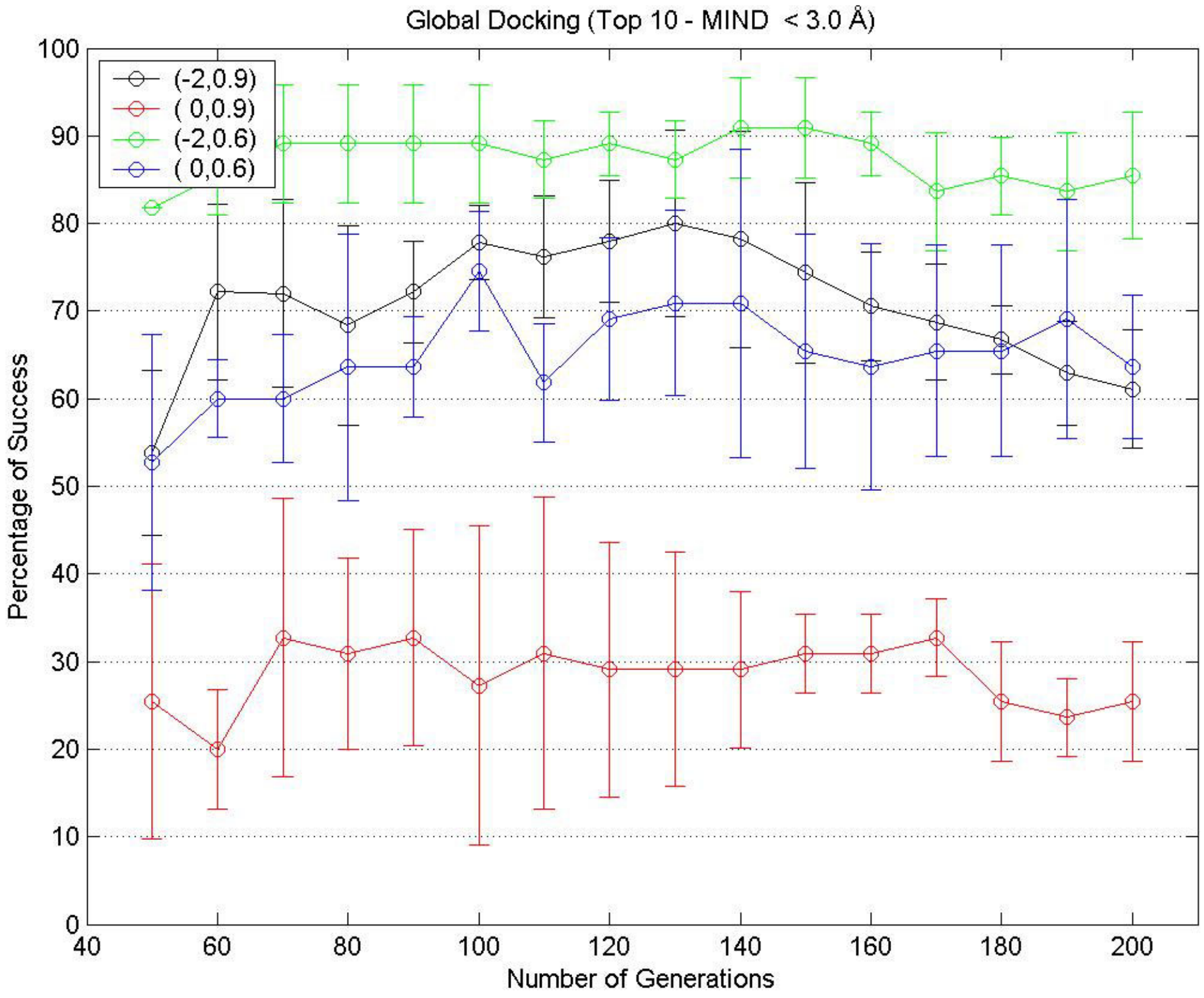

**Figure 27**. Effect of different combinations of $(\varepsilon_{sol}, k_0)$ on global rigid docking simulations considering the top 10 solutions. The relaxation of the parameter $k_0$ sharply improves the percentage of success reaching a value of 90%.

A more thorough analysis is needed of the effects of alterations of the Complementarity Function to find ways to improve the percentage of success in docking simulations. However, as shown in the present and previous sections, the Complementarity Function can be changed in way that improve the percentage of success specific for the task at hand.

The improvement brought by the consideration of unfavorable interactions with the solvent suggests that it is convenient to make as many contacts (in terms of area) as possible with protein rather than keeping a solvent accessible area. This in turn may suggest that the geometric complementarity is more important than the chemical compatibility between interacting surfaces of the ligand and protein atoms. Furthermore, considering the nature of the optimization technique used, one can rationalize that it is advantageous to restrict the advantage of easier-to-find earlier solutions found during the simulation that are partially buried vis-à-vis the more difficult to find solutions that are deeply buried in the binding site not making severe steric clashes with the protein atoms.

This conclusion is further corroborated by the results found relaxing the threshold for punishment of steric clashes which show that in rigid docking simulations, permitting steric clashes does not help in finding the exact position of the ligand but help in finding the binding site.

### *5.11.3 Comparison of different reproduction techniques*

As part of this study we have developed a new reproduction technique called Population boom without duplicates (Section 5.5). Population boom bears some qualities with respect to standard reproduction techniques making it a good choice for reproduction technique for computationally intensive applications such as docking.

Population boom was initially compared to other techniques using the function binary F6, a standard function used in the development of genetic algorithms:

$$F6 = 0.5 - \frac{\left(\sin\sqrt{x^2 + y^2}\right)^2 - 0.5}{\left(1.0 + 0.001\left(x^2 + y^2\right)\right)^2} \tag{5.11}$$

This function has a global maximum at $(x, y) = (0, 0)$ and an infinite number of local maxima separated by an equal number of local minima with the global minimum adjacent to the global maximum (Figure 28). This function is known to be a serious challenge to standard, gradient-based optimization techniques. The different reproduction techniques were compared to Population boom without duplicates in terms of the average distance between the top solution and the origin as a function of number of generations (Figure 29) on 100 independent runs.

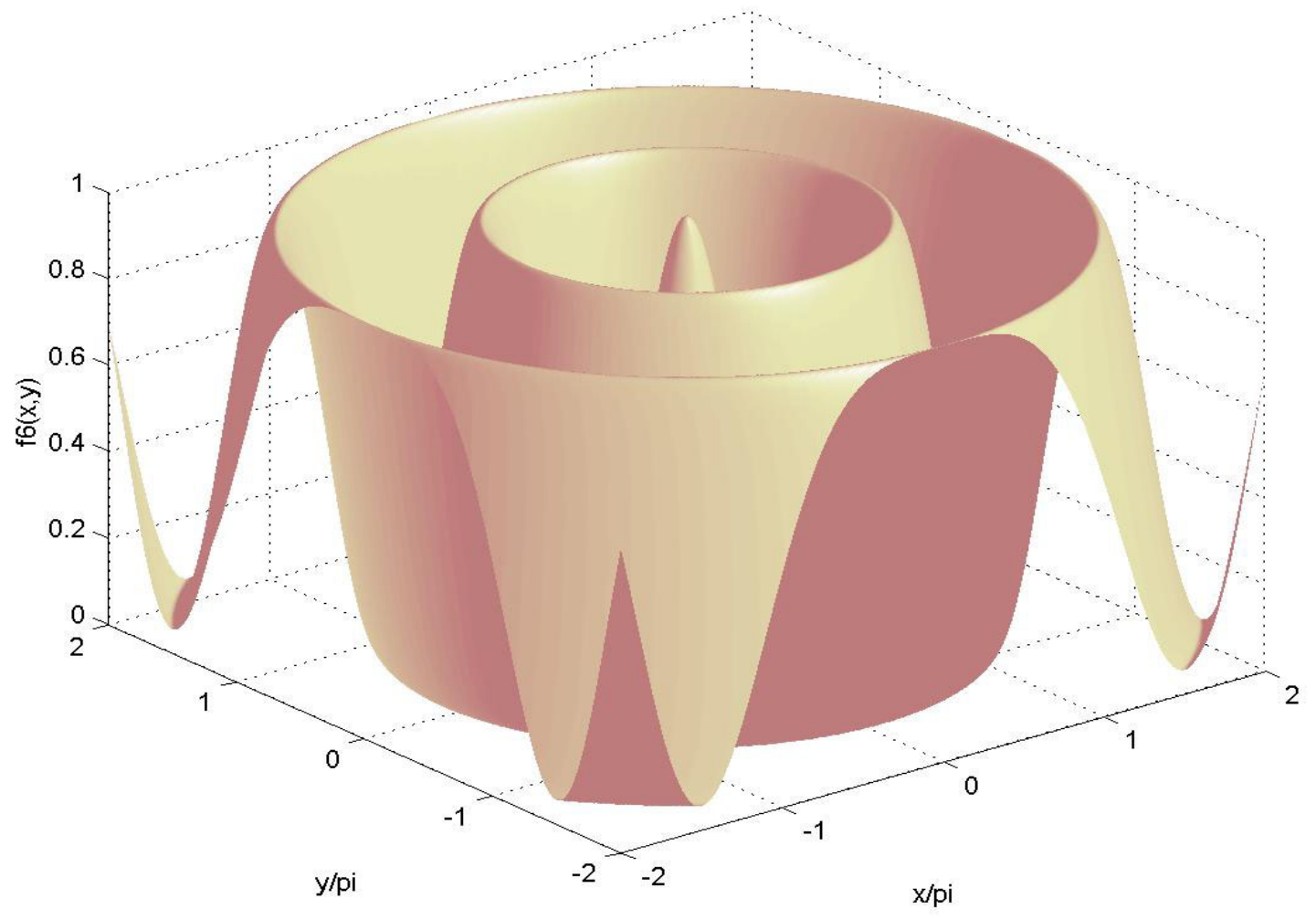

**Figure 28**. Function binary F6. This function has radial symmetry containing an infinite number of local maxima at regular intervals away from the global maximum located at the origin.

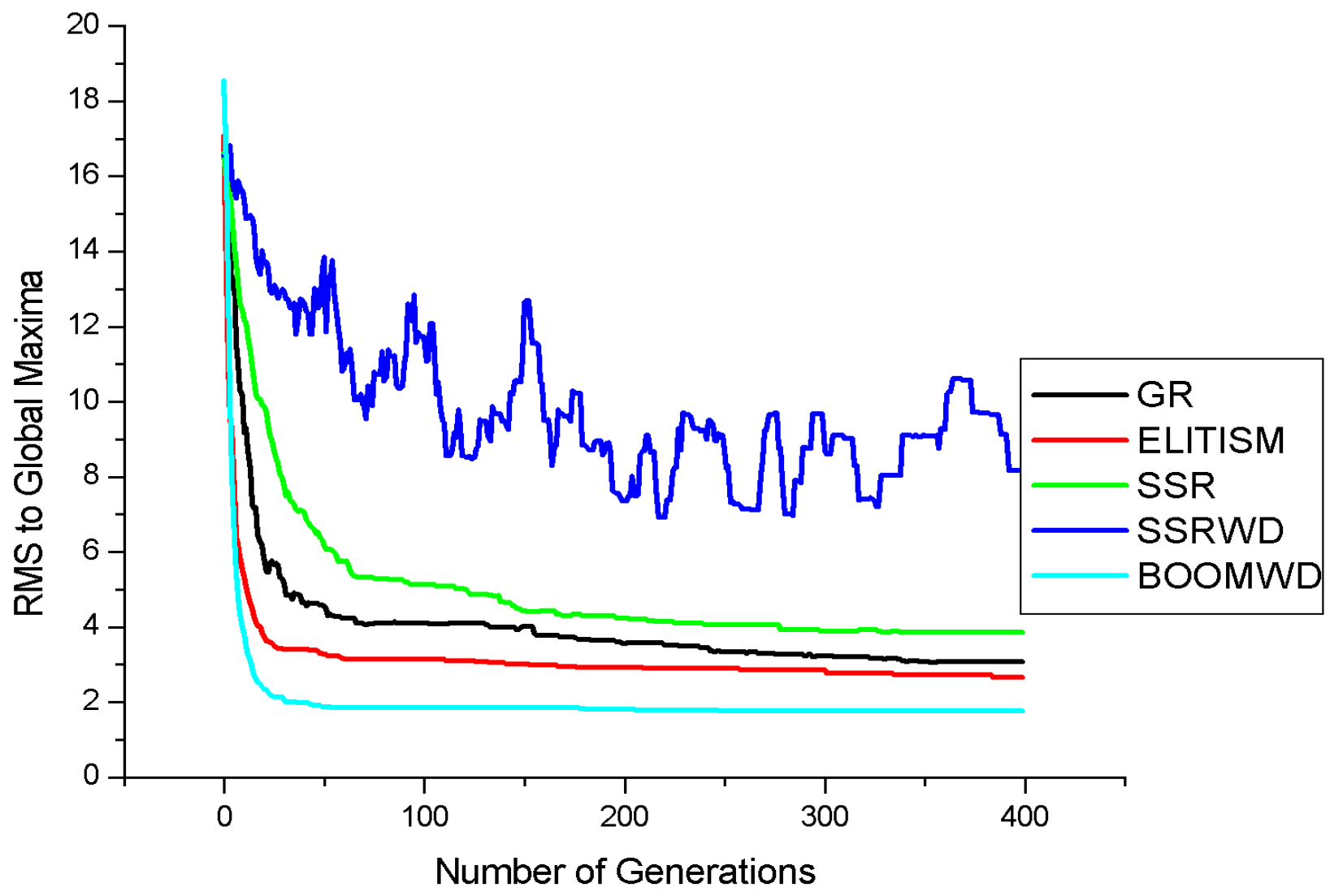

**Figure 29**. Average distance-to-origin in parameter space for different reproduction techniques as a function of number of generations. GR: Generational Replacement, ELITISM: Elitism, SSR: Steady state reproduction, SSRWD: Steady state reproduction without duplicates and BOOMWD: Population boom without duplicates.

Population boom without duplicates converges much faster and finds the global maximum more frequently than any of the other reproduction techniques. Furthermore, when analyzing which maximum is obtained after 400 generation as a fraction of 100 independent runs, one sees (Figure 30) that in the great majority of runs Population boom without duplicates finds the global maximum (peak number 0) with a frequency considerably higher than other techniques.

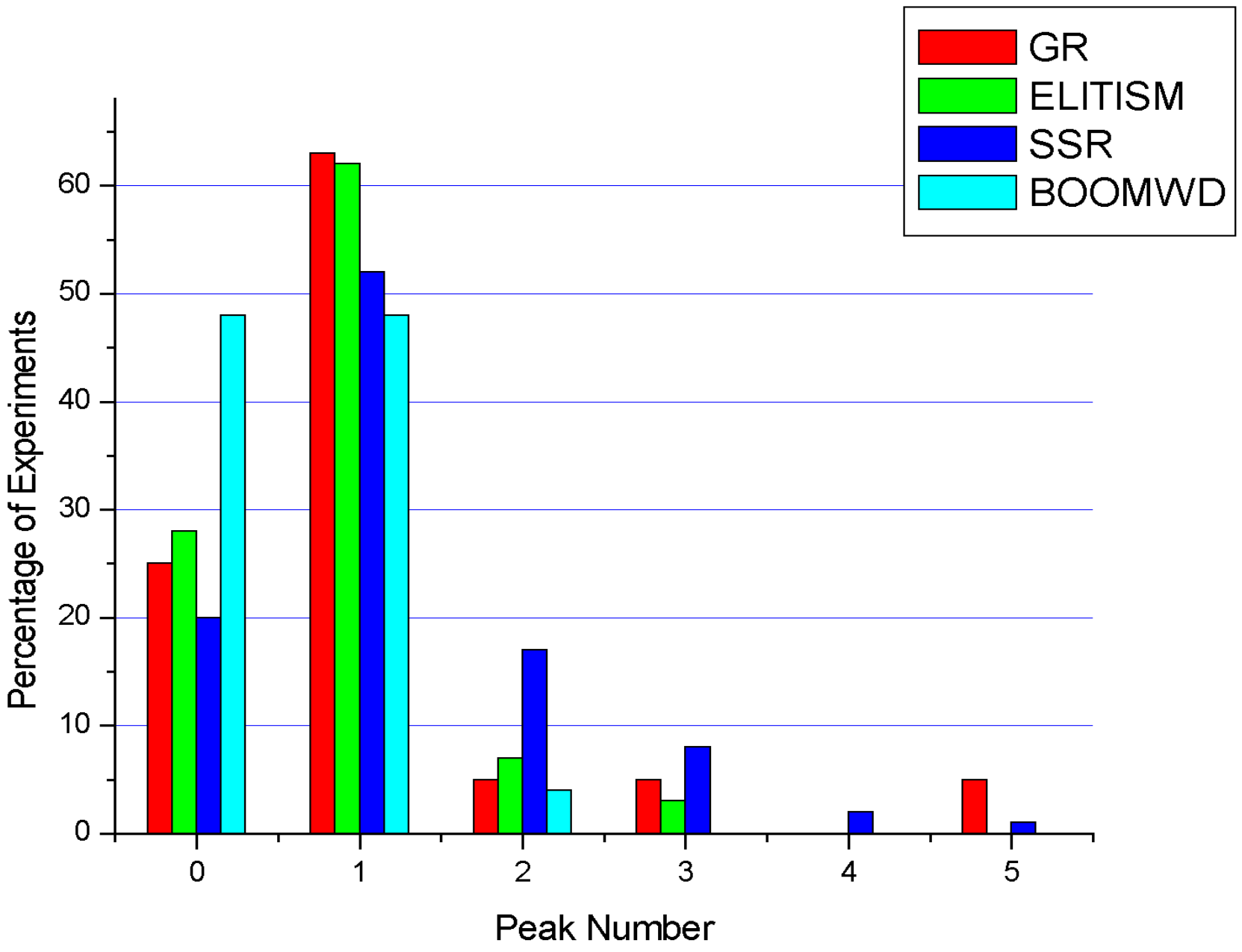

**Figure 30**. Frequency with which each maxima is found after 400 generations on 100 independent runs. GR: Generational replacement, ELITISM: Elitism, SSR: Steady state reproduction and BOOMWD: Population boom without duplicates.

The results of the comparison between Population boom without duplicates (BOOMWD) and the other reproduction techniques show that Steady state reproduction (SSD) and Elitism (ELD) are the techniques that more closely approach the accuracy obtained with BOOMWD. Therefore, when comparing the performance of the different techniques on the rigid docking case we restrict the analysis to BOOMWD, SSD and ELD.

The comparison of the performance of the three best-performing reproduction techniques is made in the same manner described in the previous section, calculating the average percentage of success on both local and global rigid docking simulations over the set of 11 complexes described in Table X and over the 5 independent GA runs as a function of the number of generations. Figure 31 shows such analysis for local rigid docking simulations.

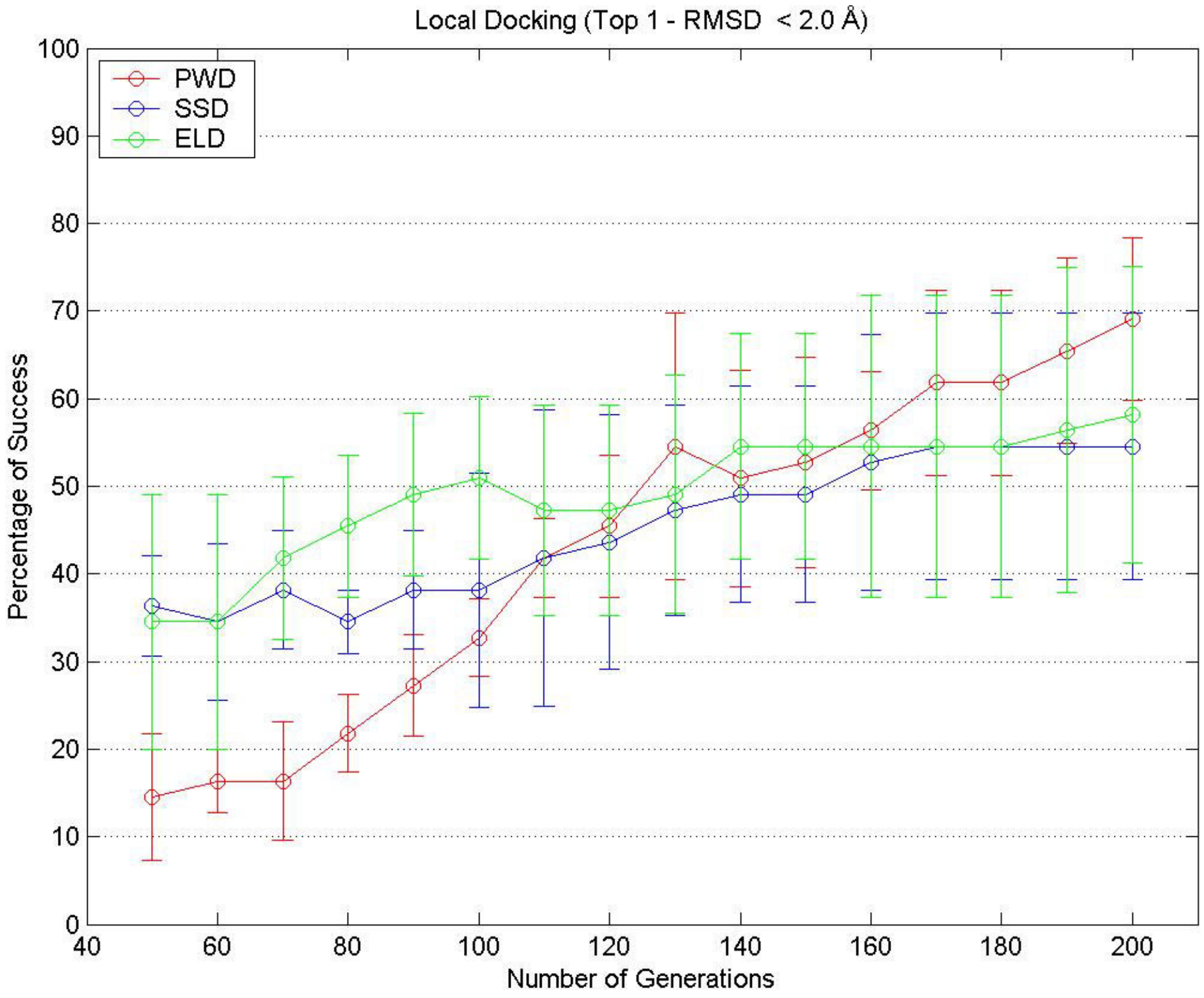

**Figure 31**. Comparison of the three best-performing reproduction techniques on local rigid docking simulations. PWD: Population boom without duplicates, SSD: Steady state reproduction, ELD: Elitism.

Population boom without duplicates shows an improvement over the other techniques at later stages of the simulation where on average 10% accuracy is gained with respect to Steady state duplicates and Elitism. When comparing the top 10 solutions instead of only the top solution, the advantage of PWD over SSD and ELD are evident throughout the simulation, irrespective of the number of generations. However, in this case, the average gain of PWD with respect to the other techniques is attenuated (Figure 32).

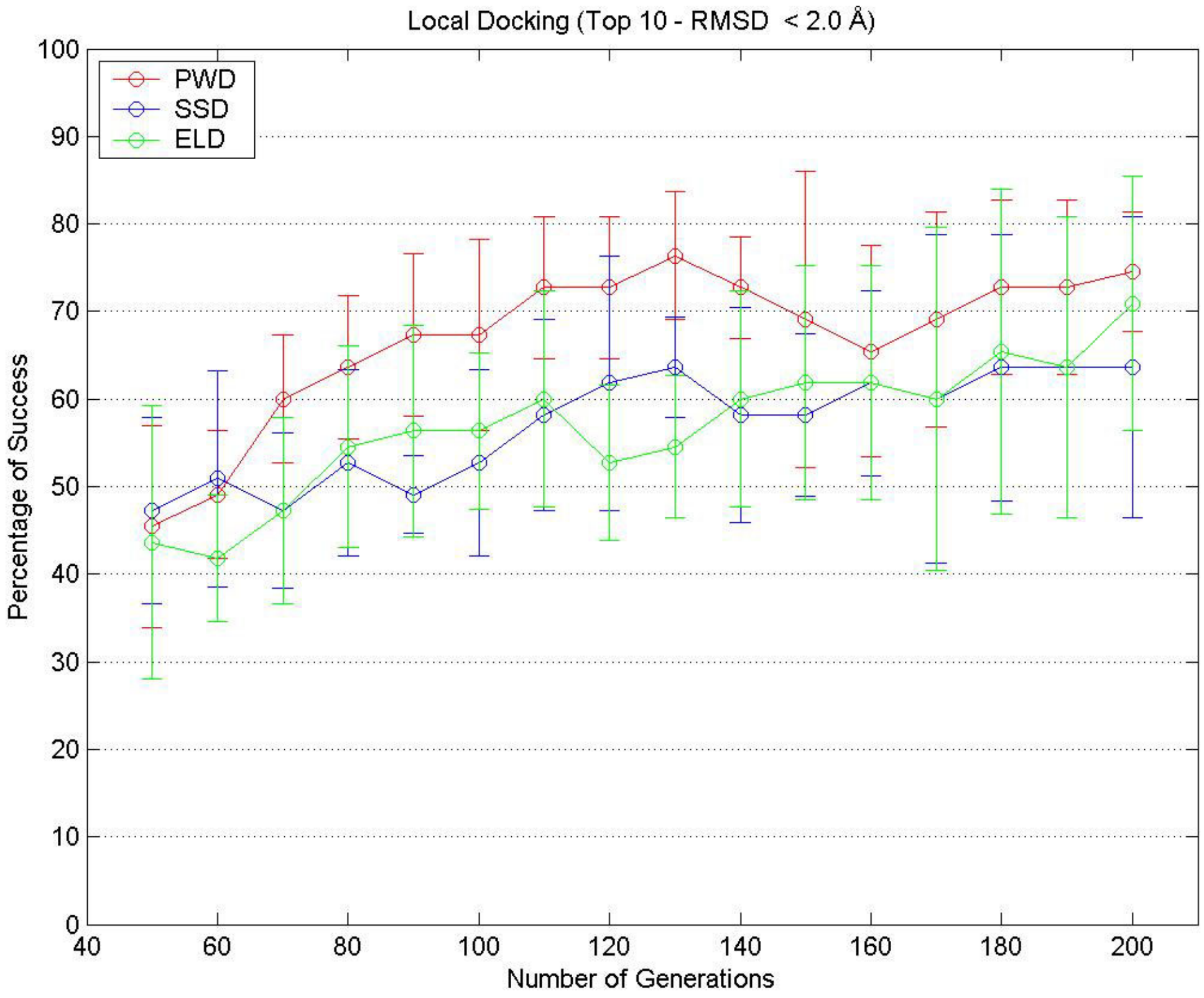

**Figure 32**. Comparison of the three best-performing reproduction techniques on local rigid docking simulations considering the top 10 solutions. PWD: Population boom without duplicates, SSD: Steady state reproduction, ELD: Elitism.

When comparing the three best-performing reproduction techniques on global rigid docking simulations we do not see any advantage of PWD over SSD and ELD when considering the top solution (Figure 33) however, when considering the top 10 solutions, PWD shows a clear advantage over SSD and ELD (Figure 34).

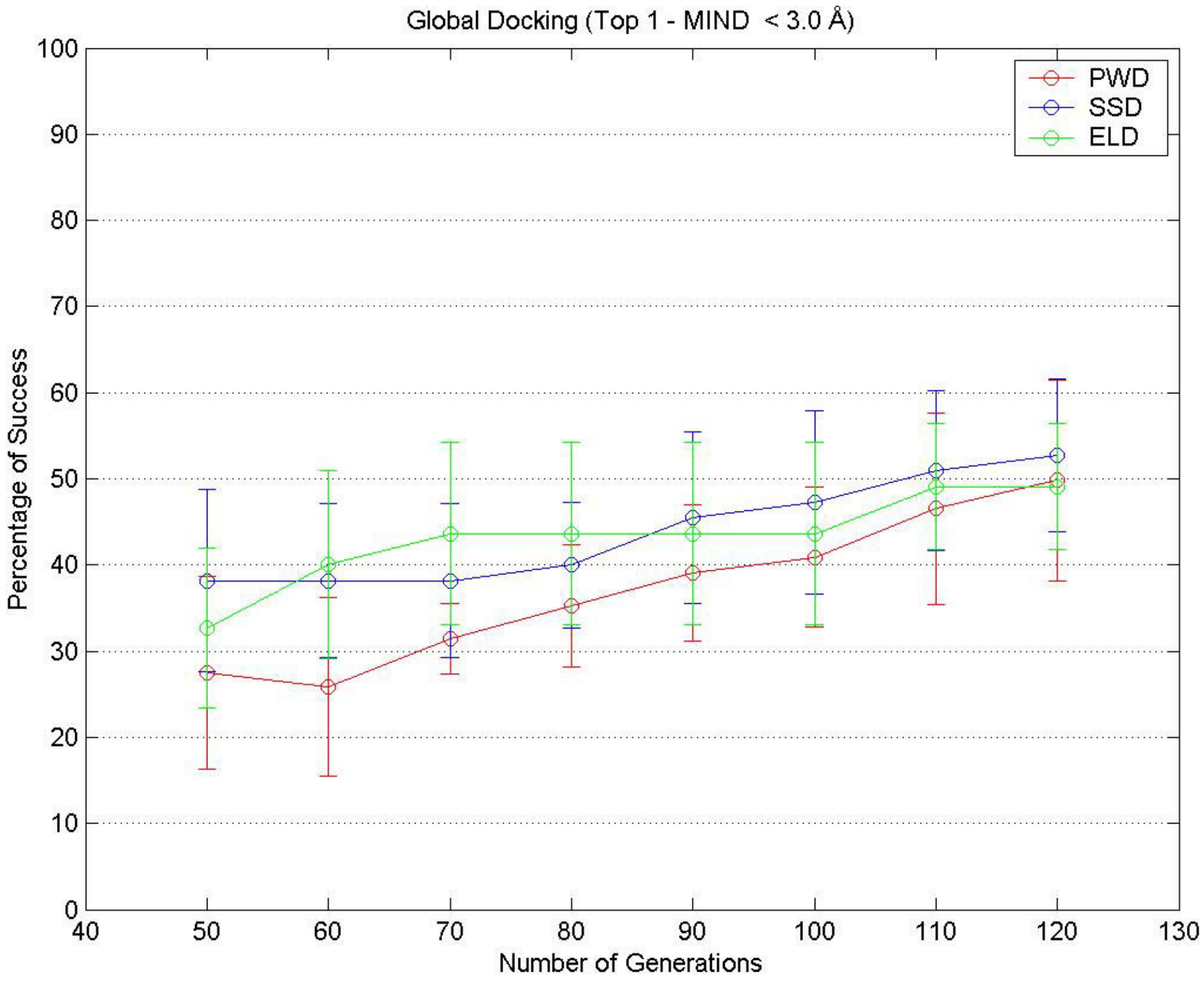

**Figure 33**. Comparison of the three best-performing reproduction techniques on global rigid docking simulations for a total of 120 generations. PWD: Population boom without duplicates, SSD: Steady state reproduction, ELD: Elitism.

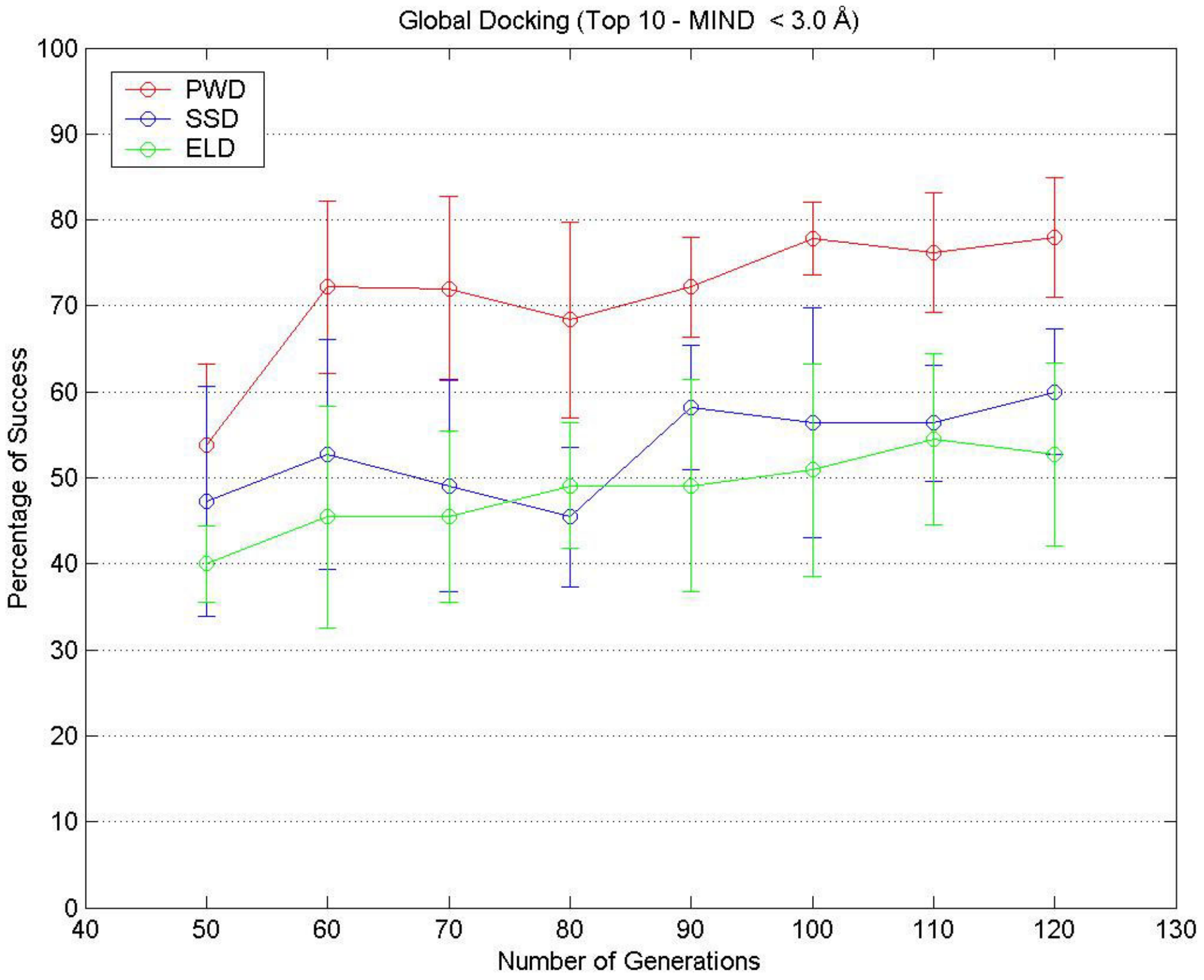

**Figure 34**. Comparison of the three best-performing reproduction techniques on global rigid docking simulations considering the top 10 solutions. PWD: Population boom without duplicates, SSD: Steady state reproduction, ELD: Elitism.

## *5.12 Flexible docking simulations unsing FlexAID*

As described in section 5.7, side chain flexibility is introduced in FlexAID by means of performing an exhaustive search considering the conformation of the flexible residues present in the protein receptor together with as many rotamer-based alternative conformations as permitted due to steric clashes. Searching through alternative side chain conformations by means of an exhaustive search assures that the introduction of side chain flexibility comes at no cost in terms of accuracy (see discussion below) with respect to the search procedure as well as little cost in terms of time as explained in section 5.7. The time spent in the exhaustive search may well be equivalent to the time increase that would be required for a genetic algorithm to achieve the same level of accuracy including the side chain dihedral angles as optimization variables.

Exhaustive search is practical only because a small number of alternative conformations are considered. However, three conditions need to be met in order that FlexAID predicts the correct conformation of the protein receptor.

The first condition is that of correctly assigning as flexible those residues whose side chain conformations need to be altered, provided that a small number of side chains (say three) need to be searched at any given time by being in contact with a putative ligand position. Three flexible side chains represents a manageable number time-wise while at the same time, as seen in chapter 2, correctly describes a great majority of binding pockets in terms of the number of flexible side chains. Therefore, when performing a global flexible search one needs to try to choose residues to be set as flexible in a way such that they are sparsely located so that the chance of more than three flexible residues being in contact to a putative ligand position is diminished. When performing a flexible local search one can choose three residues to be set flexible.

The second condition is that by using the rotamer library as described in section 5.7 at least one of the rotamers included is closely related to the correct side chain conformation to be found experimentally when the ligand binds the binding site.

The third condition is that the “correct” alternative conformation mentioned in the previous paragraph is recognized as such by the Complementarity Function by being assigned the highest score among the alternatives. All flexible docking simulations described in the following sections utilize the altered Complementarity Function definition (equation 5.9) which includes the repulsive term for atom-solvent interactions.

The three conditions described above need to be subjected to fine scrutiny in order to get a full picture of the factors influencing the accuracy of FlexAID, especially for flexible protein/rigid ligand docking. Such a level of analysis has not been performed as part of this work. Nevertheless, as a proof of concept, four different complexes were simulated obtaining promising results as described below.

### *5.12.1 Flexible docking test set*

The main condition for choosing a complex is that two different PDB files exist, one displaying the ligand in contact in the binding site and the other displaying the empty binding site, much in the spirit of the datasets used in chapter 2. Such pairs of PDB files can be used in advance to determine which side chains need to be set flexible in the apo-form protein in order to resemble the holo-form binding site side chain conformations. By choosing as flexible the residues known in advance to undergo conformational changes we are eliminating the first condition as a source of inaccuracy in our testing of the feasibility of the approach. The Complexes tested are described in Table XI.

Table XI. Flexible docking test set

| Protein Receptor | | Ligand | | | | |
|---|---|---|---|---|---|---|
| Name | Code | ID[a] | | | Name | Source[b] |
| Adipocite lipid binding protein | 1lib | STE | - | 132 | Stearic Acid | 1lif |
| Met-repressor | 1cmb | SAM | A | 105 | S-adenosyl methonine | 1cmc |
| Carboxypeptidase A | 1arl | DCY | - | 308 | D-cysteine | 1f57 |
| tRNA-Guanine Transglycosylase | 1pud | APQ | - | 900 | 2,6-diamino-8-propylsulfanylmethyl-3h-quinazoline-4-one | 1k4h |

[a] Ligands are described by their three-letter PDB code, chain identifier and residue number. The symbol `-` describes the cases for which no chain identifier is associated to the ligand under consideration.

[b] The PDB codes cited are those from which the ligands were extracted.

The ligand coordinates were extracted from the holo protein form and used to generate the ligand internal coordinates. Furthermore, the coordinates of the ligand are also used to evaluate the success of flexible docking solutions in terms of RMSD or MIND. Each of the complexes described in Table XI will be referred to by means of the protein receptor code (apo-form protein). Table XII presents a list of the binding site residues undergoing side chain conformational changes for each of the complexes present in Table XI.

Table XII. Binding-pocket side-chain dihedral-angle changes

| Complex[a] | Residue[b] | | | Side chain dihedral angle changes[c] | | | |
|---|---|---|---|---|---|---|---|
| | | | | $\Delta\chi_1$ | $\Delta\chi_2$ | $\Delta\chi_3$ | $\Delta\chi_4$ |
| 1lib | PHE* | - | 57 | 142.7 | 67.7 | - | - |
| | LYS | - | 58 | 91.6 | 90.5 | 6.6 | 18.6 |
| 1cmb | PHE* | A | 65 | 102.1 | 20.1 | - | - |
| 1arl | ARG | A | 127 | 90.2 | 12.2 | 125.7 | 159.4 |
| | ARG | A | 145 | 22.8 | 119.3 | 29.4 | 167.1 |
| | HIS* | A | 196 | 15.3 | 64.3 | - | - |
| | ILE | A | 247 | 123.6 | 1.2 | - | - |
| | THR* | A | 268 | 124.4 | - | - | - |
| | GLU* | A | 270 | 20.4 | 75.6 | 21.7 | - |
| 1pud | TYR* | A | 106 | 109.0 | 14.7 | - | - |
| | VAL* | A | 233 | 135.0 | - | - | - |

[a] Each complex is described by means of the protein receptor code (apo-form protein) present in Table XI.

[b] Residues are identified by their residue name, chain identifier and residue number. The symbol `-` describes the cases for which no chain identifier is associated to the residue under consideration. Residue names marked with `*` denote residues set flexible during docking simulations.

[c] Dihedral angle changes shown represent the absolute difference between corresponding angles in the apo and holo forms. The symbol `-` is used when a side chain does not contain the given dihedral angle.

The utilization of an exhaustive search in the space of alternative conformations excludes the possibility that the correct side chain conformations is missed due to poor sampling during searching. Having chosen the correct flexible residues (first condition) and assuming that the large number of rotamers for each side chain present in the rotamer library precludes the possibility of missing the correct side chain conformations (second condition), one is left with the third condition, namely that the Complementarity Function favors the correct combination of flexible residue rotamers over other combinations. Thus, a fair test of the percentage of success of FlexAID in flexible simulations would require a larger test set than that presented in Table XI. Nonetheless, as a proof of concept, each of the four complexes was simulated for both local and global flexible simulations permitting flexibility on those residues marked with `*` in Table XII. The time needed for performing the exhaustive search was main concern in choosing which residues were set flexible during the simulations. Thus, residues with large numbers of rotamers were not chosen as flexible despite being known to undergo side chain conformational changes upon ligand binding. The results are presented in Table XIII.

Table XIII. RMSD or MIND for the top solution on flexible simulations

| Complex | RMSD for Local simulations[a] | | MIND for Global simulations[b] | |
|---|---|---|---|---|
| | Flexible | Rigid | Flexible | Rigid |
| 1lib | 0.70 (218.0) | 2.39 (-22.6) | 0.24 (209.6)[c] | 1.5 (-234.5) |
| 1cmb | 1.47 (-93.0) | 7.70 (-254.3) | 6.11 (-211.5) | 6.5 (-214.0) |
| 1arl | 6.22 (211.5) | 3.44 (148.6) | 0.75 (171.1) | 0.68 (197.2) |
| 1pud | 3.80 (301.7) | 3.45 (33.5) | 0.57 (243.98) | 1.53 (116.3) |

[a] RMSD values in Å are shown for flexible as well as rigid local simulations, the values in parenthesis are the corresponding Complementarity Function values.

[b] MIND values in Å are shown for flexible as well as rigid global simulations, the values in parenthesis are the corresponding Complementarity Function values.

[c] This global simulation found not only the binding site but also the ligand pose with RMSD=0.75 Å.

The data presented in Table XIII have not been averaged over several independent simulations but is indicative of the kind of results obtained. Table XIII shows that for local flexible simulations, FlexAID was able to successfully find the correct ligand position (RMSD < 2.0 Å) for the first two complexes (1lib and 1cmb) while missing the correct ligand position for the last two complexes (1arl and 1pud). We note that when performing local rigid docking FlexAID is not able to find the correct ligand position (RMSD < 2.0 Å) for any of the four complexes. For the second pair of complexes (1arl and 1pud), despite reaching RMSD values larger than 2.0 Å, the values are smaller than those obtained when flexibility is allowed. However, when one looks at the Complementarity Function (CF) values in parenthesis, one immediately sees that the CF values obtained in local rigid docking are considerable smaller than those obtained in the flexible case. This is points to the possibility that the inclusion of flexibility for the second couple of complexes has created new opportunities for favorable interactions between the protein atoms and ligand atoms in positions different than those seen experimentally.

The results presented in Table XIII can be rationalized when we consider the different effect of the side chain conformational changes occurring in the first two complexes as compared to the last two complexes. In complexes 1lib and 1cmb the residues PHE – 57 and PHE A 65 respectively, block the binding site in a way that any ligand position with RMSD < 2.0 Å would make considerable steric clashes. For complexes 1arl and 1pud the alterations are of an opposite nature, closing in on the ligand during docking. As an example, Figure 35 shows the different conformations of PHE – 57 in the holo, apo and flexible local docking solution protein structures as well as the positions of the ligand STE in the holo protein (1lif) and in the docking solution.

For global simulations, except for complex 1cmb, all simulations were successful in finding the binding site. For complex 1lib, FlexAID was also able to find the correct ligand position in the binding site, a feat not possible without considering side chain flexibility.

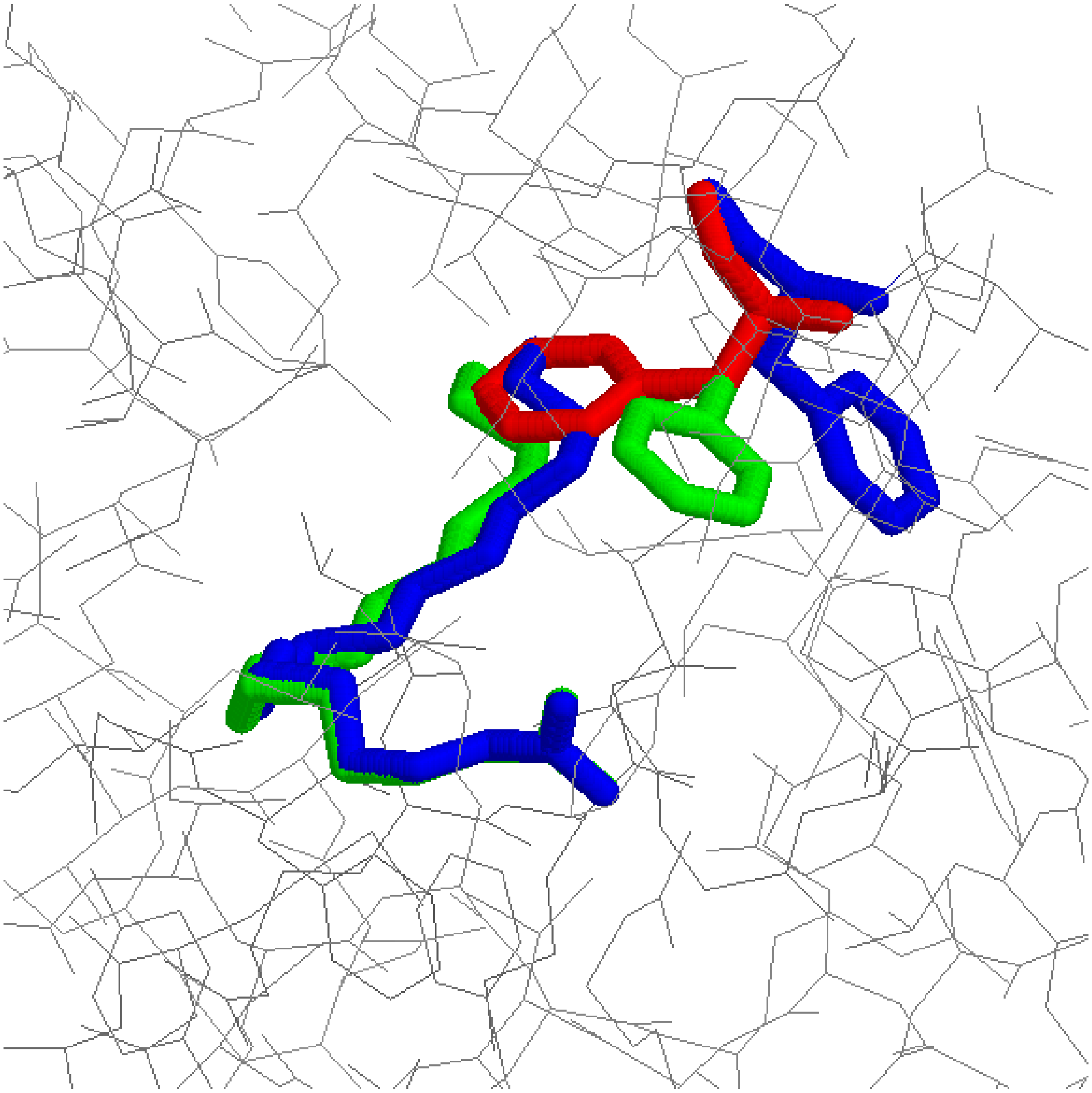

**Figure 35**. Local flexible docking solution of complex 1lib. Different conformations of PHE – 57 in the holo (PDB code 1lif, blue), apo (PDB code 1lib, red) and flexible local docking solution (green) structures as well as the positions of the ligand STE in the holo protein (blue) and in the docking solution (green). Despite the RMSD obtained of 0.7 Å, the conformation of PHE – 57 in the holo protein is not predicted correctly.

# 6. Conclusions

Side chain flexibility is an important event in ligand binding. Sixty percent of binding sites contain at least one side chain that undergoes conformational changes upon ligand binding. Consideration of up to three side chains undergoing a conformational change accounts for 85% of cases, a fact that points to the possibility of including side chain flexibility in ligand binding docking simulations. The 20 naturally occurring amino acids vary in their propensities of undergoing conformational changes. The flexibility scale obtained by analysing the side chains in binding sites is very similar to that obtained for the vicinity of point mutations as well as that obtained by comparing cases of proteins that were crystallized independently more than once, thus giving rise to the idea (put forward by Eran Eyal and myself) that the flexibility scale is an intrinsic property of amino acids.

The rationalization of which side chains undergo side chain conformational changes upon ligand binding is not simple. Classification is possible to a certain extent (approximately 70% accuracy) using support vector machines. It is difficult, so far, to pass the 70% mark.

Knowledge-based techniques were utilized to generate scoring functions for a ligand-protein docking scoring function (Complementarity Function) as well as for a pairwise amino acid contact potential. These are initial steps in further developing the Complementarity Function.

Docking simulations utilizing genetic algorithms have given promising results for the case of rigid docking. An average accuracy level of 70% is obtained for both local and global docking simulations. The introduction of a repulsive interaction solvent term considerably improves the accuracy and may be related to the dynamics of the genetic algorithm optimization. A novel genetic algorithm optimization technique (Population boom) was developed. The utilization of Population boom does not bring the same drastic improvement to GA ligand-docking optimizations as it does for other optimization problems such as the binary F6 function optimization problem.

Side chain flexibility has been introduced into docking simulations through the use of side chain rotamer libraries. A user-defined list of flexible residues is used to build alternative side chain conformations. The space of alternative side chain conformations is exhaustively searched. The exhaustive search does not increase the time required for a simulation due its application on a limited number of residues and putative ligand positions. The accuracy of the approach for the introduction of side chain flexibility has not been determined but the approach has been shown, in principle, to be able to generate consistent solution for local as well as global simulations.

The docking approach described above is implemented in the software called FlexAID using the C programming language. FlexAID needs to be tested more thoroughly in order to obtain a complete understanding of it performance, especially when considering side chain flexibility. FlexAID is also able to optimize, through the generic algorithm search procedure, any number of side chain as well as ligand dihedral angles. However, in order to test this capability one needs to alter the Complementarity Function to consider intra-molecular interactions as well as protein-protein interactions (for flexible residues).

# APPENDIX A. SVM REDUNDANCY ANALYSIS TABLES

## TABLE A.I. REDUNDANCY ANALYSIS FOR FLEXIBILITY[a]

| Number of neighbors | SVM Set Size | $\overline{\mathrm{NME}}$ $(\sigma)$<br>$\overline{\mathrm{NME}}_+$ $(\sigma)$<br>$\overline{\mathrm{NME}}_-$ $(\sigma)$ | Total non-redundant examples (NRE) (OE/NRE) | NRE+/ NRE- | NOE+/ OE/ NOE- | RE+/ RE- | Total redundant examples (TRE) |
|---|---|---|---|---|---|---|---|
| 1 | 100 | 0.48 (0.03)<br>0.30 (0.10)<br>0.18 (0.09) | 315<br>(0.37) | 123<br><br>192 | 5<br>118<br>74 | 3248<br><br>24843 | 28091 |
| 2 | 100 | 0.39 (0.02)<br>0.26 (0.05)<br>0.13 (0.03) | 1414<br>(0.25) | 415<br><br>999 | 58<br>357<br>642 | 3248<br><br>24843 | 28091 |
| 3 | 500 | 0.35 (0.01)<br>0.25 (0.03)<br>0.09 (0.02) | 3441<br>(0.16) | 760<br><br>2681 | 219<br>541<br>2140 | 3248<br><br>24843 | 28091 |
| 4 | 500 | 0.33 (0.01)<br>0.25 (0.01)<br>0.08 0.01) | 5305<br>(0.10) | 963<br><br>4342 | 426<br>537<br>3805 | 3237<br><br>24828 | 28065 |
| 5 | 500 | 0.34 (0.02)<br>0.24 (0.02)<br>0.10 (0.01) | 6324<br>(0.08) | 1043<br><br>5281 | 563<br>480<br>4801 | 3207<br><br>24624 | 27831 |
| 6 | 500 | 0.31 (0.03)<br>0.23 (0.02)<br>0.08 (0.02) | 6673<br>(0.06) | 1043<br><br>5630 | 615<br>428<br>5202 | 2990<br><br>23876 | 26866 |
| 7 | 500 | 0.30 (0.01)<br>0.21 (0.02)<br>0.09 (0.01) | 6670<br>(0.06) | 980<br><br>5690 | 602<br>378<br>5312 | 2688<br><br>22709 | 25397 |
| 8 | 500 | 0.30 (0.02)<br>0.19 (0.03)<br>0.10 (0.02) | 6450<br>(0.05) | 902<br><br>5548 | 561<br>341<br>5207 | 2271<br><br>21267 | 23538 |
| 9 | 500 | 0.28 (0.01)<br>0.18 (0.02)<br>0.10 (0.01) | 5911<br>(0.05) | 754<br><br>5157 | 476<br>278<br>4879 | 1812<br><br>19043 | 20855 |
| 10 | 500 | 0.31 (0.02)<br>0.22 (0.01)<br>0.09 (0.02) | 5192<br>(0.04) | 604<br><br>4588 | 385<br>219<br>4369 | 1436<br><br>16730 | 18166 |

[a] SVM learning and testing averaged over 3 runs using C=1.0 and linear kernel. NRE=non-redundant examples. OE=overlapping examples; RE=redundant examples. Plus and minus signs describe flexible and rigid examples respectively.

## TABLE A.II. REDUNDANCY ANALYSIS FOR SURFACE ACCESSIBLE AREA[a]

| Number of neighbors | SVM Set Size | $\overline{\text{NME}}$ ($\sigma$)<br>$\overline{\text{NME}}_+$ ($\sigma$)<br>$\overline{\text{NME}}_-$ ($\sigma$) | Total non-redundant examples (NRE) (OE/NRE) | NRE+/ NRE- | NOE+/ OE/ NOE- | RE+/ RE- | Total redundant examples (TRE) |
|---|---|---|---|---|---|---|---|
| | | 0.37 (0.01) | | 1864 | 1336 | 3248 | |
| 1 | 600 | 0.21 (0.02) | 12937 (0.04) | | 528 | | 28091 |
| | | 0.17 (0.03) | | 11073 | 10545 | 24843 | |
| | | 0.37 (0.02) | | 1869 | 1357 | 3248 | |
| 2 | 600 | 0.20 (0.01) | 13083 (0.04) | | 512 | | 28091 |
| | | 0.17 (0.01) | | 11214 | 10702 | 24843 | |
| | | 0.38 (0.03) | | 1869 | 1357 | 3248 | |
| 3 | 600 | 0.24 (0.04) | 13086 (0.04) | | 512 | | 28091 |
| | | 0.14 (0.02) | | 11217 | 10705 | 24843 | |
| | | 0.38 (0.02) | | 1864 | 1353 | 3237 | |
| 4 | 600 | 0.23 (0.03) | 13069 (0.04) | | 511 | | 28065 |
| | | 0.15 (0.03) | | 11205 | 10694 | 24828 | |
| | | 0.40 (0.01) | | 1836 | 1336 | 3207 | |
| 5 | 600 | 0.35 (0.05) | 12926 (0.04) | | 500 | | 27831 |
| | | 0.15 (0.04) | | 11090 | 10590 | 24624 | |
| | | 0.40 (0.02) | | 1754 | 1278 | 2990 | |
| 6 | 300 | 0.22 (0.04) | 12435 (0.04) | | 476 | | 26866 |
| | | 0.19 (0.05) | | 10681 | 10205 | 23876 | |
| | | 0.39 (0.05) | | 1598 | 1156 | 2688 | |
| 7 | 300 | 0.19 (0.01) | 11565 (0.04) | | 442 | | 25397 |
| | | 0.20 (0.03) | | 9967 | 9525 | 22709 | |
| | | 0.40 (0.03) | | 1352 | 955 | 2271 | |
| 8 | 300 | 0.24 (0.06) | 10527 (0.04) | | 397 | | 23538 |
| | | 0.16 (0.03) | | 9175 | 8778 | 21267 | |
| | | 0.40 (0.03) | | 1057 | 719 | 1812 | |
| 9 | 300 | 0.23 (0.02) | 9026 (0.04) | | 338 | | 20855 |
| | | 0.18 (0.03) | | 7969 | 7631 | 19043 | |
| | | 0.42 (0.03) | | 828 | 550 | 1436 | |
| 10 | 300 | 0.23 (0.03) | 7636 (0.04) | | 278 | | 18166 |
| | | 0.19 (0.02) | | 6808 | 6530 | 16730 | |

[a] SVM learning and testing averaged over 3 runs using C=1.0 and linear kernel. NRE=non-redundant examples. OE=overlapping examples; RE=redundant examples. Plus and minus signs describe flexible and rigid examples respectively.

## Table A.III. Redundancy Analysis for Hydrophobicity[a]

| Number of neighbors | SVM Set Size | $\overline{\text{NME}}$ ($\sigma$) / $\overline{\text{NME}}_{+}$ ($\sigma$) / $\overline{\text{NME}}_{-}$ ($\sigma$) | Total non-redundant examples (NRE) (OE/NRE) | NRE+/ NRE- | NOE+/ OE/ NOE- | RE+/ RE- | Total redundant examples (TRE) |
|---|---|---|---|---|---|---|---|
| 1 | 60 | 0.39 (0.11) | 299 (0.38) | 119 | 5 | 3248 | 28091 |
| | | 0.20 (0.04) | | | 114 | | |
| | | 0.19 (0.09) | | 180 | 66 | 24843 | |
| 2 | 100 | 0.42 (0.02) | 1357 (0.26) | 410 | 55 | 3248 | 28091 |
| | | 0.19 (0.04) | | | 355 | | |
| | | 0.23 (0.05) | | 947 | 592 | 24843 | |
| 3 | 100 | 0.43 (0.02) | 3334 (0.16) | 757 | 215 | 3248 | 28091 |
| | | 0.19 (0.04) | | | 542 | | |
| | | 0.24 (0.06) | | 2577 | 2035 | 24843 | |
| 4 | 100 | 0.41 (0.05) | 5250 (0.10) | 963 | 419 | 3237 | 28065 |
| | | 0.22 (0.05) | | | 544 | | |
| | | 0.19 (0.05) | | 4287 | 3743 | 24828 | |
| 5 | 100 | 0.40 (0.06) | 6310 (0.08) | 1046 | 565 | 3207 | 27831 |
| | | 0.22 (0.04) | | | 481 | | |
| | | 0.18 (0.09) | | 5264 | 4783 | 24624 | |
| 6 | 100 | 0.40 (0.03) | 6676 (0.06) | 1048 | 618 | 2990 | 26866 |
| | | 0.21 (0.07) | | | 430 | | |
| | | 0.19 (0.05) | | 5628 | 5198 | 23876 | |
| 7 | 100 | 0.42 (0.04) | 6679 (0.06) | 986 | 603 | 2688 | 25397 |
| | | 0.25 (0.04) | | | 383 | | |
| | | 0.17 (0.06) | | 5693 | 5310 | 22709 | |
| 8 | 100 | 0.43 (0.06) | 6453 (0.05) | 906 | 562 | 2271 | 23538 |
| | | 0.24 (0.04) | | | 344 | | |
| | | 0.19 (0.05) | | 5547 | 5203 | 21267 | |
| 9 | 100 | 0.40 (0.06) | 5912 (0.05) | 755 | 479 | 1812 | 20855 |
| | | 0.24 (0.05) | | | 276 | | |
| | | 0.16 (0.01) | | 5157 | 4881 | 19043 | |
| 10 | 100 | 0.42 (0.04) | 5199 (0.04) | 605 | 388 | 1436 | 18166 |
| | | 0.21 (0.04) | | | 217 | | |
| | | 0.20 (0.02) | | 4594 | 4377 | 16730 | |

[a] SVM learning and testing averaged over 5 runs using C=1.0 and linear kernel. NRE=non-redundant examples. OE=overlapping examples; RE=redundant examples. Plus and minus signs describe flexible and rigid examples respectively.

## TABLE A.IV. REDUNDANCY ANALYSIS FOR SURFACE ACCESSIBLE AREA AND HYDROPHOBICITY[a]

| Number of neighbors | SVM Set Size | $\overline{\mathrm{NME}}$ ($\sigma$) <br> $\overline{\mathrm{NME}}_+$ ($\sigma$) <br> $\overline{\mathrm{NME}}_-$ ($\sigma$) | Total non-redundant examples (NRE) (OE/NRE) | NRE+/ NRE- | NOE+/ OE/ NOE- | RE+/ RE- | Total redundant examples (TRE) |
|---|---|---|---|---|---|---|---|
| 1 | 300 | 0.39 (0.01) | 12997 | 1866 | 1348 | 3248 | 28091 |
| | | 0.20 (0.01) | (0.04) | | 518 | | |
| | | 0.19 (0.02) | | 11131 | 10613 | 24843 | |
| 2 | 300 | 0.41 (0.01) | 13085 | 1869 | 1357 | 3248 | 28091 |
| | | 0.21 (0.03) | (0.04) | | 512 | | |
| | | 0.20 (0.02) | | 11216 | 10704 | 24843 | |
| 3 | 300 | 0.40 (0.02) | 13086 | 1869 | 1357 | 3248 | 28091 |
| | | 0.20 (0.02) | (0.04) | | 512 | | |
| | | 0.20 (0.02) | | 11217 | 10705 | 24843 | |
| 4 | 300 | 0.41 (0.03) | 13069 | 1864 | 1353 | 3237 | 28065 |
| | | 0.21 (0.03) | (0.04) | | 511 | | |
| | | 0.20 (0.04) | | 11205 | 10694 | 24828 | |
| 5 | 300 | 0.38 (0.03) | 12926 | 1836 | 1336 | 3207 | 27831 |
| | | 0.20 (0.02) | (0.04) | | 500 | | |
| | | 0.18 (0.03) | | 11090 | 10590 | 24624 | |
| 6 | 300 | 0.39 (0.01) | 12435 | 1754 | 1278 | 2990 | 26866 |
| | | 0.23 (0.04) | (0.04) | | 476 | | |
| | | 0.17 (0.03) | | 10681 | 10205 | 23876 | |
| 7 | 300 | 0.41 (0.04) | 11565 | 1598 | 1156 | 2688 | 25397 |
| | | 0.28 (0.09) | (0.04) | | 442 | | |
| | | 0.13 (0.08) | | 9967 | 9525 | 22709 | |
| 8 | 300 | 0.39 (0.02) | 10527 | 1352 | 955 | 2271 | 23538 |
| | | 0.19 (0.04) | (0.04) | | 397 | | |
| | | 0.20 (0.04) | | 9175 | 8778 | 21267 | |
| 9 | 300 | 0.41 (0.02) | 9026 | 1057 | 719 | 1812 | 20855 |
| | | 0.23 (0.03) | (0.04) | | 338 | | |
| | | 0.18 (0.02) | | 7969 | 7631 | 19043 | |
| 10 | 300 | 0.41 (0.03) | 7636 | 828 | 550 | 1436 | 18166 |
| | | 0.19 (0.03) | (0.04) | | 278 | | |
| | | 0.22 (0.03) | | 6808 | 6530 | 16730 | |

[a] SVM learning and testing averaged over 5 runs using C=1.0 and linear kernel. NRE=non-redundant examples. OE=overlapping examples; RE=redundant examples. Plus and minus signs describe flexible and rigid examples respectively.

## TABLE A.V. REDUNDANCY ANALYSIS FOR FLEXIBILITY AND SURFACE ACCESSIBLE AREA[a]

| Number of neighbors | SVM Set Size | $\overline{\text{NME}}$ $(\sigma)$<br>$\overline{\text{NME}}_+$ $(\sigma)$<br>$\overline{\text{NME}}_-$ $(\sigma)$ | Total non-redundant examples (NRE) (OE/NRE) | NRE+/<br>NRE- | NOE+/<br>OE/<br>NOE- | RE+/<br>RE- | Total redundant examples (TRE) |
|---|---|---|---|---|---|---|---|
| 1 | 200 | 0.31 (0.03)<br>0.23 (0.03)<br>0.08 (0.02) | 12997<br>(0.04) | 1866<br>11131 | 1348<br>518<br>10613 | 3248<br>24843 | 28091 |
| 2 | 200 | 0.32 (0.02)<br>0.23 (0.02)<br>0.08 (0.02) | 13085<br>(0.04) | 1869<br>11216 | 1357<br>512<br>10704 | 3248<br>24843 | 28091 |
| 3 | 200 | 0.31 (0.02)<br>0.22 (0.02)<br>0.09 (0.02) | 13086<br>(0.04) | 1869<br>11217 | 1357<br>512<br>10705 | 3248<br>24843 | 28091 |
| 4 | 200 | 0.32 (0.02)<br>0.24 (0.02)<br>0.08 (0.02) | 13069<br>(0.04) | 1864<br>11205 | 1353<br>511<br>10694 | 3237<br>24828 | 28065 |
| 5 | 200 | 0.30 (0.01)<br>0.22 (0.03)<br>0.08 (0.02) | 12926<br>(0.04) | 1836<br>11090 | 1336<br>500<br>10590 | 3207<br>24624 | 27831 |
| 6 | 200 | 0.32 (0.04)<br>0.20 (0.01)<br>0.12 (0.03) | 12435<br>(0.04) | 1754<br>10681 | 1278<br>476<br>10205 | 2990<br>23876 | 26866 |
| 7 | 200 | 0.33 (0.02)<br>0.21 (0.03)<br>0.12 (0.04) | 11565<br>(0.04) | 1598<br>9967 | 1156<br>442<br>9525 | 2688<br>22709 | 25397 |
| 8 | 200 | 0.32 (0.03)<br>0.20 (0.02)<br>0.12 (0.03) | 10527<br>(0.04) | 1352<br>9175 | 955<br>397<br>8778 | 2271<br>21267 | 23538 |
| 9 | 200 | 0.32 (0.04)<br>0.19 (0.02)<br>0.17 (0.04) | 9026<br>(0.04) | 1057<br>7969 | 719<br>338<br>7631 | 1812<br>19043 | 20855 |
| 10 | 200 | 0.33 (0.03)<br>0.20 (0.03)<br>0.12 (0.04) | 7636<br>(0.04) | 828<br>6808 | 550<br>278<br>6530 | 1436<br>16730 | 18166 |

[a] SVM learning and testing averaged over 5 runs using C=1.0 and linear kernel. NRE=non-redundant examples. OE=overlapping examples; RE=redundant examples. Plus and minus signs describe flexible and rigid examples respectively.

## TABLE A.VI. REDUNDANCY ANALYSIS FOR FLEXIBILITY AND HYDROPHOBICITY[a]

| Number of neighbors | SVM Set Size | $\overline{\mathrm{NME}}$ $(\sigma)$<br>$\overline{\mathrm{NME}}_+$ $(\sigma)$<br>$\overline{\mathrm{NME}}_-$ $(\sigma)$ | Total non-redundant examples (NRE) (OE/NRE) | NRE+/ NRE- | NOE+/ OE/ NOE- | RE+/ RE- | Total redundant examples (TRE) |
|---|---|---|---|---|---|---|---|
| 1 | 100 | 0.50 (0.03) | 315 (0.37) | 123 | 5 | 3248 | 28091 |
| | | 0.27 (0.05) | | | 118 | | |
| | | 0.23 (0.03) | | 192 | 74 | 24843 | |
| 2 | 200 | 0.43 (0.04) | 1416 (0.25) | 416 | 59 | 3248 | 28091 |
| | | 0.23 (0.06) | | | 357 | | |
| | | 0.23 (0.10) | | 1000 | 643 | 24843 | |
| 3 | 200 | 0.41 (0.04) | 3442 (0.16) | 761 | 220 | 3248 | 28091 |
| | | 0.17 (0.09) | | | 541 | | |
| | | 0.23 (0.09) | | 2681 | 2140 | 24843 | |
| 4 | 200 | 0.38 (0.05) | 5309 (0.10) | 965 | 428 | 3237 | 28065 |
| | | 0.17 (0.02) | | | 537 | | |
| | | 0.22 (0.05) | | 4344 | 3807 | 24828 | |
| 5 | 200 | 0.38 (0.08) | 6328 (0.08) | 1046 | 567 | 3207 | 27831 |
| | | 0.19 (0.06) | | | 479 | | |
| | | 0.19 (0.06) | | 5282 | 4803 | 24624 | |
| 6 | 200 | 0.37 (0.06) | 6683 (0.06) | 1048 | 619 | 2990 | 26866 |
| | | 0.18 (0.06) | | | 429 | | |
| | | 0.19 (0.04) | | 5635 | 5206 | 23876 | |
| 7 | 200 | 0.34 (0.06) | 6685 (0.06) | 987 | 603 | 2688 | 25397 |
| | | 0.18 (0.06) | | | 384 | | |
| | | 0.17 (0.03) | | 5698 | 5314 | 22709 | |
| 8 | 200 | 0.34 (0.04) | 6461 (0.05) | 906 | 562 | 2271 | 23538 |
| | | 0.17 (0.03) | | | 344 | | |
| | | 0.18 (0.04) | | 5555 | 5211 | 21267 | |
| 9 | 200 | 0.35 (0.05) | 5920 (0.05) | 755 | 479 | 1812 | 20855 |
| | | 0.15 (0.04) | | | 276 | | |
| | | 0.19 (0.03) | | 5165 | 4889 | 19043 | |
| 10 | 200 | 0.36 (0.04) | 5203 (0.04) | 605 | 388 | 1436 | 18166 |
| | | 0.17 (0.04) | | | 217 | | |
| | | 0.18 (0.03) | | 4598 | 4381 | 16730 | |

[a] SVM learning and testing averaged over 5 runs using C=1.0 and linear kernel. NRE=non-redundant examples. OE=overlapping examples; RE=redundant examples. Plus and minus signs describe flexible and rigid examples respectively.

### TABLE A.VII. REDUNDANCY ANALYSIS FOR FLEXIBILITY, HYDROPHOBICITY AND SURFACE ACCESSIBLE AREA[a]

| Number of neighbors | SVM Set Size | $\overline{\text{NME}}$ $(\sigma)$<br>$\overline{\text{NME}}_+$ $(\sigma)$<br>$\overline{\text{NME}}_-$ $(\sigma)$ | Total non-redundant examples (NRE) (OE/NRE) | NRE+/ NRE- | NOE+/ OE/ NOE- | RE+/ RE- | Total redundant examples (TRE) |
|---|---|---|---|---|---|---|---|
| 1 | 200 | 0.32 (0.02)<br>0.18 (0.04)<br>0.14 (0.05) | 12997<br>(0.04) | 1866<br>11131 | 1348<br>518<br>10613 | 3248<br>24843 | 28091 |
| 2 | 200 | 0.31 (0.04)<br>0.19 (0.03)<br>0.13 (0.03) | 13085<br>(0.04) | 1869<br>11216 | 1357<br>512<br>10704 | 3248<br>24843 | 28091 |
| 3 | 200 | 0.31 (0.02)<br>0.18 (0.02)<br>0.13 (0.03) | 13086<br>(0.04) | 1869<br>11217 | 1357<br>512<br>10705 | 2348<br>13086 | 28091 |
| 4 | 200 | 0.30 (0.02)<br>0.18 (0.02)<br>0.13 (0.03) | 13069<br>(0.04) | 1864<br>11205 | 1353<br>511<br>10694 | 3237<br>24828 | 28065 |
| 5 | 200 | 0.29 (0.02)<br>0.15 (0.02)<br>0.14 (0.03) | 12926<br>(0.04) | 1836<br>11090 | 1336<br>500<br>10590 | 3207<br>24624 | 27831 |
| 6 | 200 | 0.34 (0.03)<br>0.18 (0.03)<br>0.16 (0.02) | 12435<br>(0.04) | 1754<br>10681 | 1278<br>476<br>10205 | 2990<br>23876 | 26866 |
| 7 | 200 | 0.29 (0.05)<br>0.15 (0.03)<br>0.14 (0.03) | 11565<br>(0.04) | 1598<br>9967 | 1156<br>442<br>9525 | 2688<br>22709 | 25397 |
| 8 | 200 | 0.27 (0.01)<br>0.15 (0.02)<br>0.11 (0.02) | 10527<br>(0.04) | 1352<br>9175 | 955<br>397<br>8778 | 2271<br>21267 | 23538 |
| 9 | 200 | 0.30 (0.02)<br>0.17 (0.03)<br>0.13 (0.04) | 9026<br>(0.04) | 1057<br>7969 | 719<br>338<br>7631 | 1812<br>19043 | 20855 |
| 10 | 200 | 0.33 (0.05)<br>0.17 (0.02)<br>0.16 (0.03) | 7636<br>(0.04) | 828<br>6808 | 550<br>278<br>6530 | 1436<br>16730 | 18166 |

[a] SVM learning and testing averaged over 5 runs using C=1.0 and linear kernel. NRE=non-redundant examples. OE=overlapping examples; RE=redundant examples. Plus and minus signs describe flexible and rigid examples respectively.

# APPENDIX B. STATEMENT OF COLLABORATORS CONTRIBUTIONS TO THE PRESENT WORK

The work presented chapter 2 (Najmanovich et al., 2000a) form the basis for the type of analysis used in subsequent studies collaboration with Eran Eyal. The results of this collaboration are presented in sub-section 2.4.6 on alternative side chain flexibility scales has been performed in collaboration with Eran Eyal (Eyal et al., 2001; Eyal et al., 2003a; Eyal et al., 2003b). Eran Eyal has also contributed in the creation of the program that generates the structural superimposition of apo and holo protein pairs used by LigProt.

The work present in section 4.3 on the utilization of perceptron learning for the generation of pairwise contact potentials for protein folding (Vendruscolo et al., 1999; Vendruscolo et al., 2000) has been performed in collaboration with Michelle Vendruscolo and Eytan Domany from the Department of Physics of Complex Systems, weizmann Institute of Science.

# APPENDIX C. LIST OF PUBLICATIONS ASSOCIATED TO THE PRESENT WORK

Eyal, E., **Najmanovich, R.**, Edelman, M. & Sobolev, V. (2003a). Protein side-chain rearrangement in regions of point mutations. *Proteins* 50, 272-82.

Eyal, E., **Najmanovich, R.**, Sobolev, V. & Edelman, M. (2001). MutaProt: a web interface for structural analysis of point mutations. *Bioinformatics* 17, 381-2.

Eyal, E., **Najmanovich, R.**, Sobolev, V. & Edelman, M. (2003b). Importance of solvent accessibility and contact surfaces in modeling side chain conformations in proteins. *PROTEINS* submitted.

**Najmanovich, R.**, Kuttner, J., Sobolev, V. & Edelman, M. (2000). Side-chain flexibility in proteins upon ligand binding. *PROTEINS: Structure, Function and Genetics* 39, 261-268.

Vendruscolo, M., **Najmanovich, R.** & Domany, E. (1999). Protein folding in contact map space. *Physical Review Letters* 82, 656-659.

Vendruscolo, M., **Najmanovich, R.** & Domany, E. (2000). Can a pairwise contact potential stabilize native protein folds against decoys obtained by threading? *Proteins* 38, 134-48.

פונקצית ההערכה (scoring function) של FlexAID מתבססת על התאמת משטחי פנים והתאמה כימית של הליגנד והחלבון. נוסף להם בעבודה זו ביטוי המעניש אינטראקציות עם הממס. ביטוי זה שיפר משמעותית את אחוזי ההצלחה של התוכנית, דבר שייתכן וקשור לדינמיות של אלגוריתם החיפוש הגנטי. שיפורים עתידיים נוספים בפונקצית ההערכה מחויבים לצורך הוספת יכולת גמישות לליגנד (flexible docking).

## תקציר

שינויי קונפורמציה של חומצות אמיניות הם אחד המאפיינים של של קישור ליגנדים לחלבונים ומהווים חלק מהתופעה הכללית של "התאמה מושרית" (induced fit) . על מנת לקבוע את ההיקף של שינויים כאלה בעת קישור ליגנדים בוצע ניתוח סטטיסטי של בסיס הנתונים אודות מבני חלבונים (PDB). בסיס נתונים שניוני, המורכב מזוגות של קבצים הכוללים מבנים של חלבונים בנוכחות ובהעדר הליגנד, נבנה. הניתוח הסטאטיסטי שבוצע על בסיס נתונים זה, הראה שב 40% מאתרי הקישור לא מתרחשים כלל שינויי קונפורמציה משמעותיים של שיירי חומצות אמיניות. ב 85% מהמקרים מתרשים עד 3 שינויי קונפורמציה. באנליזה של הנטייה של סוגים שונים של חומצות אמיניות לשנות קונפורמציה, נמצאו שינויים משמעותיים בין חומצות אמיניות.

סקלת הגמישות של חומצות אמיניות שונות לא מאפשרת עדיין לתת חיזוי מדויק מספיק לגבי זהות שיירי החומצות האמיניות שישנו את מיקומם בעת קישור הליגנד. חיזוי כזה חשוב מאוד לצורך קביעה מדויקת של אופן ומיקום הקישור של ליגנדים בעת חיזוי קישור ליגנד (molecular docking ).

לצורך שיפור החיזוי בוצע שימוש ב support vector machine - שיטה מתקדמת של אינטלגנציה מלאכותית, על מנת לסווג חומצות כגמישות או קבועות. מלבד סקלת הגמישות שימשו לצורך הסיווג גם נתונים לגבי הנגישות לממס (solvent accessible area) של כל חומצה אמינית. וכן נתונים לגבי הגמישות ונגישות לממס של חומצות אמיניות שכנות. אחוז סווג של כ 70% הושג באמצעות השיטה.

העובדה שאחוז קטן בלבד מהחומצות האמיניות באזור אתר הקישור משנות קונפורמציה בעת קישור הליגנד מאפשרת לשלב בדיקה של שינויי קונפורמציה בעת חיזוי קישור ליגנד (molecular docking). אלגוריתם כזה פותח המשלב חיזוי של שינויי קונפורמציה של מספר מוגבל של חומצות אמיניות באתר הקישור ביחד עם חיזוי המיקום המדויק של הליגנד. החיפוש במרחב הפתרונות משלב אלגוריתם גנטי (Genetic algorithm) לצורך קביעת מיקום הליגנד עם חיפוש מקיף (exhaustive search) עבור הקונפורמציות של שיירי החומצות האמיניות שהוגדרו כגמישות. האלגוריתם כולו ממומש בכלי תוכנה מיוחד שנבנה – FlexAID. FlexAID מאפשר לבצע חיפוש מקומי (local docking) באתר קישור מוגדר מראש, כמו גם חיפוש גלובאלי (global docking) כאשר אין מידע מקדים לגבי אתר הקישור של הליגנד על פני החלבון. FlexAID משתמשת בספריית רוטמרים (rotamer library) ליצירת קונפורמציות אלטרנטיביות עבור החומצות האמיניות שהוגדרו כגמישות. ביצועי FlexAID עבור חיזוי קישור ליגנדים קשיחים (rigid docking) נבדקו ונמצא כי אחוז ההצלחה הוא 70-80% בחיפוש מקומי וגלובאלי כאחד. ביצועי FlexAID לא הוערכו עדיין בקנה מידה רחב כאשר שינויי קונפורמציה של חומצות אמיניות משולבים. עבור מקרים ספורים שנבדקו הן בחיפוש מקומי והן בגלובאלי, התוצאות מבטיחות.

מכון ויצמן למדע

# גמישות חלבונים בעת קישור ליגנדים: חזוים ואנליזה סט

חבור לשם קבלת התואר
**דוקטור לפילוסופיה**

מאת
רפאל נחמנוביץ